\documentclass[11pt]{article}
\usepackage{amsmath,amsthm,comment,amssymb}
\usepackage{mathrsfs}
\usepackage{amsfonts}
\usepackage{graphicx}
\usepackage{multirow}
\usepackage{booktabs}
\usepackage{bbm}
\usepackage{enumitem}
\usepackage[top=1in, bottom=1in, left=1in, right=1in]{geometry}
\usepackage{leftindex}
\usepackage{dsfont}
\usepackage{bm}
\usepackage{color}
\usepackage{xcolor}
\usepackage{subcaption}
\usepackage{caption}
\usepackage[colorlinks=true,citecolor=blue,linkcolor=red,urlcolor=blue]{hyperref}
\usepackage[round,authoryear]{natbib}
\usepackage{indentfirst}
\usepackage{algorithm}
\usepackage{algorithmic}
\allowdisplaybreaks
\usepackage{appendix}
\usepackage{tikz}
\usepackage{caption}
\numberwithin{equation}{section}
\newtheorem{theorem}{Theorem}[section]

\newtheorem{lemma}{Lemma}[section]
\newtheorem{proposition}{Proposition}[section]
\theoremstyle{definition}

\newtheorem{remark}{Remark}[section]
\newtheorem{assumption}{\textbf{Assumption}}[section]

\newcommand{\var}{\mathrm{VaR}}
\newcommand{\covar}{\mathrm{CoVaR}}
\newcommand{\es}{\mathrm{ES}}
\newcommand{\coes}{\mathrm{CoES}}
\newcommand{\bE}{\mathbb{E}}

\newcommand{\bP}{\mathbb{P}}
\newcommand{\bbR}{\mathbb{R}}

\usepackage{xr}
\begin{document}

\title{Nonparametric Inference for Extreme CoVaR and CoES}
\author{
	Qingzhao Zhong\footnote{\scriptsize School of Data Science, Fudan University,
	Email: {\color{blue}qzzhong22@m.fudan.edu.cn}},\quad
	Yanxi Hou\footnote{\scriptsize School of Data Science, Fudan University,
	Corresponding author, Email: {\color{blue}yxhou@fudan.edu.cn}}
}
\date{\today}
\maketitle

\begin{abstract}
Systemic risk measures quantify the potential risk to an individual financial constituent arising from the distress of entire financial system. As a generalization of two widely applied risk measures, Value-at-Risk and Expected Shortfall, the Conditional Value-at-Risk (CoVaR) and Conditional Expected Shortfall (CoES) have recently been receiving growing attention on applications in economics and finance, since they serve as crucial metrics for systemic risk measurement. However, existing approaches confront some challenges in  statistical inference and asymptotic theories when estimating CoES, particularly at high risk levels. In this paper, within a framework of upper tail dependence, we propose several extrapolative methods to estimate both extreme CoVaR and CoES nonparametrically via an adjustment factor, which are intimately related to the nonparametric modelling of the tail dependence function. In addition, we study the asymptotic theories of all proposed extrapolative methods based on multivariate extreme value theory. Finally, some simulations and real data analyses are conducted to demonstrate the empirical performances of our methods. 
\end{abstract}

\section{Introduction}\label{sec:intro}

Quantifying systemic risk has become a central imperative for financial and insurance regulators since the 2008 global financial crisis and the 2020 COVID-19 shock.  Conventional univariate risk measures, such as  Value-at-Risk (VaR) and Expected Shortfall (ES), fail to capture cross-institutional spillovers, prompting a surge of proposals for systemic-risk measures.  Among them, the Conditional VaR (CoVaR) proposed by \cite{Adrian2016} has gained prominence: it is the quantile of an institution's loss distribution conditional on the financial system being at its own VaR.  \cite{Huang2018} extend this logic to the Conditional ES (CoES), defined as the expected institution-level loss given that (i) the system is at its VaR and (ii) the institution's loss already exceeds its CoVaR. A common issue for CoVaR and CoES is that both condition on the zero-probability event “systemic loss equals its VaR”, which complicates statistical inference methods.

Alternative definitions of CoVaR and CoES by conditioning on the event where the system loss exceeds its VaR, are proposed in \cite{Girardi2013} and \cite{Nolde2020}, respectively. This paper proceeds with CoVaR and CoES defined in \cite{Girardi2013} and \cite{Nolde2020}.
Specifically, suppose $X$ stands for the individual loss variable of interest and $Y$ is the loss variable of the financial system (or market). Denote $F_{X,Y}$ as their joint distribution function, and by $F_X(\cdot)$ and $F_Y(\cdot)$ the respective marginals. For a given (right-tail) risk level $\tau \in (0,1)$, the $\covar_{X|Y}(\tau)$ is defined by solving
\begin{equation}\label{eq:covar_X}
\bP(X \ge \covar_{X|Y}(\tau)|Y \ge \var_Y(\tau)) =  1 - \tau,
\end{equation}
where $\var_Y(\tau)$ defines the VaR of $Y$ as
\begin{equation}\label{eq:var_Y}
\var_Y(\tau) := \inf\{ y \in \bbR: F_Y(y) \ge \tau \},
\end{equation}
and $\var_X(\tau)$ can be defined in a similar way. We use subscript ``$X|Y$" to characterize the conditional order between $X$ and $Y$. It is readily to see that $\covar_{X|Y}(\tau)$ is the $\tau$-quantile of the conditional distribution $F_{X|Y \ge \var_Y(\tau) }(\cdot) = \bP(X \leq \,\cdot \, | Y \ge \var_Y(\tau))$. Then, the $\coes_{X|Y}(\tau)$ can be defined via
\begin{equation}\label{eq:coes_X}
\coes_{X|Y}(\tau) := \frac{1}{1-\tau} \int_{\tau}^{1} \covar_{X|Y}(\alpha) \,d \alpha.
\end{equation}
The assumed continuity of $F_{X,Y}$ implies that $\coes_{X|Y}(\tau)$ coincides with the following conditional expectation
\begin{equation}\label{eq:coes_X_exp}
\coes_{X|Y}(\tau) = \bE[X|X \ge \covar_{X|Y}(\tau), Y \ge \var_Y(\tau)].
\end{equation}
It is easy to check $\covar_{X|Y}(\tau) = \var_X(\tau)$ and $\coes_{X|Y}(\tau) = \es_{X}(\tau)$ if $X$ is independent of $Y$.

To perform inference on CoVaR and CoES, most existing work still relies on parametric assumptions.
\cite{Capponi2022} derives a closed-form expression for CoES under a multivariate Student-$t$ distribution, while \cite{Francq2025} obtain dynamic CoVaR estimates from location-scale models by computing residual quantiles. Outside a parametric model the situation changes: CoVaR and CoES are known to be \emph{non-elicitable}, i.e., neither of them can be written as the unique minimiser of an expected scoring function. This fact blocks a similar  M-estimation approach route as the standard ``quantile/expectile'' estimator for VaR and ES. \cite{Fissler2024} addresses this issue by introducing \emph{multi-objective elicitability}, where the triple (VaR,CoVaR,CoES) can be jointly elicited with a single vector scoring rule, enabling a one-step M-estimator. Yet the joint estimation imposes stringent statistical requirements, and a simpler multi-step route---first VaR, then CoVaR, finally CoES, each using the previous estimate---remains attractive in practice. Moreover, while early work on CoVaR and CoES focused on a fixed risk level, recent research has shifted to estimating these measures at extreme levels. Building on the tail-dependence literature, \cite{Cai2015} proposes an extrapolation estimator for Marginal Expected Shortfall (MES) when the probability level tends to one; MES, introduced by \cite{Acharya2017}, is the expected loss of an institution given that the system return suffers an extreme loss. Similarly, \cite{Nolde2020} develops a semi-parametric framework for extreme CoVaR that embeds extreme-value theory in a bivariate GARCH model: the conditional probability in \eqref{eq:covar_X} is approximated by the tail of a skew-elliptical distribution whose parameters are updated dynamically. This approach, however, carries two strong restrictions: (i) the vector $(X,Y)$ must be multivariate regularly varying with identical tail indices for $X$ and $Y$, and (ii) the dependence structure is fully dictated by the skew-elliptical family.

Recently, \cite{Nolde2022} proposed a semi-parametric method for extreme CoVaR that exploits tail dependence without invoking the multi-objective elicitability of \cite{Fissler2024}.
They introduce an adjustment factor $\eta_{\tau}$ (see \eqref{eq:adj_factor}) that links the extreme $\covar_{X|Y}(\tau)$ to the marginal $\var_{X}(\tau)$ via
\begin{equation}\label{eq:covar_var}
\covar_{X|Y}(\tau)=\var_X\bigl(1-(1-\tau)\eta_{\tau}\bigr).
\end{equation}
This one-to-one mapping reduces extrapolation to estimating a high quantile of the marginal distribution of $X$.
Although the theory only requires upper-tail regular variation of $X$, which is weaker than the joint regular variation assumed in \cite{Nolde2020}, the procedure still hinges on a correctly specified parametric tail-dependence model, leaving it exposed to misspecification. More importantly, neither \cite{Nolde2020} nor \cite{Nolde2022} provides asymptotic normality for the estimators, even though \cite{Nolde2020} provides consistency.
As a result, extreme CoES remains an open theoretical challenge: no analogue of $\eta_{\tau}$ exists for CoES, and its uncertainty has yet to be quantified.

Motivated by the limitations of existing methods, this paper extends the upper-tail-dependence framework previously used by \cite{Cai2015} and \cite{Nolde2022} to provide fully nonparametric inference on extreme CoVaR and CoES. We first establish existence and uniqueness of the adjustment factor and propose two nonparametric estimators at an intermediate level: one using empirical distributions and the other using rank statistics, both proven to be consistent and asymptotically normal.
Next, we derive the first asymptotic results for intermediate CoVaR and CoES, and then introduce two extrapolation approaches for extreme CoVaR and CoES: one based on the estimated adjustment factor and one directly extending the intermediate estimates, without imposing parametric assumptions on either margins or dependence. Simulations confirm that the methods are robust and perform well in finite samples, offering reliable tools for empirical applications.

We structure the remainder of this paper as follows. In Section \ref{sec:meth}, we first present the basic description of the framework of tail dependence and set down the fundamental assumptions that we use throughout this paper. We next provide the theoretical justifications for adjustment factor and several extrapolative estimators for $\covar_{X|Y}(\tau'_n)$ and $\coes_{X|Y}(\tau'_n)$ via distinct nonparametric estimations of the adjustment factor and intermediate CoVaR, CoES, in Sections \ref{sec:af}, \ref{sec:ext_af} and \ref{sec:ext_int}. The corresponding asymptotic theories are established therein. Section \ref{sec:simulation} contains a simulation study to compare the empirical performances of different extrapolations, and Section \ref{sec:realanalysis} is a real data analysis that illustrates CoVaR and CoES as systemic risk measures in a real-world context. The supplementary materials collect some additional simulated and empirical results and all the theoretical proofs involved in this article.

\section{Methodologies}\label{sec:meth}


Let $(X,Y)$ be a two-dimensional random vector with joint distribution $F_{X,Y}$, with marginal distributions  $F_X(\cdot)=F_{X,Y}(\cdot,\infty)$ and $F_Y(\cdot)=F_{X,Y}(\infty,\cdot)$. Denote $\overline{F}_X(\cdot):=1-F_X(\cdot)$ and $\overline{F}_Y(\cdot):=1-F_Y(\cdot)$ as the survival distribution of $F_X$ and $F_Y$, respectively. Suppose the bivariate samples $\{(X_i,Y_i)\}_{i=1}^n$ are independent and identically distributed generated from the distribution $F_{X,Y}$. We further denote the empirical distributions for $X_1,...,X_n$ and $Y_1,...,Y_n$ as 
\begin{equation*}
\widehat{F}_X(s) := \frac{1}{n} \sum_{i=1}^{n} I(X_i \leq s)\quad\text{and}\quad \widehat{F}_Y(t) := \frac{1}{n} \sum_{i=1}^{n} I(Y_i \leq t).
\end{equation*}
Define $k=k(n)$ as an intermediate sequence of integers, such that $k \to \infty$ and $k/n \to 0$ as $n \to \infty$. Let $A^{\top}$ be the transpose of a matrix or vector $A$. 

Our aim is to establish the nonparametric statistical methodologies for estimating $\covar_{X|Y}(\tau'_n)$ and $\coes_{X|Y}(\tau'_n)$, when the confidence level $\tau'_n$ is \emph{extreme}, that is, $\tau'_n \to 1$ and $n(1-\tau'_n) \to c \in [0,\infty)$ as $n\to\infty$. 
It is necessary to consider the modelling conditions for both the tails of marginal distribution and the tail dependence structure.
From a practical perspective, as we expect that high values of $Y$ correspond to high values of $X$, it is reasonable to impose assumption on the right-hand upper tail dependence of $(X,Y)$ to deal with the tail event. Specifically, we suppose that, for all $(x,y) \in [0,\infty)^2/\{(0,0)\}$, the following limit exists,
\begin{equation}\label{eq:rutd}
\lim_{t \to \infty} t \bP(\overline{F}_X(X) \leq x /t, \overline{F}_Y(Y) \leq y/t) =: R(x,y).
\end{equation}
The bivariate function $R$ is called {\it tail copula}, which characterizes the tail dependence structure of $F_{X,Y}$;
see \cite{Drees1998}, \cite{Beirlant2006} and \cite{Cai2015} for details. Besides, $R$ also share some desirable properties, such as monotonicity, continuity, homogeneity, and etc. We summarize these properties of $R$ in Proposition A.1 of the supplementary material. 
Moreover, denote
\begin{equation*}
R_1(x,y) = \frac{\partial}{\partial x} R(x,y), ~\text{and}~ R_2(x,y) = \frac{\partial}{\partial y} R(x,y).
\end{equation*}
as the partial derivatives of $R(x,y)$.

One the other hand, for the marginal distribution, we assume $X$ follows a distribution with a heavy righted-hand tail: there exists a $\gamma_1 > 0$ such that, for $x > 0$,
\begin{equation}\label{eq:rv_UX}
\lim_{t \to \infty} \frac{U_1(tx)}{U_1(t)} = x^{\gamma_1},
\end{equation}
where $U_1 = (1/(\overline{F}_X))^{-1}$ serves as the left-continuous inverse of $1/\overline{F}_X(\cdot)$, (\emph{that is}, tail quantile function). This implies that $\overline{F}_X(\cdot)$ is regularly varying with index $-1/\gamma_1$,
\begin{equation*}
\lim_{t \to \infty}\frac{\overline{F}_X(tx)}{\overline{F}_X(t)} = x^{-1/\gamma_1},
\end{equation*}
and $\gamma_1$ is the \emph{extreme value index} (EVI); see \cite{dehaan2006}.

To study the asymptotic properties of statistical inference methods, it is essential to consider the second-order regular variation conditions. Hence, we first list some conditions which are crucial for establishing the corresponding asymptotic properties.
\begin{assumption}\label{ass:conditions}
~
\begin{enumerate}[label=(\alph*)]
  \item The partial derivative $R_i(x,y)$ for $i=1,2$ are continuous with respect to $x$ in the neighborhood of $(0,1)$ and $R_i(0,1)>0$.
  \item There exist $\beta > \gamma_1$ and $\alpha > 1$ such that, as $t \to \infty$
  \begin{equation*}
  \sup_{\substack{0<x<\infty \\ 1/2 \leq y \leq 2}} \frac{| t\bP(\overline{F}_X(X) \leq x/t, \overline{F}_Y(Y) \leq y/t) - R(x,y)|}{x^{\beta} \wedge 1} = O(t^{-\alpha}).
  \end{equation*}
  \item There exist $\rho_1 < 0$ and an eventually positive or negative function $A_1$ such that, for all $x > 0$,
  \begin{equation*}
  \lim_{t \to \infty} \frac{\frac{U_1(tx)}{U_1(t)} - x^{\gamma_1}}{A_1(t)} = x^{\gamma_1} \frac{x^{\rho_1}-1}{\rho_1},
  \end{equation*}
  where auxiliary function $A_1$ is also regularly varying with index $\rho_1$, \emph{that is}, $A_1(tx)/A_1(t) \to x^{\rho_1}$ as $t \to \infty$ and $\sqrt{k}A_1(n/k) \to \lambda_1 \in \bbR$ as $n \to \infty$.
  \item As $n \to \infty$, $k = O(n^{\iota})$ with $\frac{2}{3} < \iota < \frac{2(\alpha+\beta)}{2(\alpha + \beta) + 1}$.
  \item Let $d_n = k/(n(1-\tau'_n))$. As $n \to \infty$, $\sqrt{k}/\log d_n \to \infty$ and $\frac{k}{n} \log d_n \to 0$ when $n(1-\tau'_n) \to 0$.
\end{enumerate}
\end{assumption}

Assumption \ref{ass:conditions}(a) is a necessary smoothness assumption of the tail dependence function $R$. Assumptions \ref{ass:conditions}(b) and (c) are second-order strengthenings of \eqref{eq:rutd} and \eqref{eq:rv_UX}, respectively (see, for example, conditions (7.2.8) and (3.2.4) in \cite{dehaan2006}), which quantify the rates of convergence in \eqref{eq:rutd} and \eqref{eq:rv_UX}. We further require conditions (d) and (e) on the intermediate sequence $k$ and extreme level $\tau'_n$. It is worth noting that, on the one hand, the condition $\alpha > 1$ in condition (b) is essential and simulations show that it leads to poor distributional performance when violating this condtion; on the other hand, we relax the upper bound on $\iota$ to $\frac{2(\alpha+\beta)}{2(\alpha+\beta) + 1}$, while previous literature typically constrain it as $\iota \leq 2\alpha/(2\alpha + 1)$, for example, see \cite{Einmahl2006,Einmahl2008,Einmahl2012}, etc.

The following proposition shows the limiting behavior between $\covar_{X|Y}(\tau)$ and $\coes_{X|Y}(\tau)$ as $\tau \to 1$ and provides an important relationship to obtain the extrapolative approaches for $\coes_{X|Y}(\tau)$ in the following sections.
\begin{proposition}\label{pro:coes_covar}
Under \eqref{eq:rutd}, \eqref{eq:rv_UX}, and Assumption \ref{ass:conditions}(a), we have that for $\gamma_1 \in (0,1)$,
\begin{equation}\label{eq:coes_covar}
\lim_{\tau \uparrow 1} \frac{\coes_{X|Y}(\tau)}{\covar_{X|Y}(\tau)} = \frac{1}{1-\gamma_1}.
\end{equation}
\end{proposition}
The above proposition only requires relatively mild conditions to hold, including first-order conditions for tail dependence function $R$ and regular variation, as well as smoothness on $R$. 

\subsection{On adjustment factor}\label{sec:af}

As mentioned above, to estimate $\covar_{X|Y}(\tau)$ with $\tau$ tending to 1, \cite{Nolde2022} proposes an approach via an adjustment factor $\eta_\tau$ associated with $\tau$, given as \eqref{eq:covar_var}. Thus, $\covar_{X|Y}(\tau)$ intuitively becomes a quantile of the unconditional distribution of $X$ with the adjusted quantile level $1-(1-\tau)\eta_\tau$. The adjust factor admits a closed expression as,
\begin{equation}\label{eq:adj_factor}
\eta_\tau: = \frac{\bP(X \ge \covar_{X|Y}(\tau))}{\bP(X \ge \covar_{X|Y}(\tau) | Y \ge \var_Y(\tau))},
\end{equation}
provided that the marginal $F_X$ is continuous and strictly monotonic.
However, \cite{Nolde2022} assumes that $X$ and $Y$ exhibit positive quadrant dependence (PQD), i.e., $\bP(X \ge x, Y \ge y) > \bP(X \ge x) \bP(Y \ge y)$ for any $x,y \in \bbR$. This is generally regarded as a rather strong condition, especially since our primary interest lies in tail modeling. Moreover, \cite{Nolde2022} failed to validate $\eta_\tau$ in \eqref{eq:covar_var}, such as existence and uniqueness, leaving the statistical methodologies unguaranteed. Building on exploring the theoretical underpinnings for $\eta_\tau$, it is sufficient to relax the PQD condition by instead considering $\bP(X \ge \var_X(\tau), Y \ge \var_Y(\tau)) > (1-\tau)^2$ and constrain the analysis on a tail region $\tau \in (\tau_0,1)$, rather than $(0,1)$. Therefore, the extrapolations for extreme CoVaR and CoES can be developed based on the validation of $\eta_\tau$.


Denote $\var_X(1-)$ and as $\var_Y(1-)$ the right-endpoints of $X$ and $Y$, respectively, $\mathbb D_{X,Y}(\tau):=[\var_X(\tau),\var_X(1-)]\times[\var_Y(\tau),\var_Y(1-)]$, and $\mathbb D_{X}(\tau):=[\var_X(\tau),\var_X(1-)]$. The following mild assumption can be satisfied by most meaningful models used in practice.

\begin{assumption}\label{ass:ext_uni}
There exists a $\tau_0\in(0,1)$ such that
\begin{enumerate}[label=(\alph*)]
   \item $(X,Y)$ is absolutely continuously distributed with joint density function $f_{X,Y}(\cdot)$ on $\mathbb D_{X,Y}(\tau_0)$;
   \item $X$ has strictly increasing distribution function $F_X(\cdot)$ on $\mathbb D_X(\tau_0)$;
   \item $\covar_{X|Y}(\tau) < \var_X(1-)$ for all $\tau \in (\tau_0,1)$.
\end{enumerate}
\end{assumption}

\begin{proposition}[Existence and uniqueness]\label{pro:ext_uni}
  Under Assumption \ref{ass:ext_uni} (a) and (b), for all $\tau \in (\tau_0,1)$, $\covar_{X|Y}(\tau) > \var_X(\tau)$ if and only if $\bP(X \ge \var_X(\tau), Y \ge \var_Y(\tau)) > (1-\tau)^2$. Moreover, if Assumption \ref{ass:ext_uni} (c) is further satisfied, then there exists unique $\eta_\tau \in (0,1)$ such that \eqref{eq:covar_var} holds if and only if $\covar_{X|Y}(\tau) > \var_X(\tau)$, for all $\tau \in (\tau_0,1)$.
\end{proposition}

\begin{remark}
Given $\tau\in(\tau_0,1)$, it is readily to check that $\eta_\tau \equiv 1$ if and only if $\bP(X \ge \var_X(\tau), Y \ge \var_Y(\tau)) = (1-\tau)^2$, in which case $\covar_{X|Y}(\tau) = \var_X(\tau)$ always holds. Note also that the case when $\bP(X \ge \var_X(\tau), Y \ge \var_Y(\tau)) < (1-\tau)^2$ is incompatible with tail dependence, i.e.
\begin{equation*}
t \bP(\overline{F}_X(X) \leq x /t, \overline{F}_Y(Y) \leq y/t) = t \bP(X \ge \var_X(1-x /t), Y \ge \var_Y(1- y/t)) < t \cdot \frac{x}{t} \cdot \frac{y}{t} \to 0,
\end{equation*}
as $t$ diverges, implying $R(x,y) \equiv 0$ for any $x,y \in \mathbb{R}$, which contradicts with \eqref{eq:rutd}.
\end{remark}


In the framework of tail dependence \eqref{eq:rutd}, we can assert that $\eta_\tau \in (0,1)$ as soon as the tail probability $\bP(X \ge \var_X(\tau), Y \ge \var_Y(\tau))>(1-\tau)^2$. This obviously implies that the level $1-(1-\tau)\eta_\tau$ is higher than $\tau$. Using definitions \eqref{eq:covar_X} and \eqref{eq:covar_var}, we further observe,
\begin{equation}\label{eq:prob_eta}
1-\tau = \frac{\bP(X \ge \covar_{X|Y}(\tau), Y \ge \var_Y(\tau))}{1-\tau} = \frac{ \bP( \overline{F}_X(X) \leq (1-\tau)\eta_\tau, \overline{F}_Y(Y) \leq  1-\tau ) }{1-\tau}.
\end{equation}
From a limiting viewpoint, as $\tau \to 1$, both \eqref{eq:rutd} and \eqref{eq:prob_eta} suggest a possibility of approximating the true adjustment factor $\eta_\tau$ with an approximation, denoted $\eta_\tau^*$, which is defined implicitly via
\begin{equation}\label{eq:solve_eta}
R(\eta_\tau^*,1) = 1-\tau.
\end{equation}
The limit relationship between $\eta_\tau$ and $\eta^*_\tau$ can be guaranteed by conditions (a) and (b) in Assumption \ref{ass:conditions}. We report it in the following lemma without proof as it is analogous to that of Lemmas S1.1 and S1.2 in \cite{Nolde2022}.
\begin{lemma}\label{lem:rela_eta}
Under Assumptions \ref{ass:conditions}(a) - (b) and \ref{ass:ext_uni}, we have that as $\tau \to 1$,
\begin{equation*}
\frac{\eta^*_{\tau}}{\eta_\tau} \to 1, ~ \text{and} ~ \frac{1-\tau}{\eta^*_\tau} \to R_1(0,1).
\end{equation*}
\end{lemma}

In other words, as $\tau \to 1$, Assumptions \ref{ass:conditions}(a) and (b) ensure that $\eta^*_\tau \to 0$ with the same speed as $1-\tau$. Moreover, as the bivariate function $R$ is monotonically non-decreasing in both arguments, $R(\eta,1)$ is increasing from 0 to $R(1,1)$ with $\eta$ varying from 0 to 1. Thus, there exists a unique solution $\eta^*_\tau$ to \eqref{eq:solve_eta} provided that $1-\tau < R(1,1)$, which holds for $\tau$ close to 1. Hence, an estimator of $\eta^*_\tau$ with statistical guarantees will serve as a good estimator for $\eta_\tau$ as well.

\subsection{Extrapolative estimations based on adjustment factor}\label{sec:ext_af}

We then propose nonparametric estimation methods for $\covar_{X|Y}(\tau'_n)$ via \eqref{eq:covar_var}. Notice first that the  $1-(1-\tau'_n)\eta_{\tau'_n}$ represents a more extreme level than $\tau'_n$, and hence by the regular variation \eqref{eq:rv_UX}, we have that
\begin{equation*}
\frac{\covar_{X|Y}(\tau'_n)}{\var_X(1-k/n)} = \frac{\var_X(1 - (1-\tau'_n)\eta_{\tau'_n})}{\var_X(1-k/n)} \sim \left( \frac{k}{n(1-\tau'_n)}  \right)^{\gamma_1} \eta_{\tau'_n}^{-\gamma_1},
\end{equation*}
which implies that
\begin{equation}\label{eq:extra_covar}
\covar_{X|Y}(\tau'_n) \sim \left( \frac{k}{n(1-\tau'_n)} \right)^{\gamma_1} \eta_{\tau'_n}^{-\gamma_1} \var_X(1-k/n).
\end{equation}
Recall that $k$ is an intermediate order sequence satisfying $k \to \infty$ and $k/n \to 0$ as $n\to\infty$. From \eqref{eq:extra_covar}, there are three quantities, $\gamma_1$, $\var_X(1-k/n)$ and $\eta_{\tau'_n}$, which are suppposed to estimate. Throughout this paper, the estimations of $\var_X(1-k/n)$ and $\gamma_1$ will be chosen as the $(n-k)$-th order statistic of $X_1,...,X_n$ and the Hill estimator (\cite{Hill1975}) such that
\begin{equation}\label{eq:int_varX}
\widehat{\var}_X(1-k/n):=X_{n-k,n},
\end{equation}
and
\begin{equation}\label{eq:Hill}
\hat{\gamma}_1 = \frac{1}{k} \sum_{i=1}^{k} \log X_{n-i+1,n} - \log X_{n-k,n},
\end{equation}
where $X_{1,n}\le\ldots\le X_{n,n}$ are the order statistics of samples $X_1,...,X_n$. The asymptotic behaviors of $\widehat{\var}_X(1-k/n)$ and $\hat\gamma_1$ have been well-studied in Theorems 2.4.8 and 3.2.5 of \cite{dehaan2006}, respectively. In order to be compatible with our framework, we alternatively provide a tail-dependence-based version of asymptotic normality for $\widehat{\var}_X(1-k/n)$ in Lemma A.3 of the supplementary material. 


Then, it remains to study the estimation of $\eta_{\tau'_n}$. Given an intermediate level $1-k/n$, following Lemma \ref{lem:rela_eta}, we have the asymptotic relationship,
\begin{equation}\label{eq:extra_eta}
\eta_{\tau'_n} \sim \frac{n(1-\tau'_n)}{k} \eta_{1-k/n},
\end{equation}
as $n \to \infty$. Therefore, \eqref{eq:extra_eta} suggests that $\eta_{\tau'_n}$ can be estimated via an extrapolative from $\eta_{1-k/n}$ and it is hence sufficient to estimate $\eta_{1-k/n}$. Instead of the semi-parametric method used in \cite{Nolde2022}, we now propose two full nonparametric estimators for $\eta_{1-k/n}$ based on observation in \eqref{eq:prob_eta}. Let $R_i^X$ and $R_i^Y$ denote the rank of $X_i$ among $X_1,...,X_n$ and the rank of $Y_i$ among $Y_1,...,Y_n$, respectively. We define two nonparametric estimators for the tail dependence function $R$ as follows,
\begin{equation}\label{eq:nonp_R_emp}
\widehat{R}^{(1)}_n(x,y) := \frac{1}{k} \sum_{i=1}^{n} I \left( 1 - \widehat{F}_X(X_i) \leq \frac{xk}{n}, 1 - \widehat{F}_Y(Y_i) \leq \frac{yk}{n}  \right),
\end{equation}
and
\begin{equation}\label{eq:nonp_R_rank}
\widehat{R}^{(2)}_n(x,y) := \frac{1}{k} \sum_{i=1}^{n} I\left( R_i^X \ge n + \frac{1}{2} - kx, R_i^Y \ge n + \frac{1}{2} - ky  \right).
\end{equation}
These two estimators may have slight numerical diffenceces. Our simulation experiments show that the latter estimator usually performs slightly better, see \cite{Einmahl2008} or \cite{Einmahl2012} for arbitrary dimensions.

It is also worth noting that both \eqref{eq:nonp_R_emp} and \eqref{eq:nonp_R_rank} can be regarded as the nonparametric estimators of the probability in the right-hand side of \eqref{eq:prob_eta}. Thus, according to \eqref{eq:nonp_R_emp} and \eqref{eq:nonp_R_rank}, two nonparametric estimators of $\eta_{1-k/n}$ can be defined via
\begin{equation}\label{eq:nonp_eta}
\hat{\eta}^{(i)}_{1-k/n} := \inf\{ \eta \in (0,1): \widehat{R}^{(i)}_n(\eta,1) \ge k/n \}, ~\text{for}~ i=1,2.
\end{equation}
In other words, $\hat{\eta}^{(i)}_{1-k/n}$ can be seen as the left-continuous inverse functions of $\widehat{R}^{(i)}_n(\cdot,1)$ at the level $1-k/n$. 
The straightforward procedure in Algorithm \ref{alg:eta_est} below provides closed-form solutions of $\hat{\eta}^{(i)}_{1-k/n}$ with $i=1,2$ defined in \eqref{eq:nonp_eta}.
\begin{algorithm}[htbp]
\caption{Procedure to estimate $\hat{\eta}^{(j)}_{1-k/n}$ for $j=1,2$.}
\label{alg:eta_est}
\begin{algorithmic}[1]
\REQUIRE Sample size $n$, intermediate $k$, and bivariate samples $\{(X_i,Y_i)\}_{i=1}^n$.
\STATE Generate empirical distributions $\widehat{F}_X(\cdot)$ and $\widehat{F}_Y(\cdot)$; Generate samples of ranks statistics $\{R_i^X,R_i^Y\}_{i=1}^n$;
\STATE Calculate the positive integer $m: = \lceil \frac{k^2}{n} \rceil$, where $\lceil \cdot \rceil$ denotes the ceiling function;
\IF{$j=1$}
    \STATE Denote $\{Z_i^X,Z_i^Y \}:= \{ 1 - \widehat{F}_X(X_i), 1 - \widehat{F}_Y(Y_i) \}$ for $i=1,...,n$;
    \STATE Filter samples by the indictor $I(Z_i^Y \leq k/n)$ and obtain a sub-samples
    \begin{equation*}
    \{(\widetilde{Z}^X_i,\widetilde{Z}^Y_i)\}:=\{(Z^X_iI(Z_i^Y \leq k/n),Z^Y_iI(Z_i^Y \leq k/n)\},
    \end{equation*}
    with sample size $k+1$. Note the original samples will be removed when the indictor takes 0.
    \STATE Then,
    \begin{equation*}
    \hat{\eta}^{(1)}_{1-k/n} := \frac{n}{k} \widetilde{Z}^X_{m,k+1},
    \end{equation*}
    where $\widetilde{Z}^X_{m,k+1}$ denotes the $m$-th order statistic of the sub-samples.
\ELSE
    \STATE Filter samples by the indictor $I(R_i^Y \ge n + 1/2 - k)$ and obtain a sub-samples
    \begin{equation*}
    \{(\widetilde{R}^X_i,\widetilde{R}^Y_i)\}:=\{(R^X_iI(R_i^Y \ge n + 1/2 - k),R^Y_iI(R_i^Y \ge n + 1/2 - k)\},
    \end{equation*}
    with sample size $k$. Note the original samples will be removed when the indictor takes 0.
    \STATE Then,
    \begin{equation*}
    \hat{\eta}^{(2)}_{1-k/n} := \frac{n + 1/2 - \widetilde{R}^X_{k+1-m,k}}{k},
    \end{equation*}
    where $\widetilde{R}^X_{k+1-m,k}$ denotes the $(k+1-m)$-th order statistic of the sub-samples.
\ENDIF
\end{algorithmic}
\end{algorithm}

Firstly, based on the two sub-samples with sample sizes $k+1$ and $k$, we can certainly determine such ordered statistics $\widetilde{Z}^X_{m,k+1}$ and $\widetilde{R}^X_{k+1-m,k}$ since $1 < m < k$ and $1 < k+1-m < k$ by assuming $k = O(n^{\iota})$ in Assumption \ref{ass:conditions}(d). Secondly, for example, if a smaller order statistic is chosen in Step 6, it would result in $\widehat{R}^{(1)}_n(\hat{\eta}^{(1)}_{1-k/n},1) < k/n$, violating the definition \eqref{eq:nonp_eta}. This confirms that the estimator $\hat{\eta}^{(1)}_{1-k/n}$ obtained from the above procedure indeed attains the minimum. Thirdly, subsequent simulations demonstrate that this procedure consistently yields $\hat{\eta}^{(i)}_{1-k/n}$ strictly bounded within $(0,1)$ across all four models considered.


\begin{proposition}\label{pro:asy_nonpeta}
Under Assumptions \ref{ass:conditions}(a), (b), (d) and \ref{ass:ext_uni}, we have that  for $i = 1,2$ and as $n \to \infty$, there exists a centered Wiener process $W_R$ on $[0,\infty]^2/\{ \infty, \infty \}$ with covariance structure,
\begin{equation*}
\mathbb{E}[W_R(x_1,y_1)W_R(x_2,y_2)] = R(x_1 \wedge x_2, y_1 \wedge y_2),
\end{equation*}
suh that
\begin{equation}\label{eq:asyn_eta}
\sqrt{k} \left( \hat{\eta}^{(i)}_{1-k/n} - \eta_{1-k/n} \right) \xrightarrow{d} \frac{ R_2(0,1) }{R_1(0, 1)} W_R(\infty,1).
\end{equation}
\end{proposition}

Motivated by \eqref{eq:extra_covar} and \eqref{eq:extra_eta}, we propose the following extrapolative estimators for $\covar_{X|Y}(\tau'_n)$ that
\begin{equation}\label{eq:extest_covar}
\widetilde{\covar}^{(i)}_{X|Y}(\tau'_n) = \left( \frac{k}{n(1-\tau'_n)}  \right)^{2\hat{\gamma}_1} \left(\hat\eta_{1-k/n}^{(i)} \right)^{-\hat{\gamma}_1} \widehat{\var}_X(1-k/n), 
\end{equation}
for $i=1,2$, respectively.
Moreover, by Proposition \ref{pro:coes_covar}, the corresponding extrapolative methods for estimating $\coes_{X|Y}(\tau'_n)$ folllows that
\begin{equation}\label{extcoes_covar}
\widetilde{\coes}^{(i)}_{X|Y}(\tau'_n)  = \frac{1}{1-\hat{\gamma}_1} \widetilde{\covar}^{(i)}_{X|Y}(\tau'_n)  = \frac{1}{1-\hat{\gamma}_1} \left( \frac{k}{n(1-\tau'_n)}  \right)^{2\hat{\gamma}_1} \left(\hat\eta_{1-k/n}^{(i)}\right)^{-\hat{\gamma}_1} \widehat{\var}_X(1-k/n), 
\end{equation}
for $i=1,2$.
Note that it will leads to the same results if we implement extrapolation by considering the asymptotic connection between $\coes_{X|Y}(\tau'_n)$ and $\var_X(\tau'_n)$, which follows
\begin{equation*}
\frac{\coes_{X|Y}(\tau'_n)}{\var_X(1-k/n)} = \frac{\coes_{X|Y}(\tau'_n)}{\covar_{X|Y}(\tau'_n)} \frac{\covar_{X|Y}(\tau'_n)}{\var_X(1-k/n)} \sim \frac{1}{1-\gamma_1} \left( \frac{k}{n(1-\tau'_n)} \right)^{2\gamma_1} \eta_{1-k/n}^{-\gamma_1}.
\end{equation*}

We present the joint asymptotic normalities for $\widetilde{\covar}^{(i)}_{X|Y}(\tau'_n)$ and $\widetilde{\coes}^{(i)}_{X|Y}(\tau'_n)$ with $i=1,2$ in the following theorem. They share the same asymptotic properties; it is primarily because $\hat\eta_{1-k/n}^{(i)}$ have identical asymptotic behavior.
\begin{theorem}\label{th:asyn_extreme}
Under Assumptions \ref{ass:conditions} - \ref{ass:ext_uni} and $\gamma_1 \in (0,1)$, we have that for $i=1,2$ and as $n \to \infty$, 
\begin{equation}\label{eq:asyn_covarcoes}
\frac{k^{3/2}}{n} \left( \frac{\widetilde{\covar}^{(i)}_{X|Y}(\tau'_n)}{\covar_{X|Y}(\tau'_n)} - 1,  \frac{\widetilde{\coes}^{(i)}_{X|Y}(\tau'_n)}{\coes_{X|Y}(\tau'_n)} - 1 \right)^{\top} \xrightarrow{d} (1,1)^{\top} \gamma_1 R_2(0,1) W_R(\infty,1).
\end{equation}
\end{theorem}

\subsection{Extrapolative estimations based on intermediate $\covar_{X|Y}$ and $\coes_{X|Y}$}\label{sec:ext_int}

In this section, we propose another class of extrapolative approaches for $\covar_{X|Y}(\tau'_n)$ and $\coes_{X|Y}(\tau'_n)$ without involving $\eta_{1-k/n}$ but considering intermediate $\covar_{X|Y}(1-k/n)$ and $\coes_{X|Y}(1-k/n)$. First of all, applying \eqref{eq:covar_var} and \eqref{eq:extra_eta}, it follows that,
\begin{equation*}
\frac{\covar_{X|Y}(\tau'_n)}{\covar_{X|Y}(1-k/n)} = \frac{\var_X(1-(1-\tau'_n)\eta_{\tau'_n})}{\var_X(1-k\eta_{1-k/n}/n)} \sim \left( \frac{k}{n(1-\tau'_n)}  \right)^{2\gamma_1},
\end{equation*}
which suggests the third extrapolation,
\begin{equation}\label{eq:extra3_covar}
\widetilde{\covar}^{(3)}_{X|Y}(\tau'_n) = \left( \frac{k}{n(1-\tau'_n)}  \right)^{2\hat{\gamma}_1} \widehat{\covar}_{X|Y}(1-k/n),
\end{equation}
and
\begin{equation}\label{eq:extra3_coes}
\widetilde{\coes}^{(3)}_{X|Y}(\tau'_n) = \frac{1}{1-\hat{\gamma}_1} \widetilde{\covar}^{(3)}_{X|Y}(\tau'_n),
\end{equation}
where $\widehat{\covar}_{X|Y}(1-k/n)$, the estimator of $\covar_{X|Y}(1-k/n)$, can be defined as, by the definition in \eqref{eq:covar_X},
\begin{equation}\label{eq:covar_fix}
\widehat{\covar}_{X|Y}(1-k/n) := \sup\left\{ s \in (0,\infty): C_n(s) \ge (k/n)^2 \right\},
\end{equation}
where
\begin{equation*}
C_n(s) := \frac{1}{n} \sum_{i=1}^{n} I(X_i \ge s, Y_i \ge \widehat{\var}_Y(1-k/n)),
\end{equation*}
with $\widehat{\var}_Y(1-k/n)$ corresponding to the $(n-k)$-th order statistic $Y_{n-k,n}$ of $Y_1,...,Y_n$.
Therefore, $\widehat{\covar}_{X|Y}(1-k/n)$ can be regarded as the right-continuous inverse function of $C_n(\cdot)$ at the level $1-k/n$. Similar to Algorithm \ref{alg:eta_est}, a brief procedure below provides the explicit solution of $\widehat{\covar}_{X|Y}(1-k/n)$ defined in \eqref{eq:covar_fix}.
\begin{algorithm}[htbp]
\caption{Procedure to estimate $\covar_{X|Y}(1-k/n)$.}
\label{alg:covar_int}
\begin{algorithmic}[1]
\REQUIRE Sample size $n$, intermediate $k$, and bivariate samples $\{(X_i,Y_i)\}_{i=1}^n$.
\STATE Calculate the positive integer $m: = \lceil \frac{k^2}{n} \rceil$, where $\lceil \cdot \rceil$ denotes the ceiling function;
\STATE Calculate the empirical quantile estimator for $Y$, $\widehat{\var}_Y(1-k/n):= Y_{n-k,n}$;
\STATE Filter samples by the indictor $I(Y_i \ge \widehat{\var}_Y(1-k/n))$ and obtain a sub-samples
\begin{equation*}
\{(X^*_i,Y^*_i)\}=\{(X_iI(Y_i \ge \widehat{\var}_Y(1-k/n)),Y_iI(Y_i \ge \widehat{\var}_Y(1-k/n)))\},
\end{equation*}
with sample size $k+1$. Note the original samples will be removed when the indictor takes 0.
\STATE Output,
\begin{equation*}
\widehat{\covar}_{X|Y}(1-k/n) := X^*_{k+2-m,k+1},
\end{equation*}
where $X^*_{k+2-m,k+1}$ denotes as the $(k+2-m)$-th order statistic of the sub-samples.
\end{algorithmic}
\end{algorithm}

Moreover, by \eqref{eq:coes_X_exp}, we can also construct a nonparametric estimator of $\coes_{X|Y}(1-k/n)$ as follows
\begin{equation}\label{eq:coes_int}
\widehat{\coes}_{X|Y}(1-k/n) = \frac{n}{k^2} \sum_{i=1}^{n} X_i I\left(  X_i \ge \widehat{\covar}_{X|Y}(1-k/n), Y_i \ge \widehat{\var}_Y(1-k/n) \right).
\end{equation}
When establishing the asymptotic normality of $\widehat{\coes}_{X|Y}(1-k/n)$, the asymptotic properties of $\frac{\widehat{\var}_Y(1-k/n)}{\var_Y(1-k/n)}$ are not required; rather, the properties of $\frac{n}{k}\overline{F}_Y(\widehat{\var}_Y(1-k/n))$ are essential. 
This approach eliminates the need for stringent assumptions on the tail behavior of $Y$. Consequently, the fourth extrapolative estimator for $\coes_{X|Y}(\tau'_n)$ is denoted as
\begin{equation}\label{eq:extra4_coes}
\widetilde{\coes}_{X|Y}^{(4)}(\tau'_n) = \left( \frac{k}{n(1-\tau'_n)} \right)^{2\hat{\gamma}_1}  \widehat{\coes}_{X|Y}(1-k/n),
\end{equation}
which follows the asymptotic relationship
\begin{equation*}
\frac{\coes_{X|Y}(\tau'_n)}{\coes_{X|Y}(1-k/n)} = \frac{\coes_{X|Y}(\tau'_n)}{\covar_{X|Y}(\tau'_n)} \frac{\covar_{X|Y}(\tau'_n)}{\covar_{X|Y}(1-k/n)} \frac{\covar_{X|Y}(1-k/n)}{\coes_{X|Y}(1-k/n)} \sim \left( \frac{k}{n(1-\tau'_n)}  \right)^{2\gamma_1}.
\end{equation*}


One important observation is that all the estimators in \eqref{eq:extra3_coes} to \eqref{eq:extra4_coes} do not involve the adjustment factor $\hat\eta_{1-k/n}^{(i)}$; however, the convergence of these estimators depend implicitly on that of $\eta_{1-k/n}$. 
The following proposition presents the joint asymptotic normality for $\widehat{\covar}_{X|Y}(1-k/n)$ and $\widehat{\coes}_{X|Y}(1-k/n)$, which share the same rate of convergence of extreme case. It is because that $\covar_{X|Y}(1-k/n)$ has a faster rate than $\var_{X}(1-k/n)$.
\begin{proposition}\label{pro:inter_covar_coes}
Under Assumptions \ref{ass:conditions} - \ref{ass:ext_uni} and $\gamma_1 \in (0,1/2)$, we have that  as $n \to \infty$, 
\begin{equation}\label{eq:asyn_inter_covar_coes}
\frac{k^{3/2}}{n}\left( \frac{\widehat{\covar}_{X|Y}(1-k/n)}{\covar_{X|Y}(1-k/n)} -1, \frac{\widehat{\coes}_{X|Y}(1-k/n)}{\coes_{X|Y}(1-k/n)} -1 \right)^{\top}  \xrightarrow{d} (1,2)^{\top} \gamma_1 R_2(0,1) W_R(\infty,1).
\end{equation}
\end{proposition}

Note that the condition $\gamma_1 \in (0,1/2)$ is necessary for $\widehat{\coes}_{X|Y}(1-k/n)$ in Proposition \ref{pro:inter_covar_coes}. However, when studying the asymptotic properties of $\widehat{\covar}_{X|Y}(1-k/n)$, this condition can be relaxed to $\gamma_1 \in (0,1)$. The following theorem establishes that both sets of extrapolative estimators, namely \eqref{eq:extra3_covar}, \eqref{eq:extra3_coes}, and \eqref{eq:extra4_coes}, exhibit joint asymptotic normality.
\begin{theorem}\label{th:extra3}
Under Assumptions \ref{ass:conditions} - \ref{ass:ext_uni} and $\gamma_1 \in (0,1)$, we have that as $n \to \infty$, 
\begin{equation}\label{eq:asyn_covarcoes3}
\frac{k^{3/2}}{n} \left( \frac{\widetilde{\covar}^{(3)}_{X|Y}(\tau'_n)}{\covar_{X|Y}(\tau'_n)} - 1,  \frac{\widetilde{\coes}^{(3)}_{X|Y}(\tau'_n)}{\coes_{X|Y}(\tau'_n)} - 1 \right)^{\top} \xrightarrow{d} (1,1)^{\top} \gamma_1 R_2(0,1) W_R(\infty,1).
\end{equation}
Moreover, if $\gamma_1 \in (0,1/2)$, then we have that
\begin{equation}\label{eq:asyn_covarcoes4}
\frac{k^{3/2}}{n} \left( \frac{\widetilde{\covar}^{(3)}_{X|Y}(\tau'_n)}{\covar_{X|Y}(\tau'_n)} - 1,  \frac{\widetilde{\coes}^{(4)}_{X|Y}(\tau'_n)}{\coes_{X|Y}(\tau'_n)} - 1 \right)^{\top} \xrightarrow{d} (1,2)^{\top} \gamma_1 R_2(0,1) W_R(\infty,1).
\end{equation}
\end{theorem}

We conclude this section by comparing differnt extrapolation methods proposed in this paper. On the one hand, for $\widetilde{\coes}^{(i)}_{X|Y}(\tau'_n)$ with $i=1,2,3,4$, the first three extrapolation estimators share identical structures, as they are all derived from the extrapolation in Proposition \ref{pro:coes_covar} and are thus determined by $\widetilde{\covar}^{(i)}_{X|Y}(\tau'_n)$ with $i=1,2,3$. As a result, their limiting distributions are the same.
On the other hand, unlike classical methods in the extreme value theory, all the extrapolation approaches mentioned above have the same convergence rate of $k^{3/2}/n$. This can be attributed to two factors: first, the expressions of extrapolative estimators with $i=1,2$, involve the estimators of $\eta_{1-k/n}$, and $\eta_{1-k/n}$ converges to zero at a rate of $k/n$, which cannot be eliminated under standard convergence rates; second, the latter two extrapolative methods for $i=3,4$, are closely associated with the estimators \eqref{eq:covar_fix} and \eqref{eq:coes_int}, which have a convergence rate of $k^{3/2}/n$. By multiplying this rate, the estimation error of $\hat\gamma_1$ can be eliminated. Therefore, it is precisely the estimators \eqref{eq:covar_fix} and \eqref{eq:coes_int} that determine their asymptotic behavior.

\section{Simulation Study}\label{sec:simulation}

In this section, we implement some simulation studies to examin the empirical performance of our proposed methods. We first generate data from the following four transformed bivariate distributions with a common extreme value index.
\begin{itemize}
  \item The first model is a transformed bivariate variables on $(0,\infty)^2$ defined as
  \begin{equation*}
    (X,Y) = (Z_1^{1/3}, Z_2),
  \end{equation*}
  where $(Z_1,Z_2)$ follows a bivariate logistic distribution with standard Fr$\acute{e}$chet margins given by
  \begin{equation*}
    F(s,t; \theta) = \exp \left\{ - \left( s^{-1/\theta} + t^{-1/\theta}  \right)^{\theta}   \right\}, ~ s,t > 0, ~\text{and}~ \theta \in (0,1].
  \end{equation*}
  We hence have $\gamma_1 = \frac{1}{3}$, $R(x,y) = x+y-\left( x^{1/\theta} + y^{1/\theta} \right)^{\theta}$ and $\alpha = 1$; see Section 4.4 in \cite{Fougères2015}. We will take $\theta = 0.6$ in the this study.
  \item The second model is a bivariate transformed Cauchy distribution on $(0,\infty)^2$ defined as
  \begin{equation*}
    (X,Y) = (|Z_1|^{1/3},|Z_2|),
  \end{equation*}
  where $(Z_1,Z_2)$ follows a standard Cauchy distribution on $\mathbb{R}^2$ with joint density
  \begin{equation*}
    f(s,t) = \frac{1}{2\pi} \frac{1}{(1+s^2+t^2)^{3/2}}.
  \end{equation*}
  We have $\gamma_1 = \frac{1}{3}$, $R(x,y)=x+y-\sqrt{x^2+y^2}$ and $\alpha = 2$; see the example of \cite{Einmahl2001} on pages 1409-1410 or Section 4.1 in \cite{Fougères2015}.
  \item The third model is a transformed bivariate variables on $(0,\infty)^2$ defined as
  \begin{equation*}
    (X,Y) = (Z_1^{1/6},Z_2),
  \end{equation*}
  where $(Z_1,Z_2)$ follows a bivariate Pareto of type II distribution, referred to $MP^{(2)}(II)(\textbf{0},\textbf{1},\theta)$ in \cite{Kozt2000} (page 603), with tail
  \begin{equation*}
  \bP(Z_1 \ge s, Z_2 \ge t) = (1+s+t)^{-\theta},~s,t > 0,
  \end{equation*}
  and PDF
  \begin{equation*}
    f(s,t) = \theta(\theta+1) (1+s+t)^{-(\theta+2)}.
  \end{equation*}
  We then have $\gamma_1 = \frac{1}{3}$, $R(x,y) = (x^{-2} + y^{-2})^{-1/2}$ and $\alpha = 2$; see Section 4.1 in \cite{Fougères2015}. We will take $\theta = 1/2$ in the following study.
  \item The fourth model is a transformed bivariate Student-$t$ distribution on $(0,\infty)^2$ defined as
  \begin{equation*}
  (X,Y) = (|Z_1|^{1/2},|Z_2|),
  \end{equation*}
  where $(Z_1,Z_2)$ follows a bivariate Student-$t$ distribution with degree of freedom $\nu = 1.5$ and correction parameter $\rho \in (0,1)$,
  \begin{equation*}
    f(\textbf{z};\nu, \rho) = \frac{\Gamma((\nu+2)/2)}{\sqrt{1-\rho^2}\nu \pi \Gamma(\nu/2)} \left( 1 + \frac{1}{\nu} \textbf{z}^{\top} \Sigma^{-1} \textbf{z}  \right)^{-(\nu+2)/2}, ~ \textbf{z} \in \mathbb{R}^2~\text{and}~ \Sigma =
    \begin{pmatrix}
      1 & \rho \\
      \rho & 1
    \end{pmatrix}
    .
  \end{equation*}
  We have $\gamma_1 = \frac{1}{3}$, and
  \begin{equation*}
    R(x,y) = yF\left( \sqrt{\frac{\nu+1}{1-\rho^2}} \left( \rho - (y/x)^{1/\nu} \right)  ;\nu+1\right) + xF\left( \sqrt{\frac{\nu+1}{1-\rho^2}} \left( \rho - (x/y)^{1/\nu} \right)  ;\nu+1\right),
  \end{equation*}
  with $\alpha = 4/3$, where $F(\cdot;\nu+1)$ is the CDF of univariate Student-$t$ distribution with $\nu+1$ degrees of freedom; see Section 4.1 in \cite{Fougères2015}. In this simulation, we always set $\rho = 0.3$.
\end{itemize}
Note that the parameter $\alpha$ mentioned in Assumption \ref{ass:conditions}(b) equals to 1 when the transformed bivariate Logistic model is considered, a little violating the condition $\alpha > 1$. This counterexample serves to highlight the importance of the constraint on the scope of $\alpha$. 

For all simulated data, we set sample size $n = 500,1000,2000$, and $5000$, extremes levels $\tau'_n = 0.99, 0.995,$ and $0.999$, and repeat the simulation $N = 100$ times. We will implement both $\widetilde{\covar}_{X|Y}^{(i)}(\tau'_n)$ for $i=1,2,3$ and $\widetilde{\coes}_{X|Y}^{(i)}(\tau'_n)$ for $i=1,2,3,4$ to compare the finite sample performances. In order to evaluate the performances of all proposed extrapolative estimators, we calculate the \emph{Mean Squared Relative Error} (MSRE) given by
\begin{equation*}
  {\rm MSRE} = \frac{1}{N} \sum_{l=1}^{N} \left( \frac{\hat{\theta}_n^{(l)}}{\theta}  - 1 \right)^2,
\end{equation*}
where $\hat{\theta}_n^{(l)}$ is the estimator given the simulated data of the $l$-th replication, and $\theta$ is the true value.
The choice of the intermediate order \( k \) in Section \ref{sec:meth} is crucial. This is because both the Hill estimator and \(\hat{\eta}^{(i)}_{1 - k/n}\), which are related to the tail dependence function \( R \), necessitate selecting an appropriate intermediate order \( k \). A conventional method for determining \( k \) involves plotting Hill plots. Typically, a larger \( k \) introduces more bias into \(\hat{\gamma}_1\), whereas a smaller \( k \) increases the variation. Additionally, according to Assumption \ref{ass:conditions}(d), there is a theoretical constraint on \( k \) such that \( k = O(n^{\iota}) \) with \(2/3 < \iota < 2(\alpha + \beta)/[2(\alpha + \beta) + 1]\). Consequently, we opt for a suitable \( k \) by plotting the Hill estimator \(\hat{\gamma}_1\) against \( k \) and then selecting a \( k \) that corresponds to the stable region of the plot, thereby satisfying Assumption \ref{ass:conditions}(d).

\begin{figure}[htbp]
\centering
\begin{minipage}[b]{0.24\textwidth}
\includegraphics[width=\textwidth,height = 0.15\textheight]{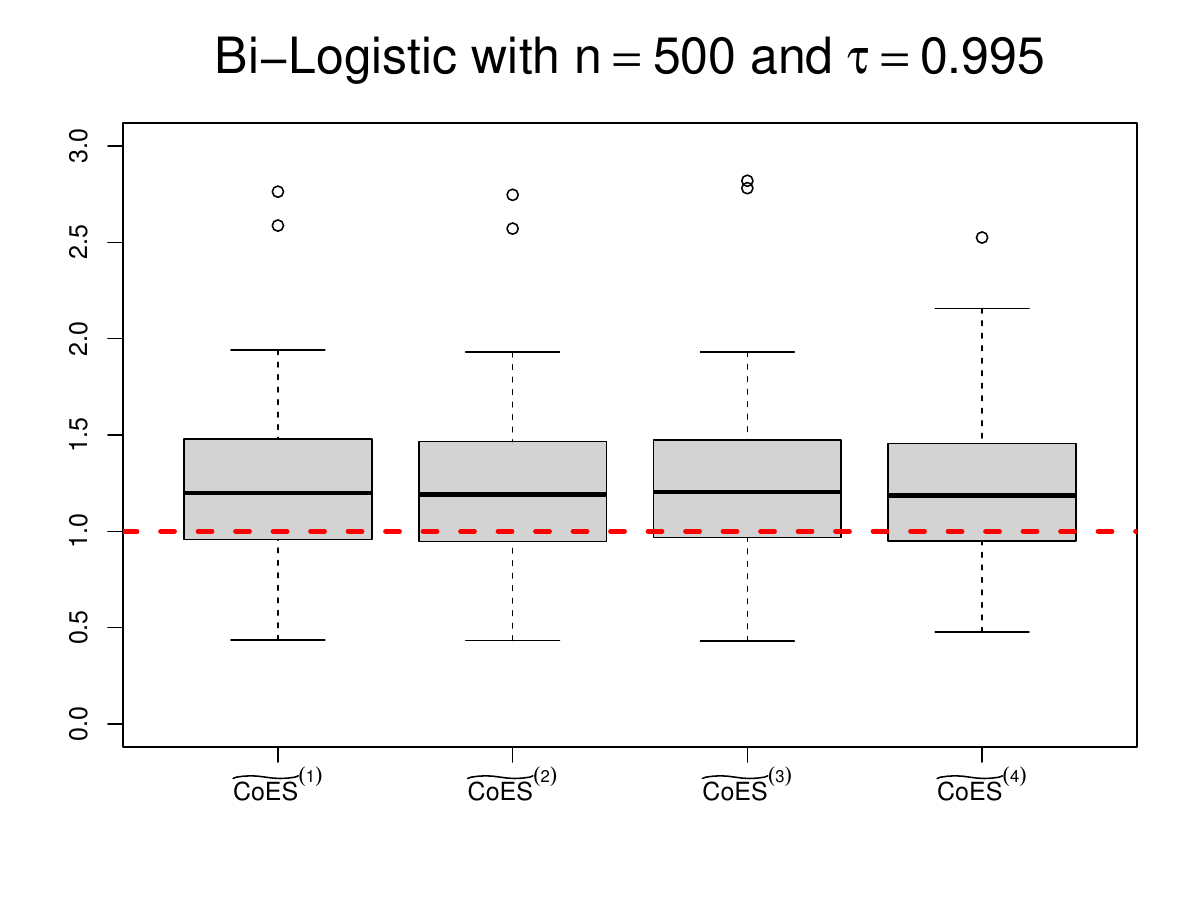}
\end{minipage}
\begin{minipage}[b]{0.24\textwidth}
\includegraphics[width=\textwidth,height = 0.15\textheight]{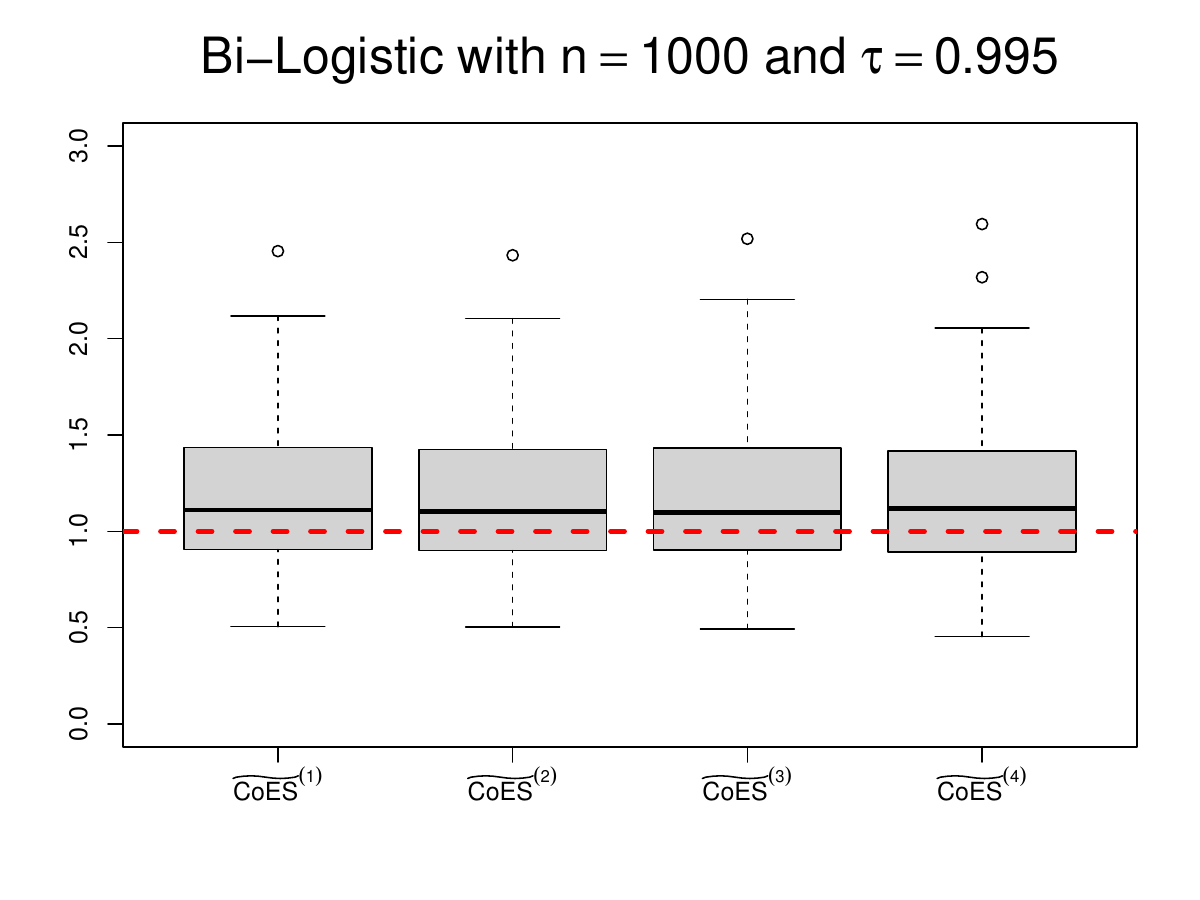}
\end{minipage}
\begin{minipage}[b]{0.24\textwidth}
\includegraphics[width=\textwidth,height = 0.15\textheight]{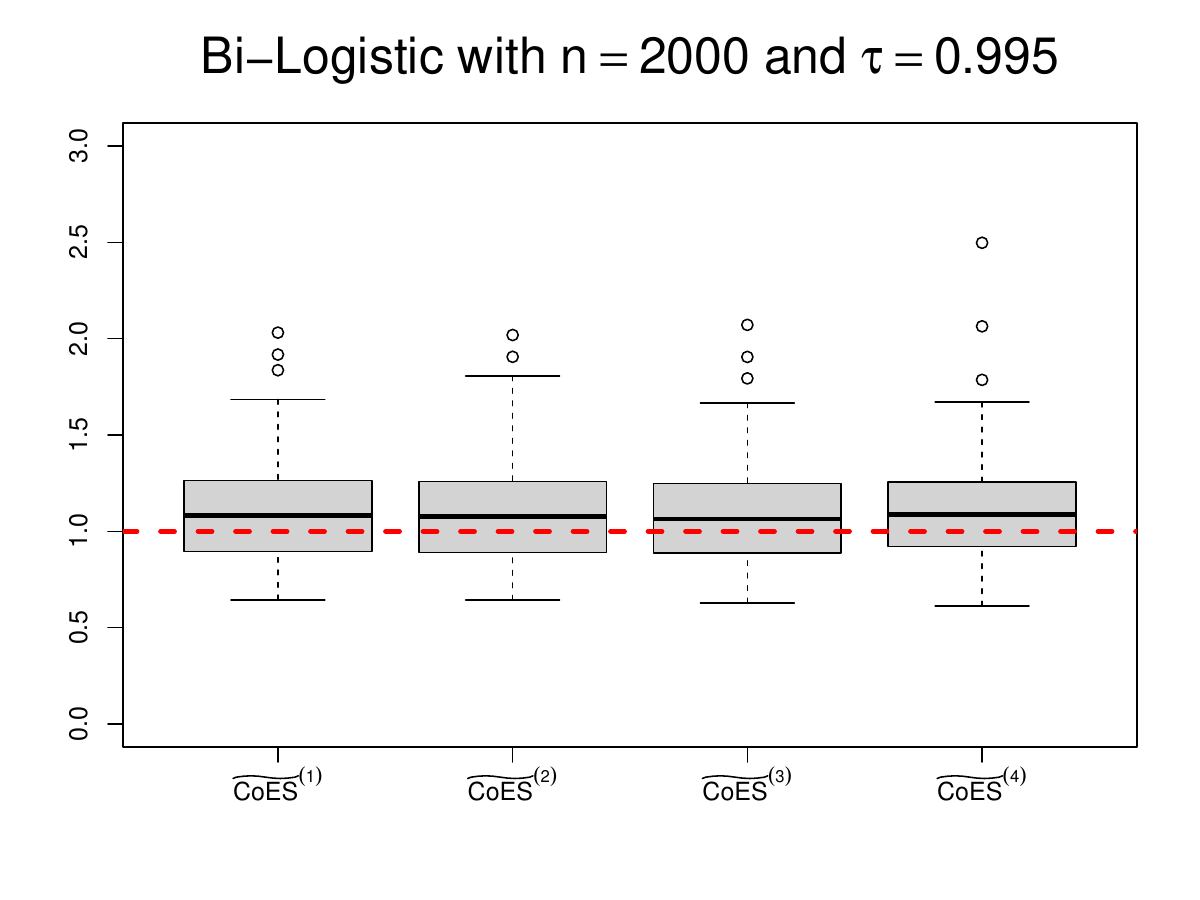}
\end{minipage}
\begin{minipage}[b]{0.24\textwidth}
\includegraphics[width=\textwidth,height = 0.15\textheight]{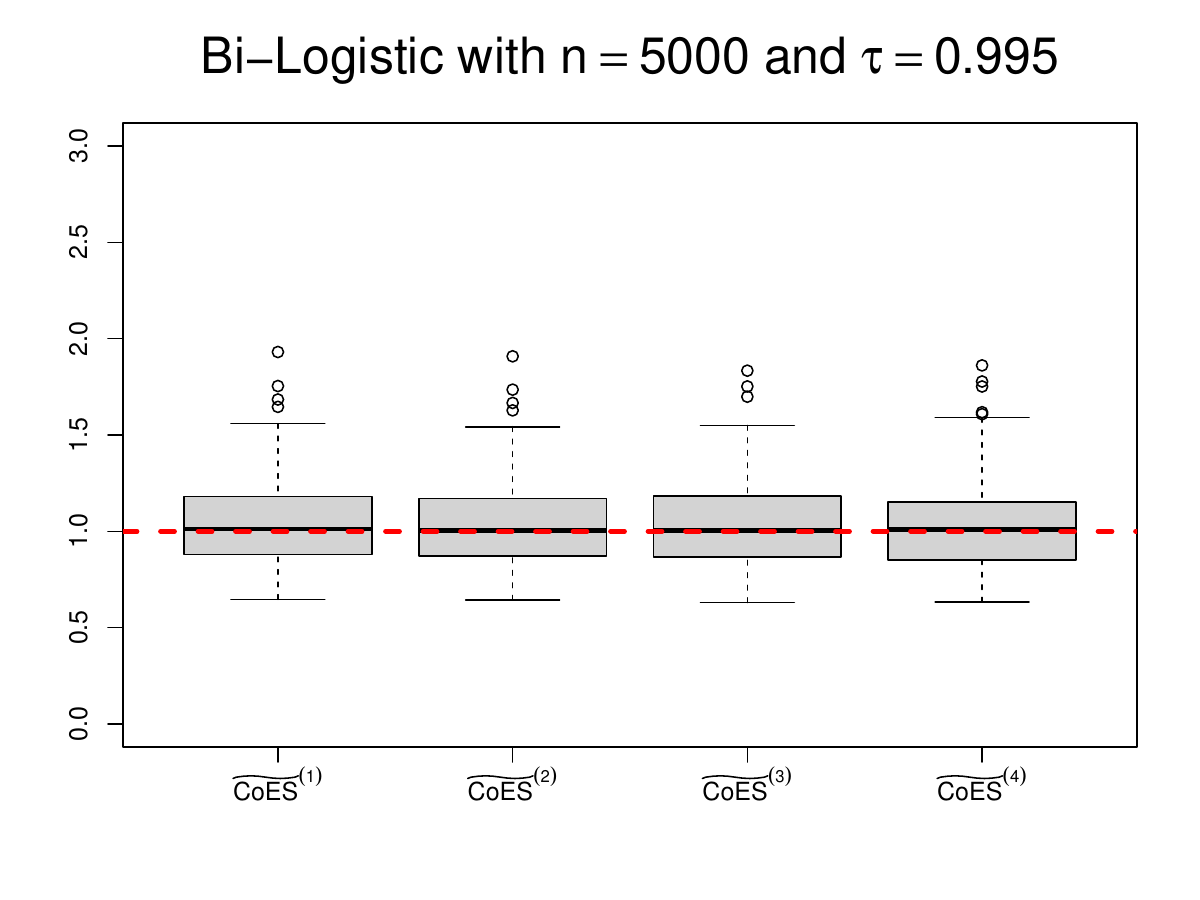}
\end{minipage}
\\
\begin{minipage}[b]{0.24\textwidth}
\includegraphics[width=\textwidth,height = 0.15\textheight]{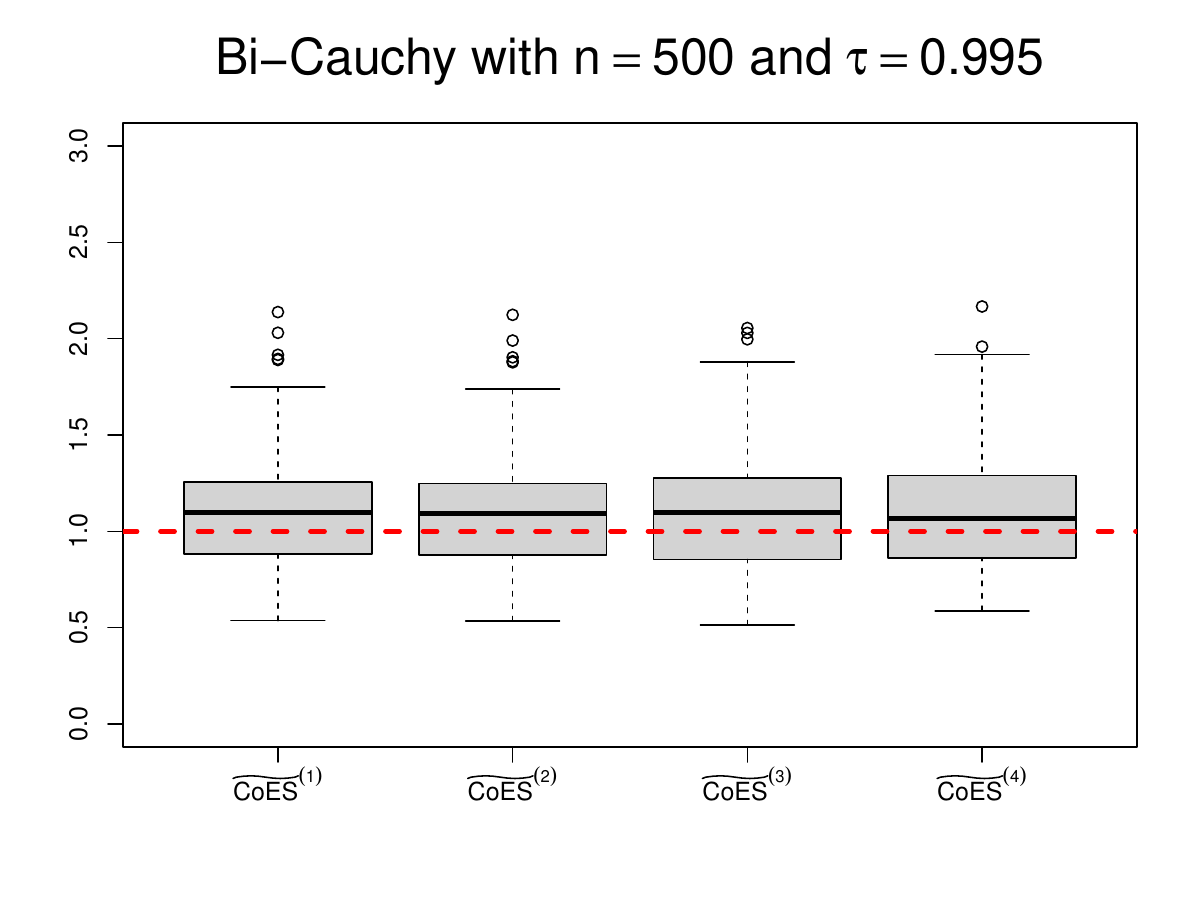}
\end{minipage}
\begin{minipage}[b]{0.24\textwidth}
\includegraphics[width=\textwidth,height = 0.15\textheight]{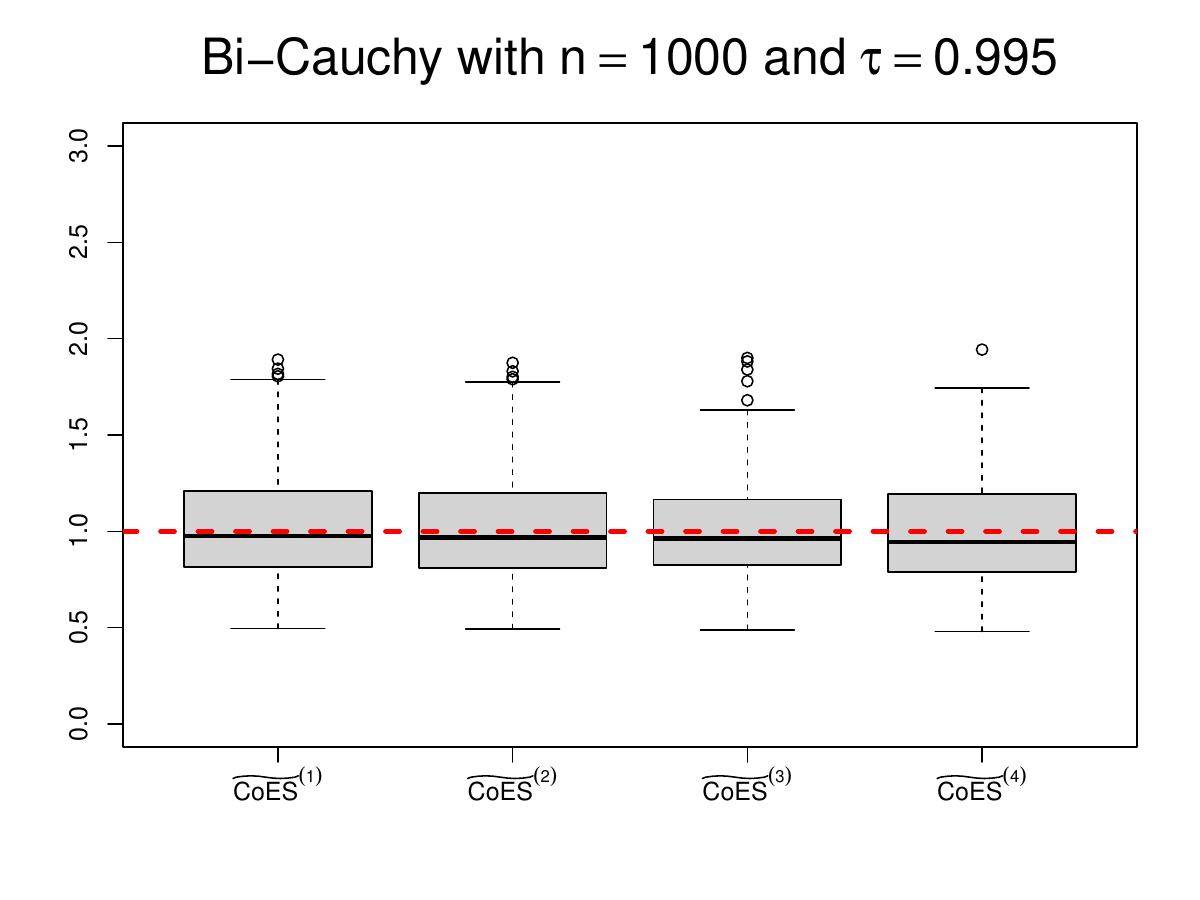}
\end{minipage}
\begin{minipage}[b]{0.24\textwidth}
\includegraphics[width=\textwidth,height = 0.15\textheight]{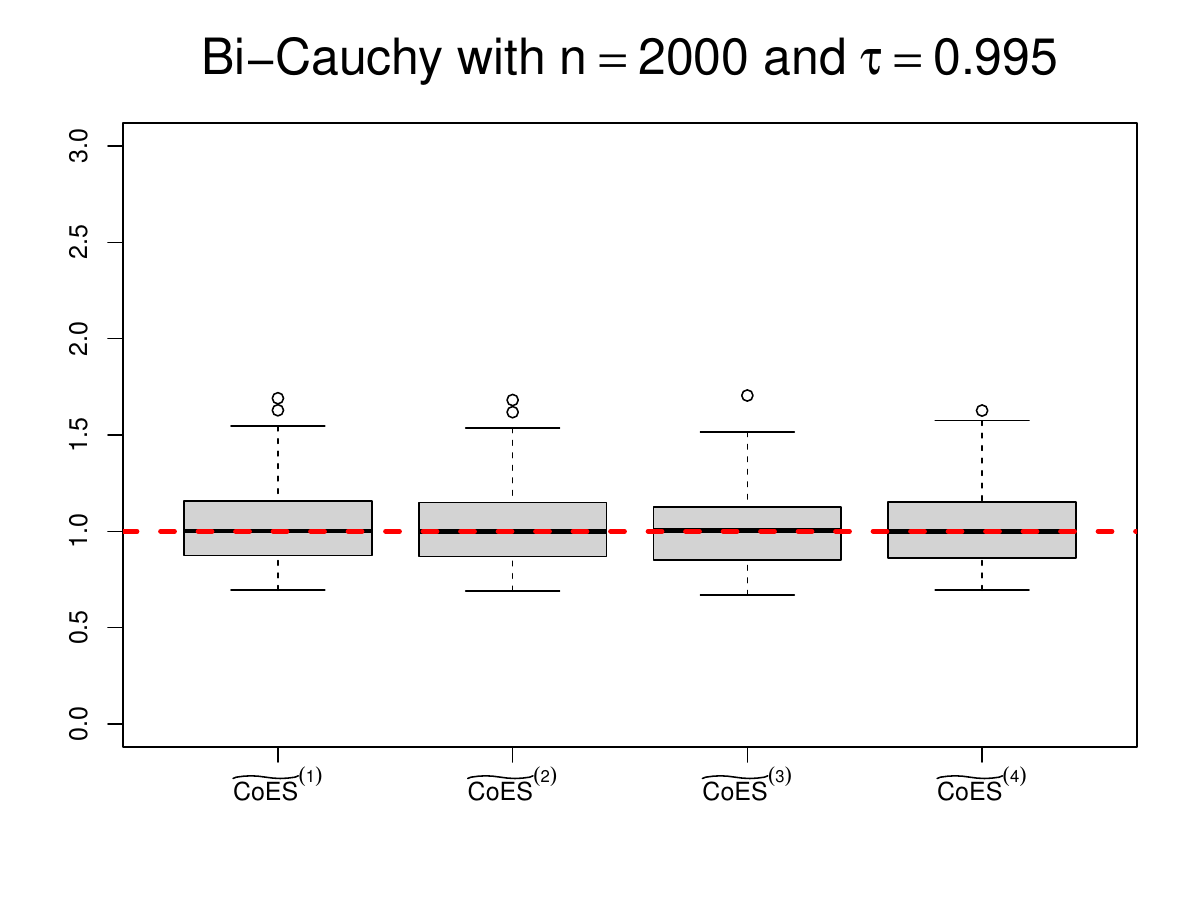}
\end{minipage}
\begin{minipage}[b]{0.24\textwidth}
\includegraphics[width=\textwidth,height = 0.15\textheight]{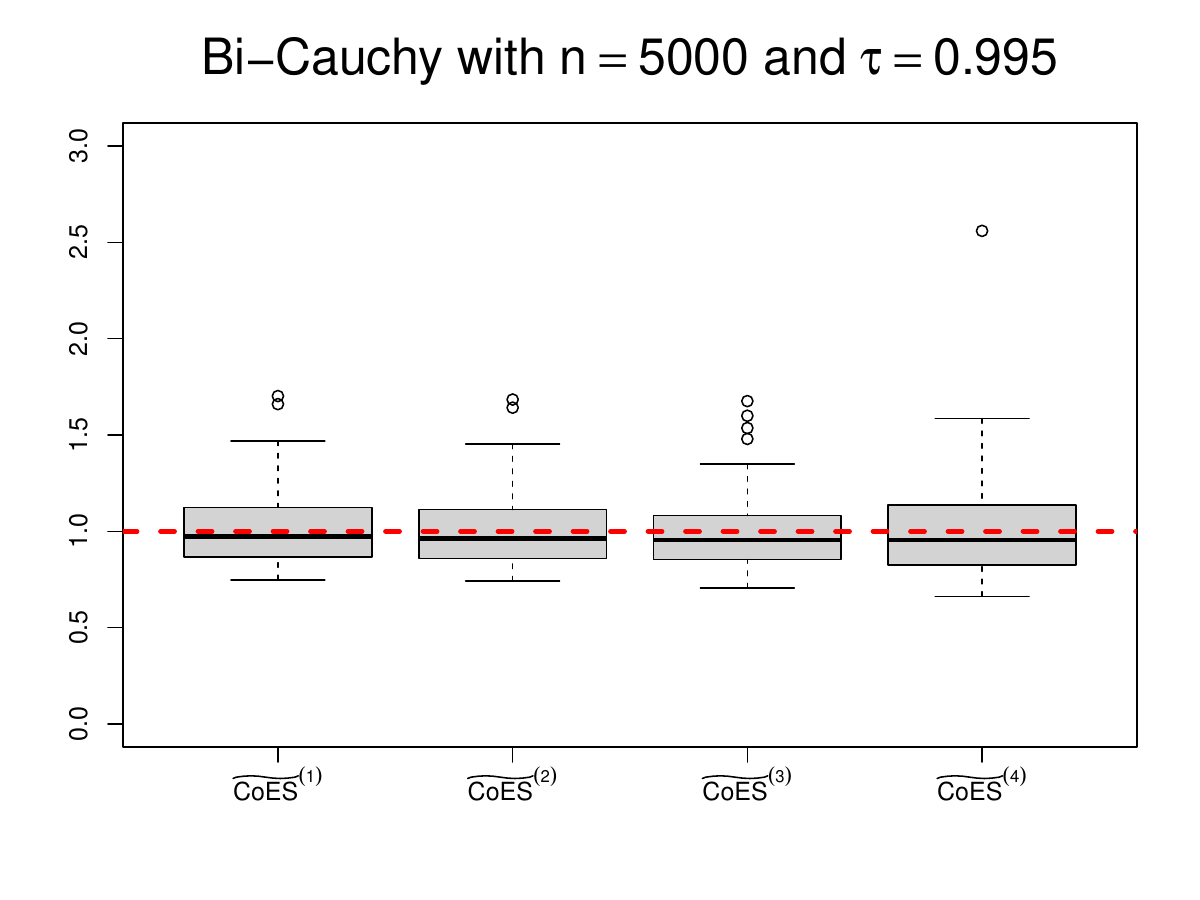}
\end{minipage}
\\
\begin{minipage}[b]{0.24\textwidth}
\includegraphics[width=\textwidth,height = 0.15\textheight]{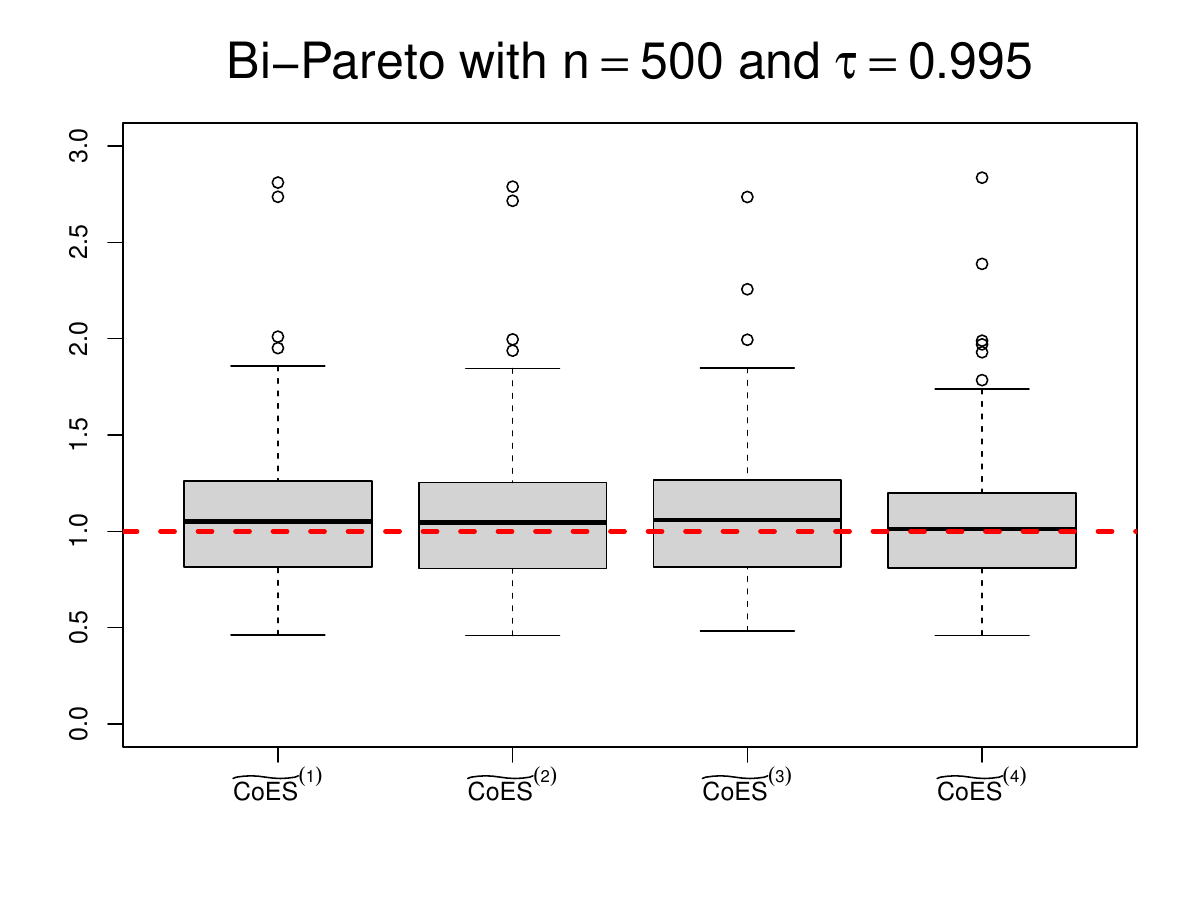}
\end{minipage}
\begin{minipage}[b]{0.24\textwidth}
\includegraphics[width=\textwidth,height = 0.15\textheight]{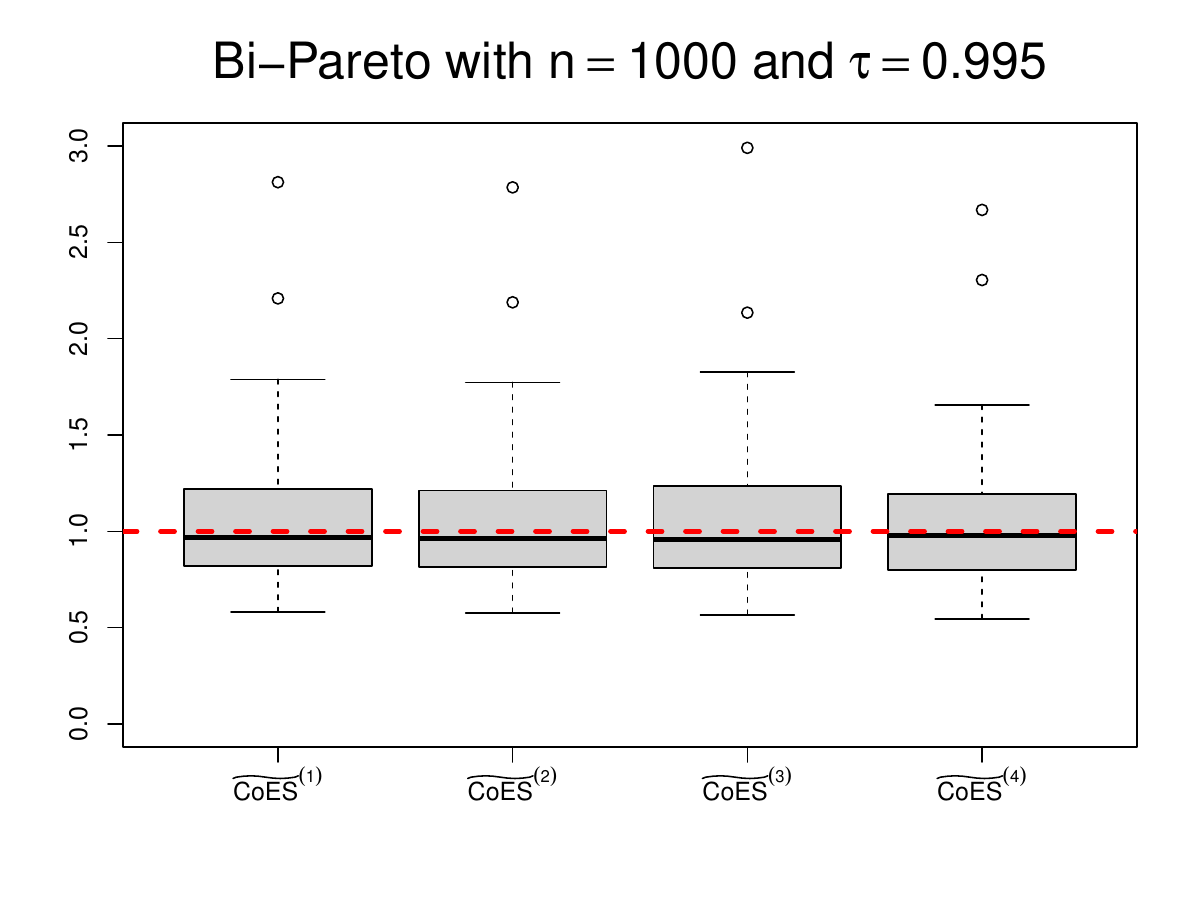}
\end{minipage}
\begin{minipage}[b]{0.24\textwidth}
\includegraphics[width=\textwidth,height = 0.15\textheight]{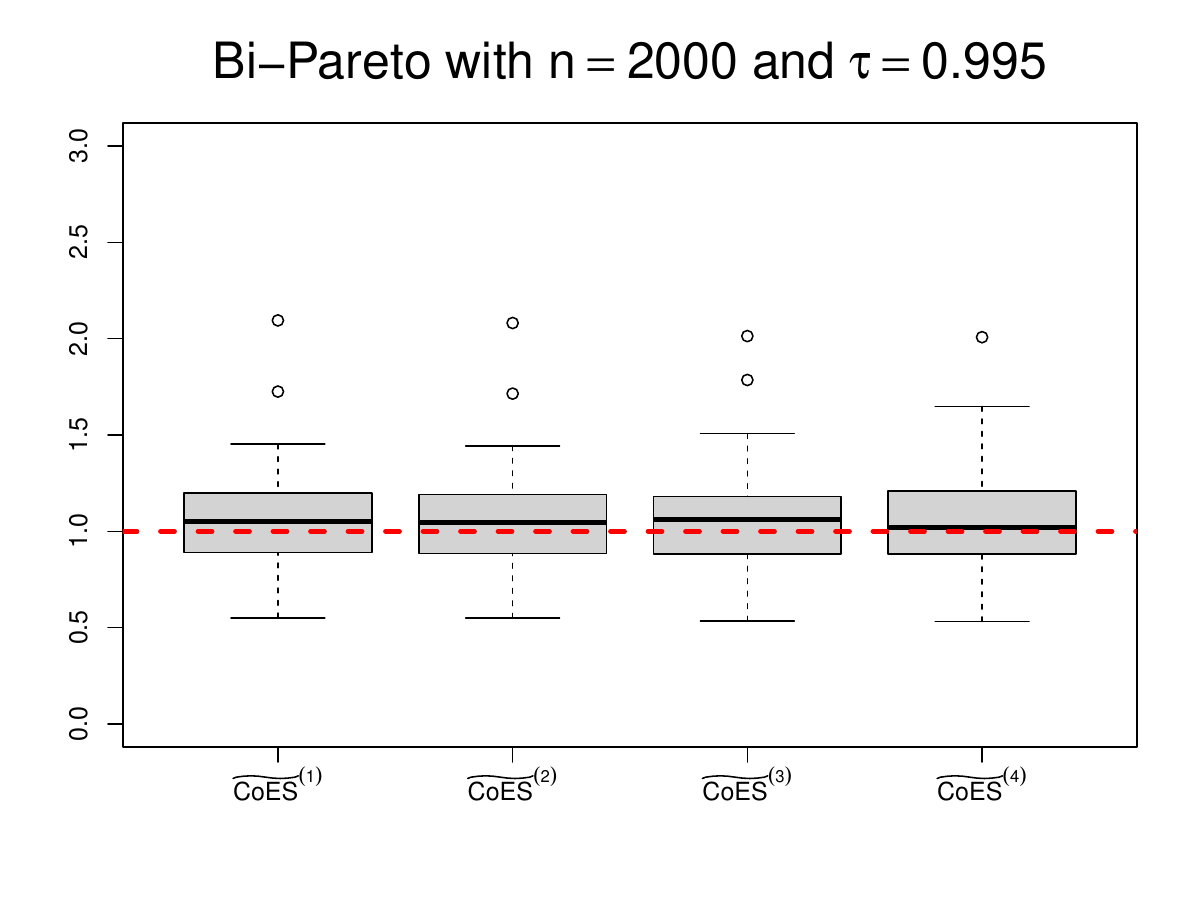}
\end{minipage}
\begin{minipage}[b]{0.24\textwidth}
\includegraphics[width=\textwidth,height = 0.15\textheight]{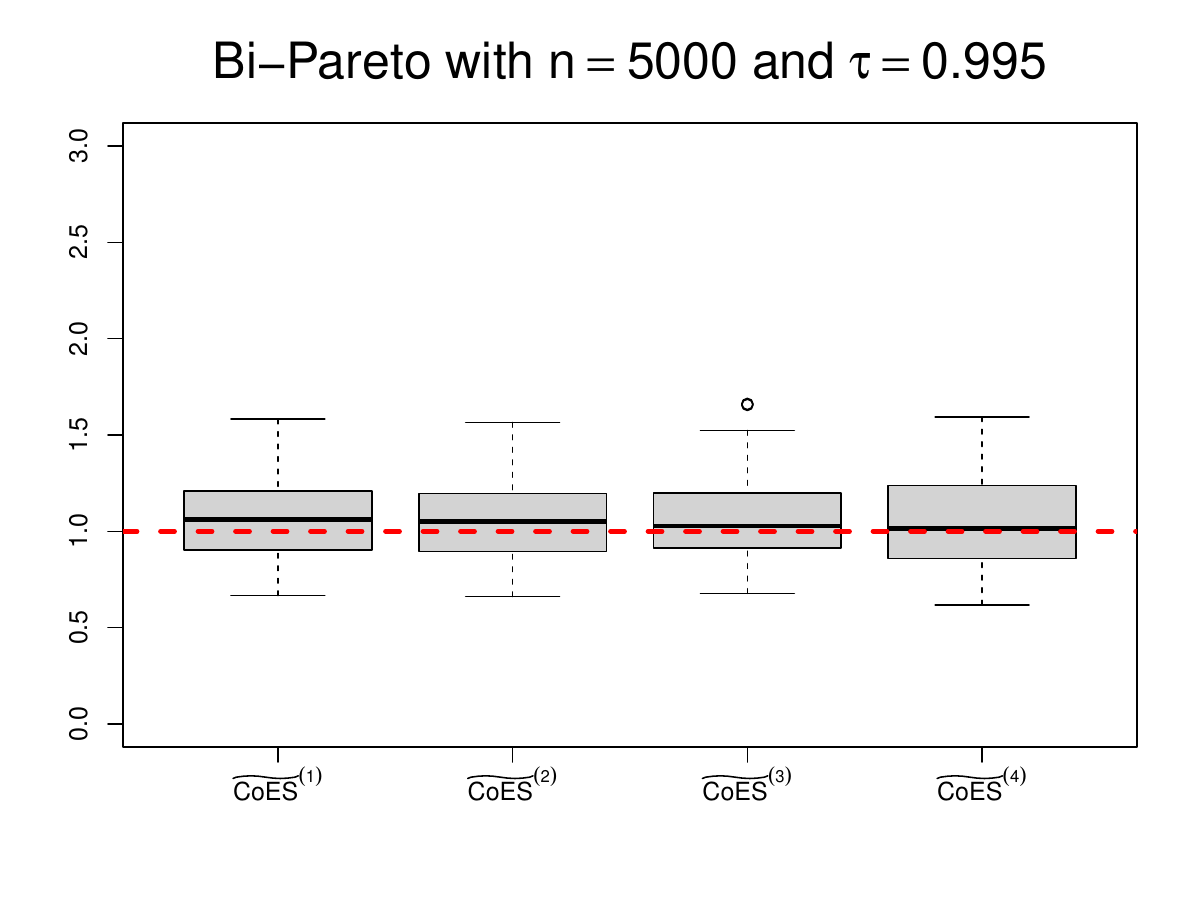}
\end{minipage}
\\
\begin{minipage}[b]{0.24\textwidth}
\includegraphics[width=\textwidth,height = 0.15\textheight]{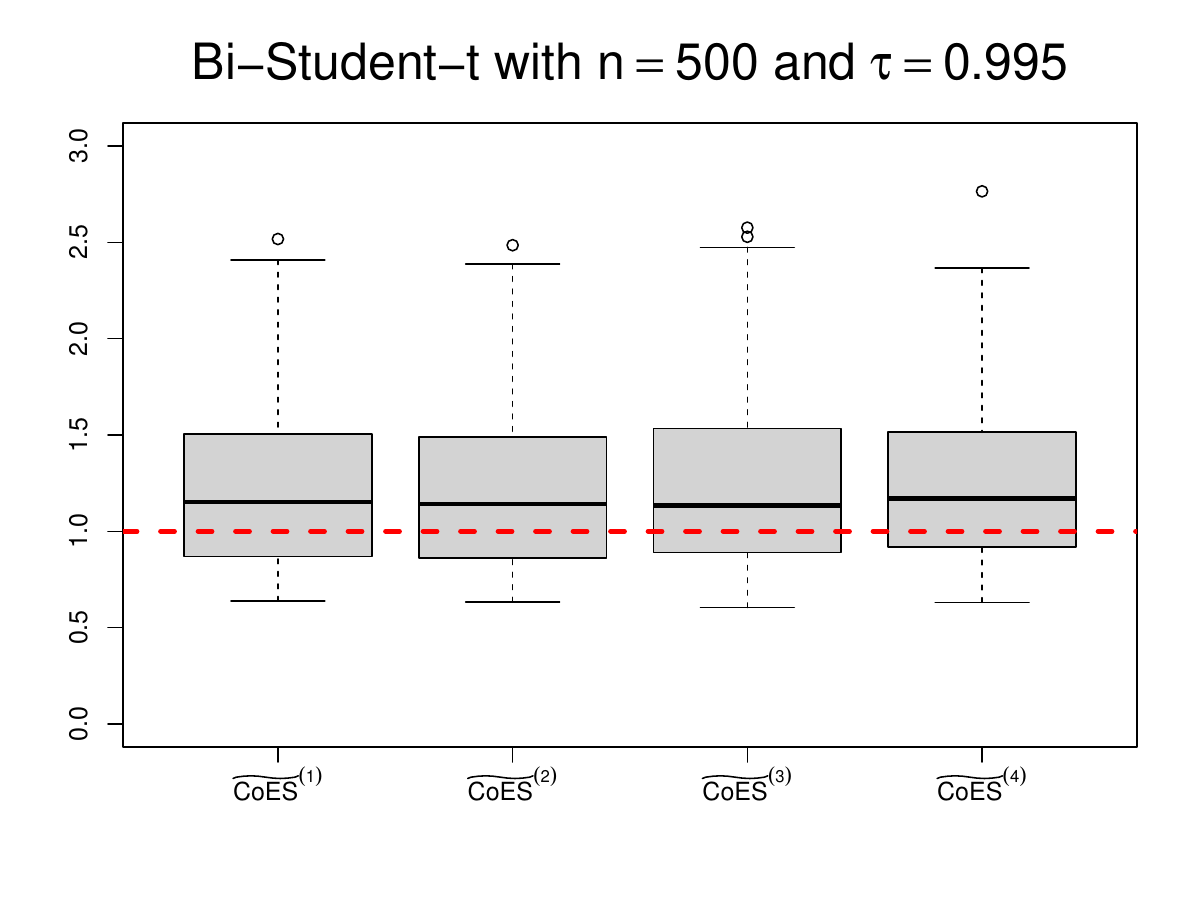}
\end{minipage}
\begin{minipage}[b]{0.24\textwidth}
\includegraphics[width=\textwidth,height = 0.15\textheight]{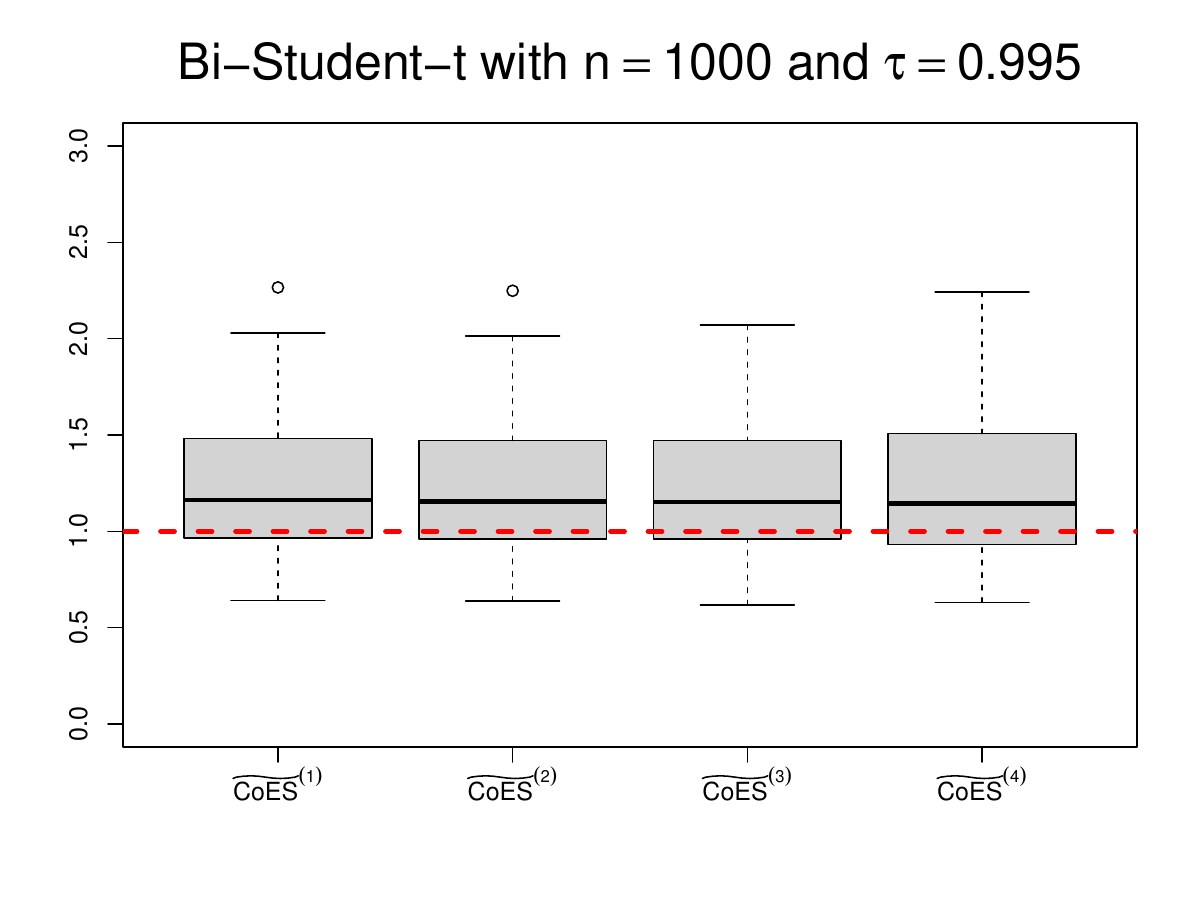}
\end{minipage}
\begin{minipage}[b]{0.24\textwidth}
\includegraphics[width=\textwidth,height = 0.15\textheight]{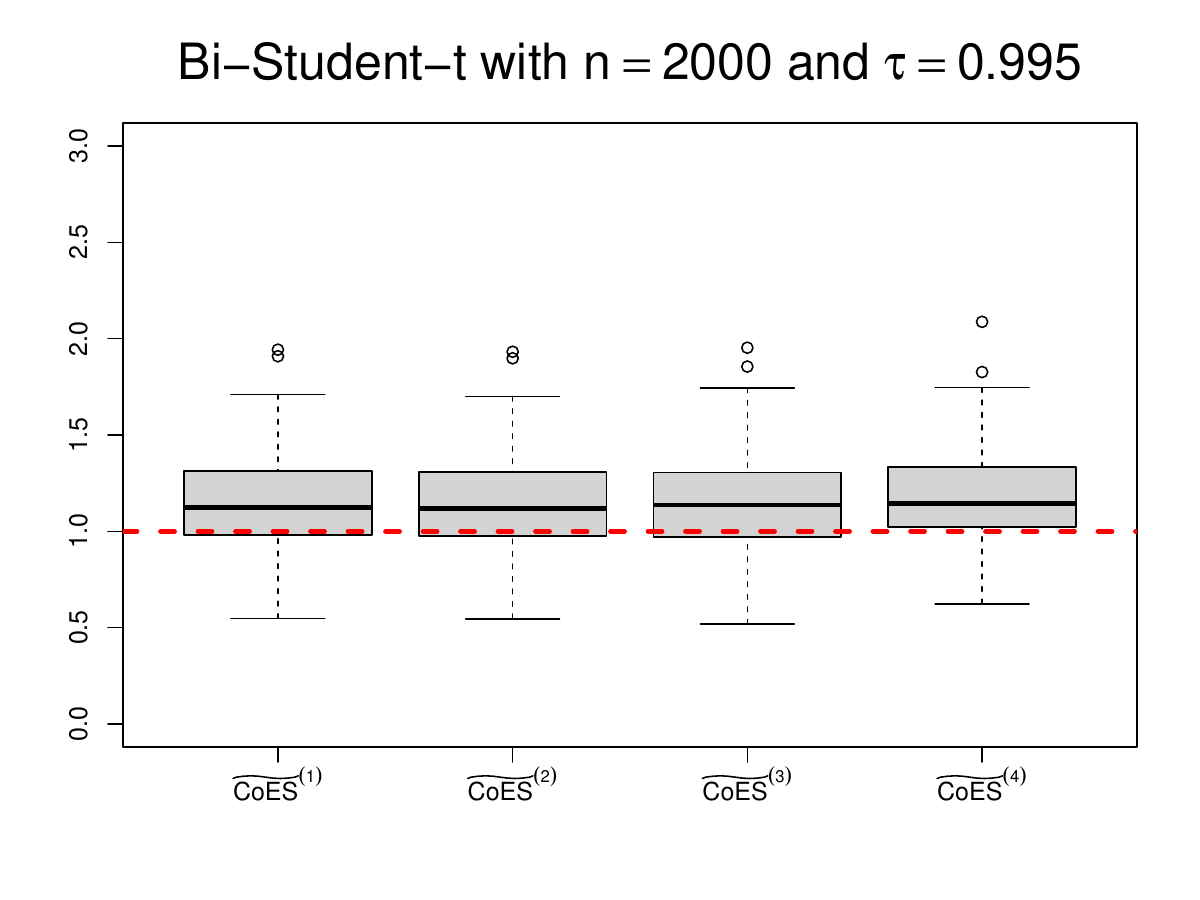}
\end{minipage}
\begin{minipage}[b]{0.24\textwidth}
\includegraphics[width=\textwidth,height = 0.15\textheight]{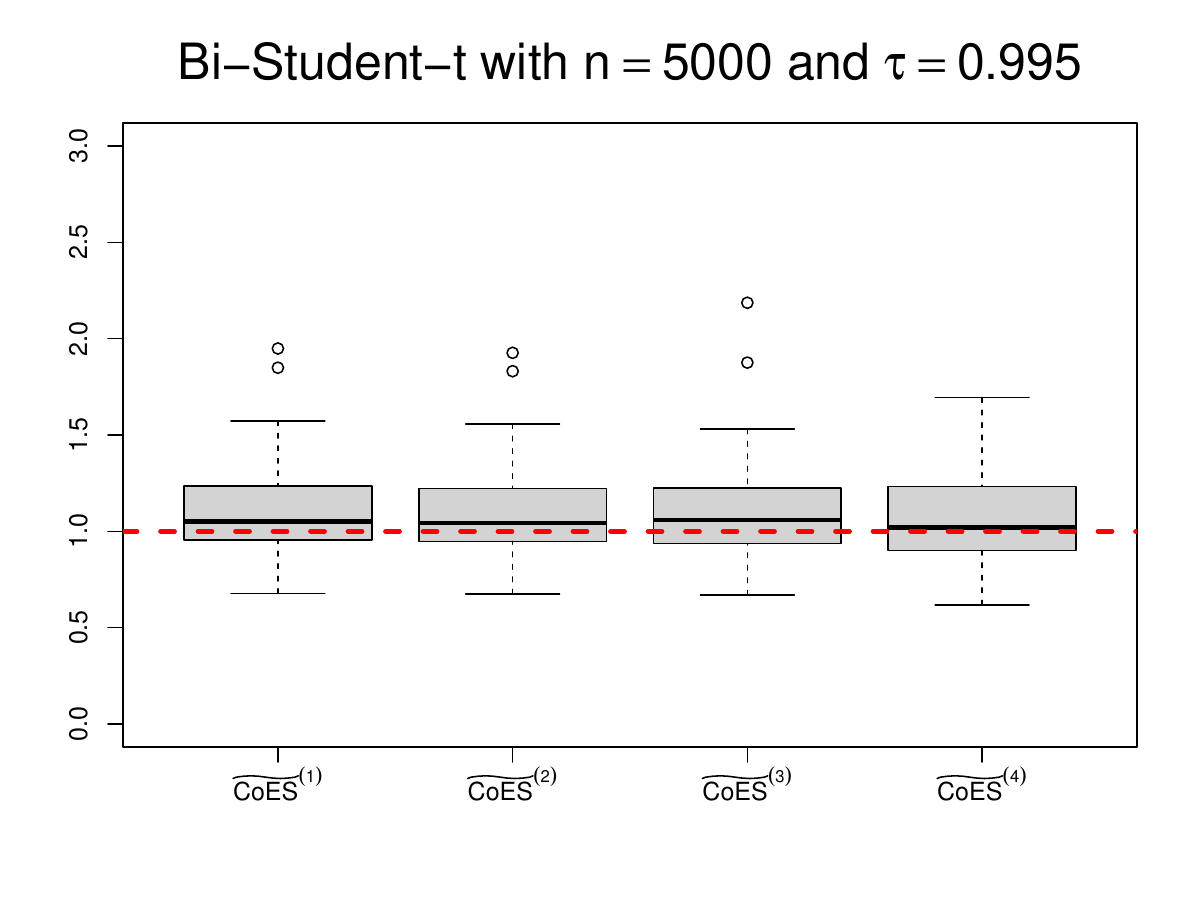}
\end{minipage}
\caption{The boxplots of $\widetilde{\coes}^{(i)}_{X|Y}(\tau'_n)$ for $i=1,2,3,4$ with $\tau'_n = 0.995$, under transformed bivariate Logistic, Cauchy, Pareto and Student-$t$ models. The boxplots from the left panel to the right panel are drawn with $n=500,1000,2000,$ and $5000$, respectively.}
\label{Fig:boxplots_CES_995}
\end{figure}

\begin{figure}[htbp]
\centering
\begin{minipage}[b]{0.24\textwidth}
\includegraphics[width=\textwidth,height = 0.15\textheight]{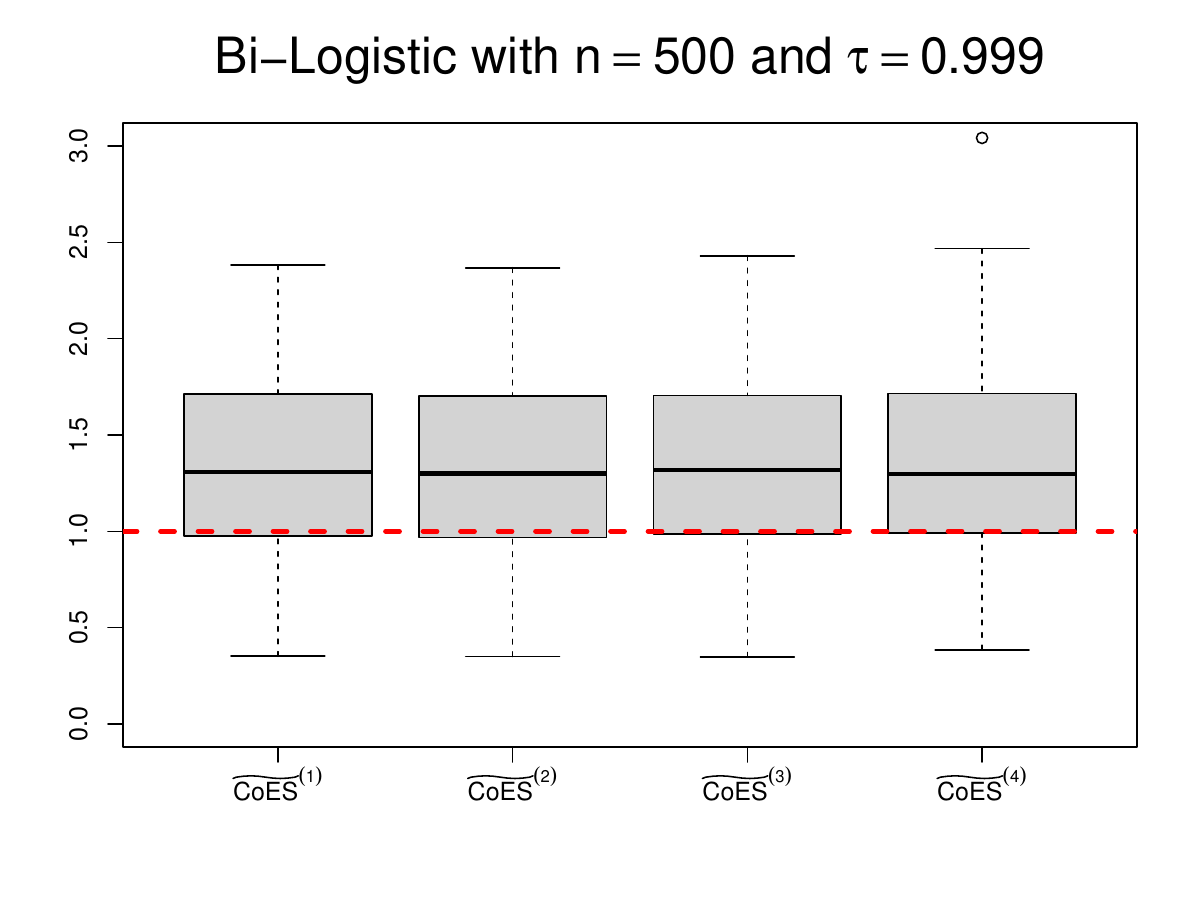}
\end{minipage}
\begin{minipage}[b]{0.24\textwidth}
\includegraphics[width=\textwidth,height = 0.15\textheight]{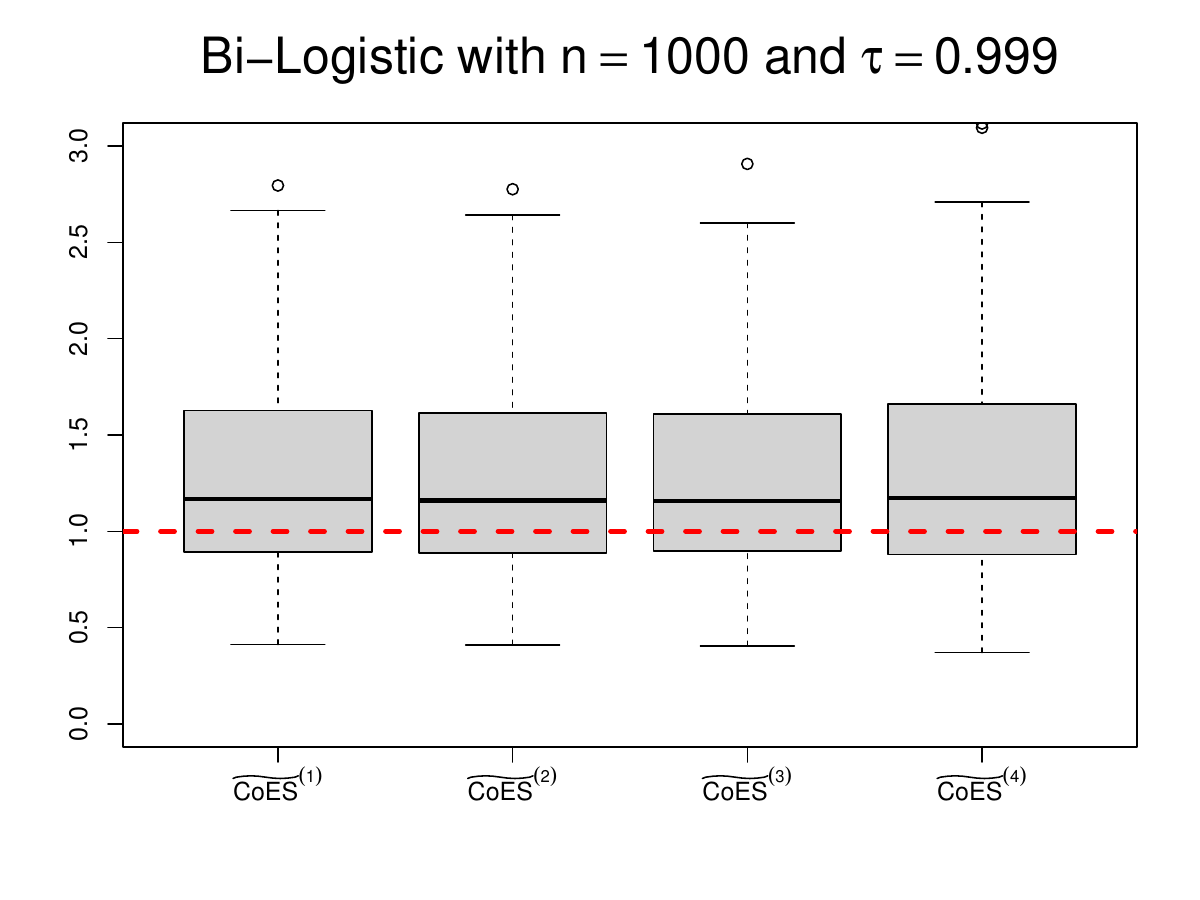}
\end{minipage}
\begin{minipage}[b]{0.24\textwidth}
\includegraphics[width=\textwidth,height = 0.15\textheight]{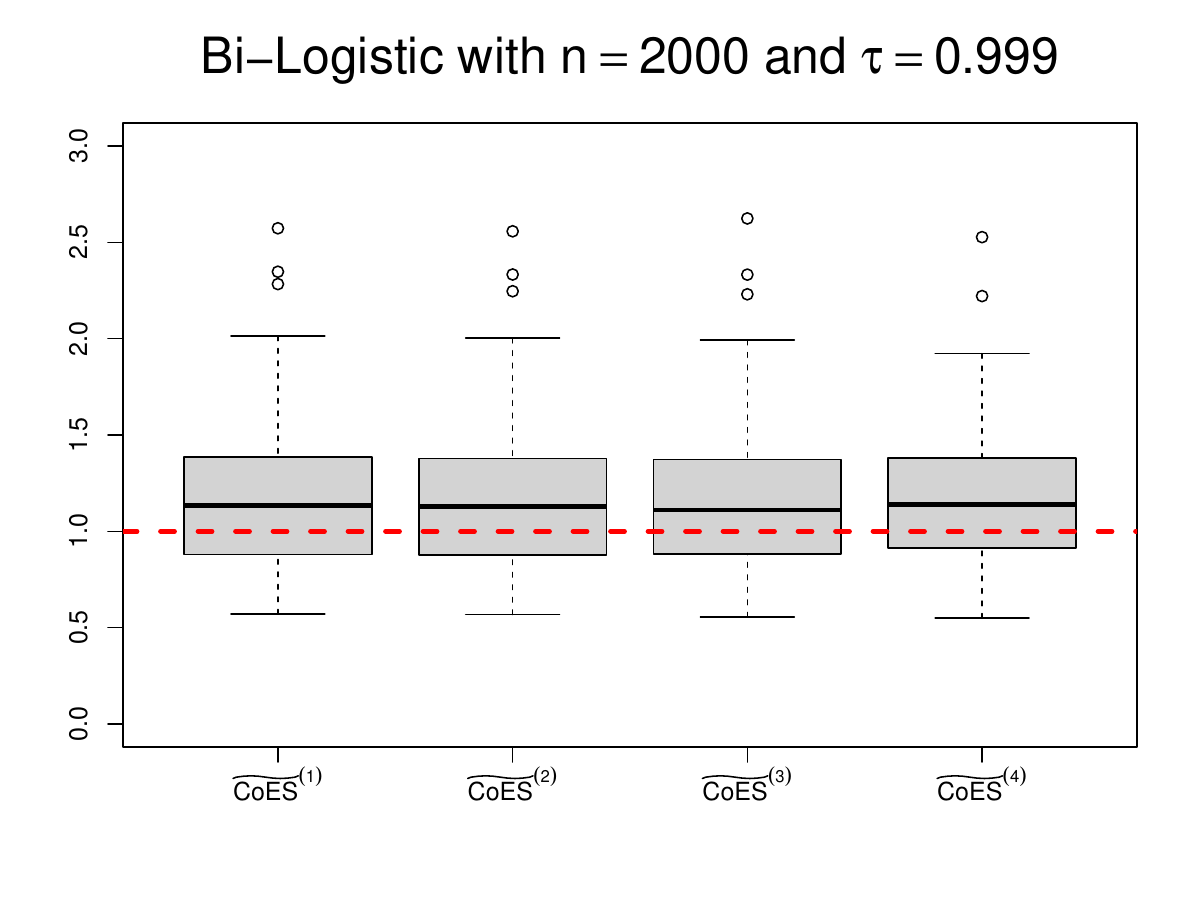}
\end{minipage}
\begin{minipage}[b]{0.24\textwidth}
\includegraphics[width=\textwidth,height = 0.15\textheight]{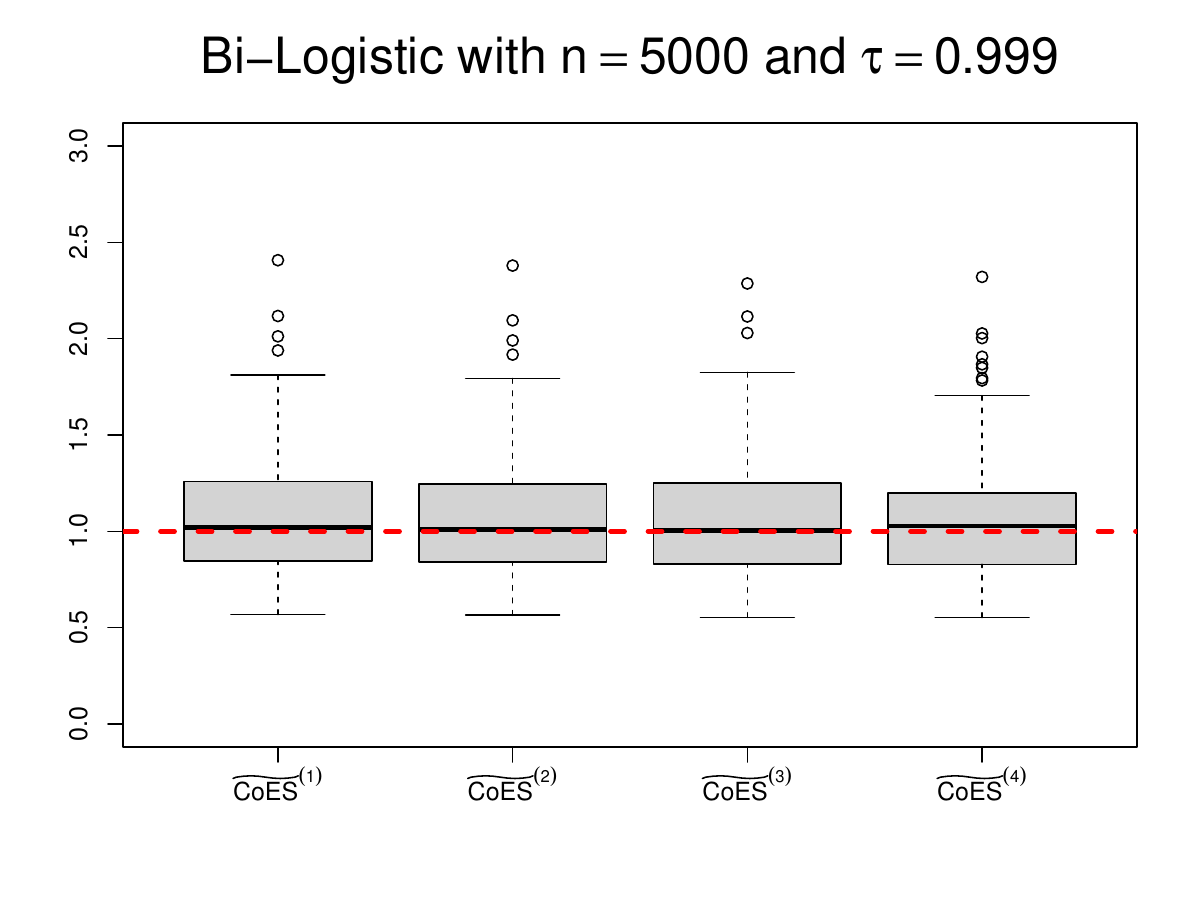}
\end{minipage}
\\
\begin{minipage}[b]{0.24\textwidth}
\includegraphics[width=\textwidth,height = 0.15\textheight]{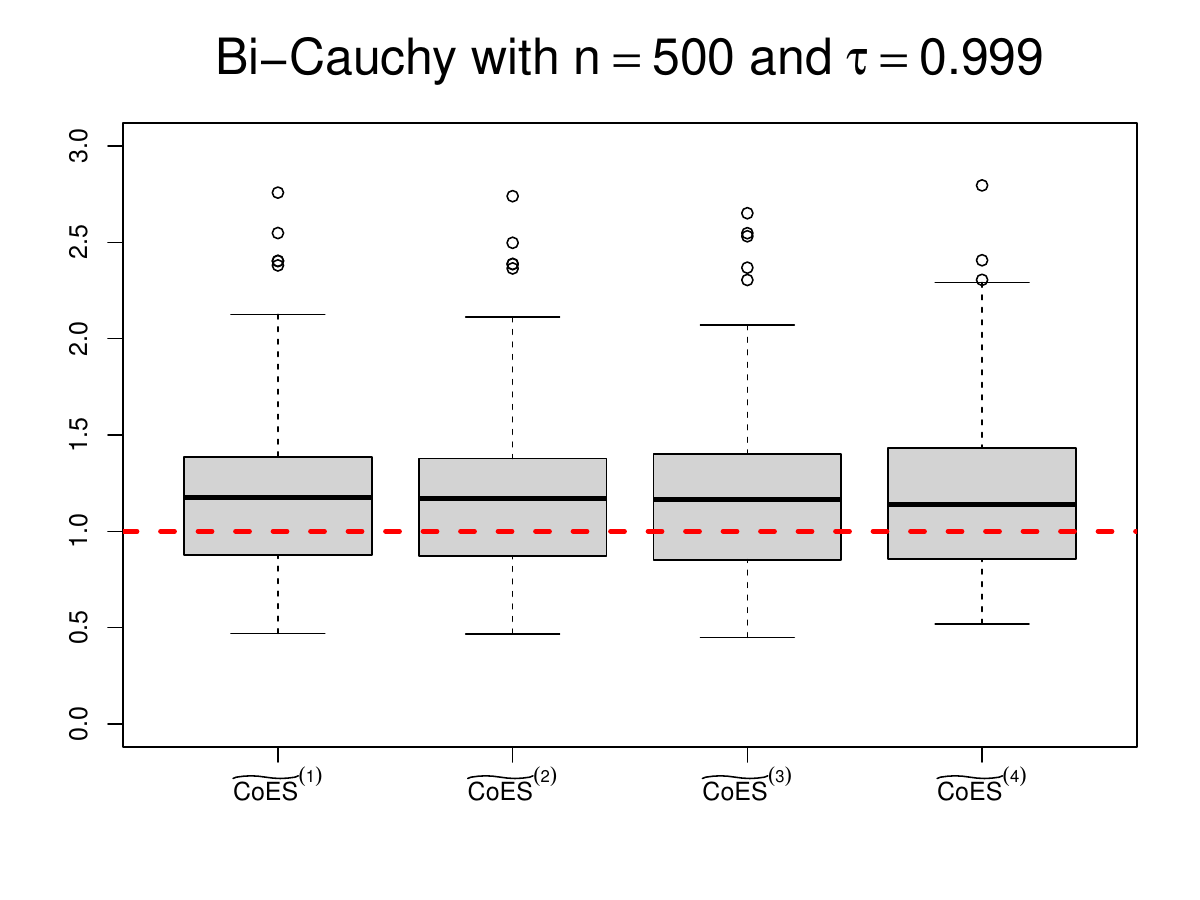}
\end{minipage}
\begin{minipage}[b]{0.24\textwidth}
\includegraphics[width=\textwidth,height = 0.15\textheight]{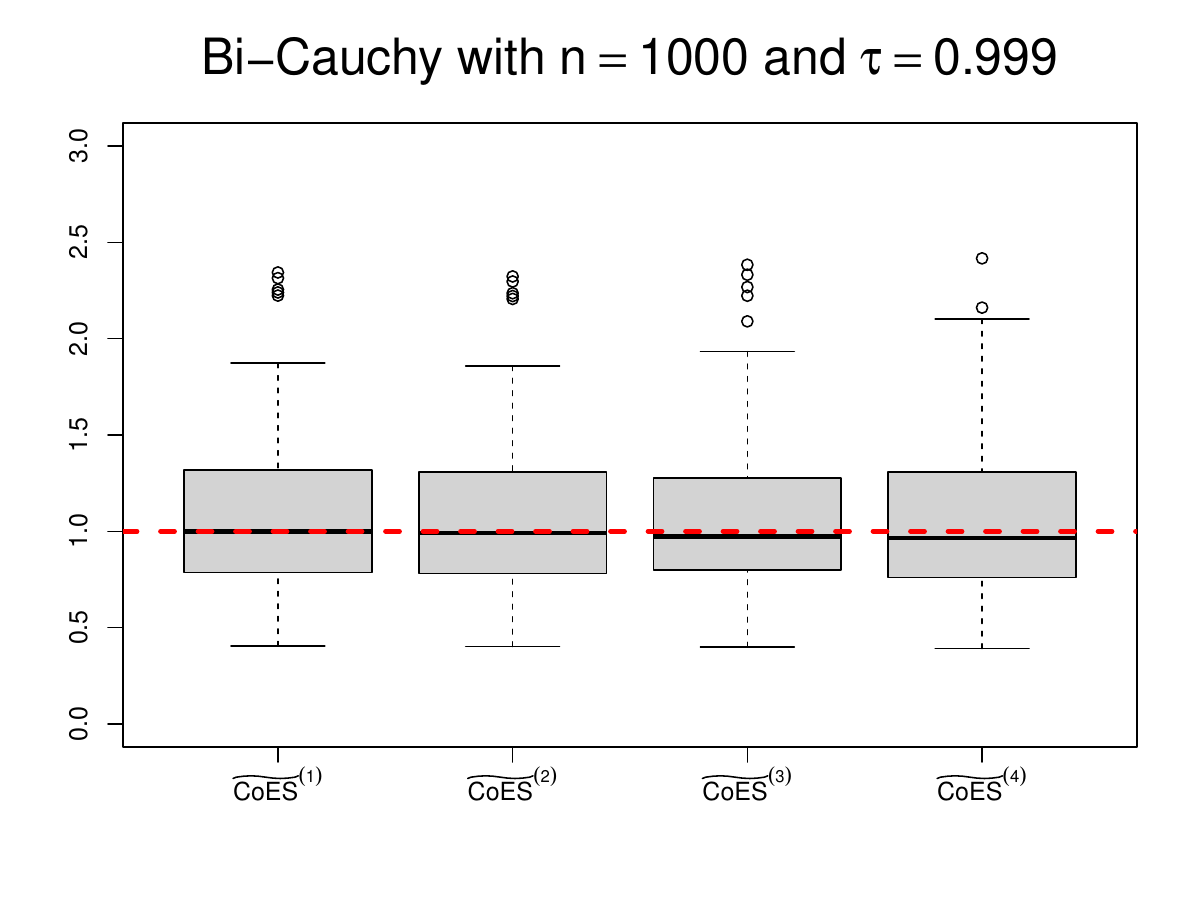}
\end{minipage}
\begin{minipage}[b]{0.24\textwidth}
\includegraphics[width=\textwidth,height = 0.15\textheight]{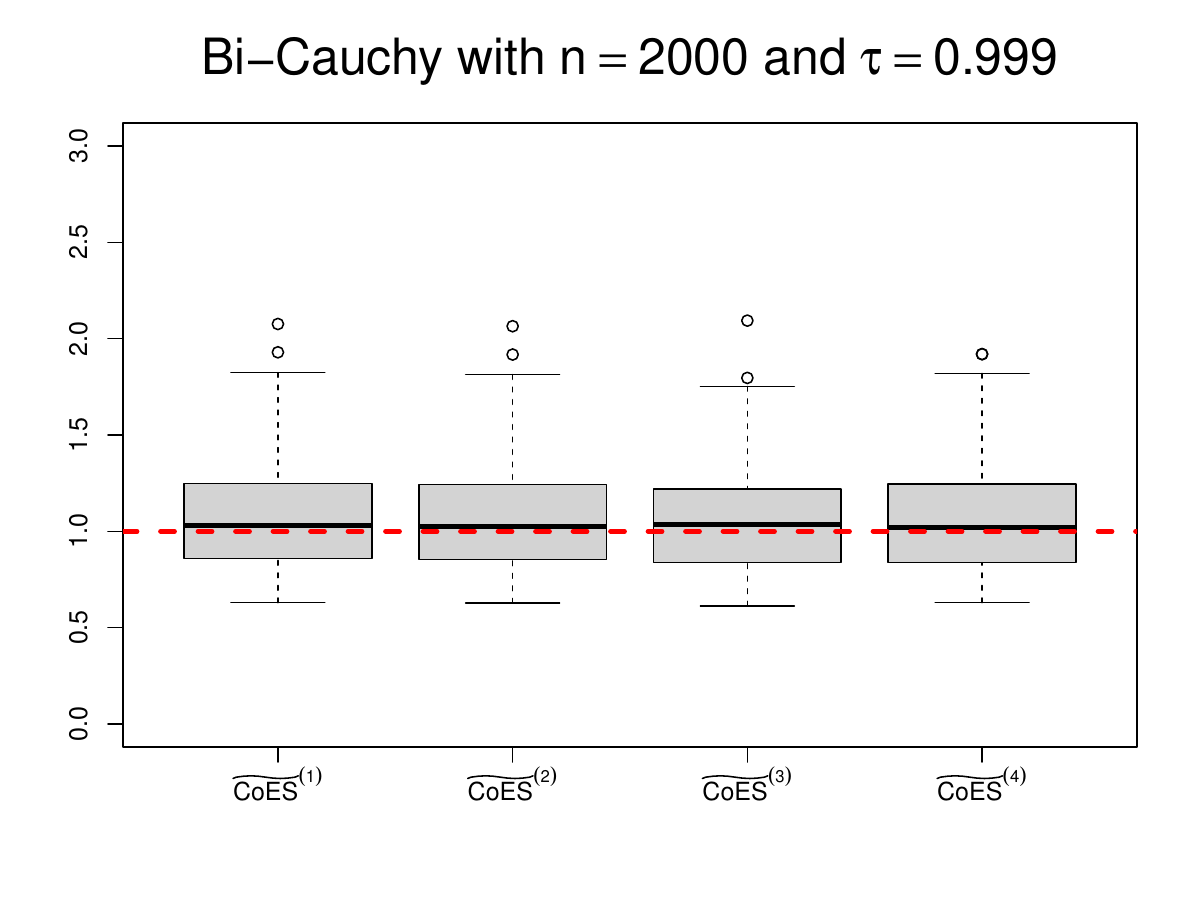}
\end{minipage}
\begin{minipage}[b]{0.24\textwidth}
\includegraphics[width=\textwidth,height = 0.15\textheight]{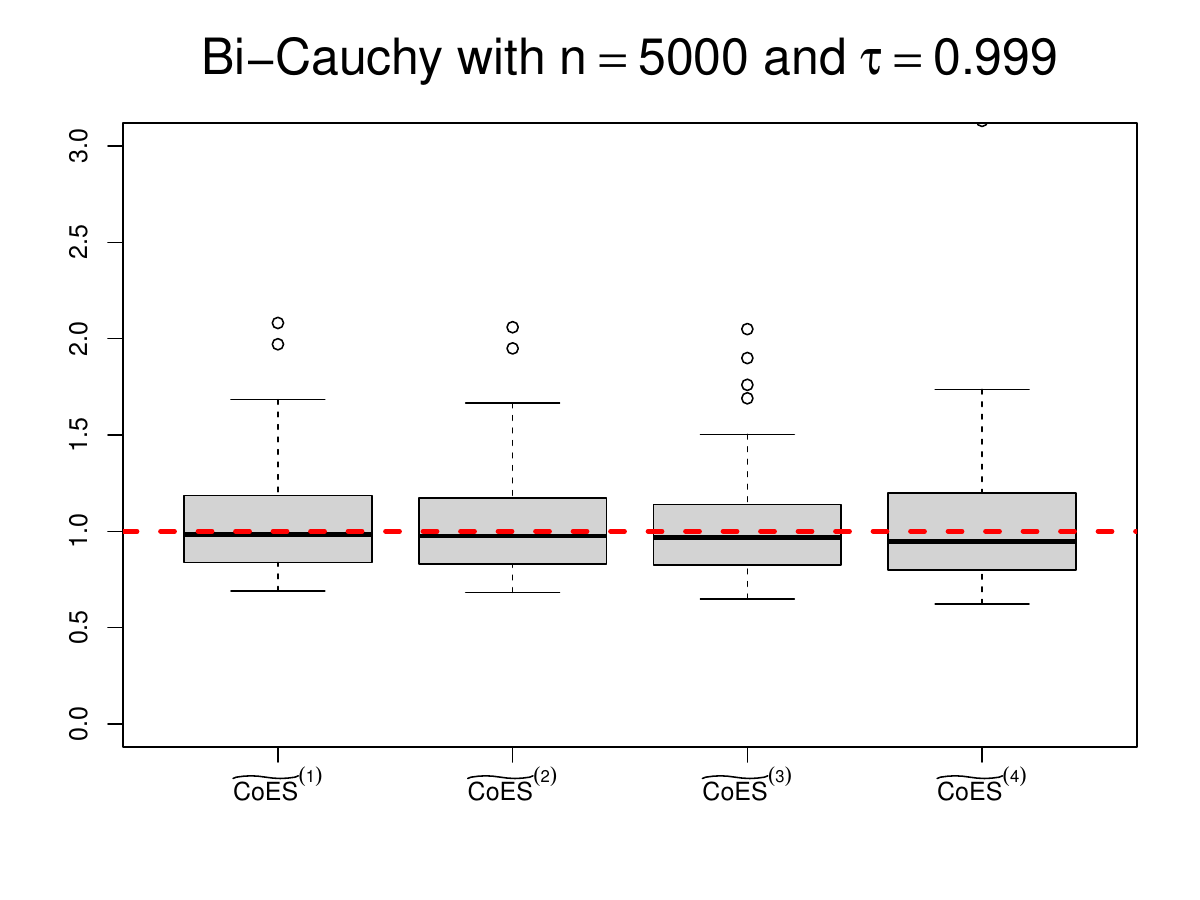}
\end{minipage}
\\
\begin{minipage}[b]{0.24\textwidth}
\includegraphics[width=\textwidth,height = 0.15\textheight]{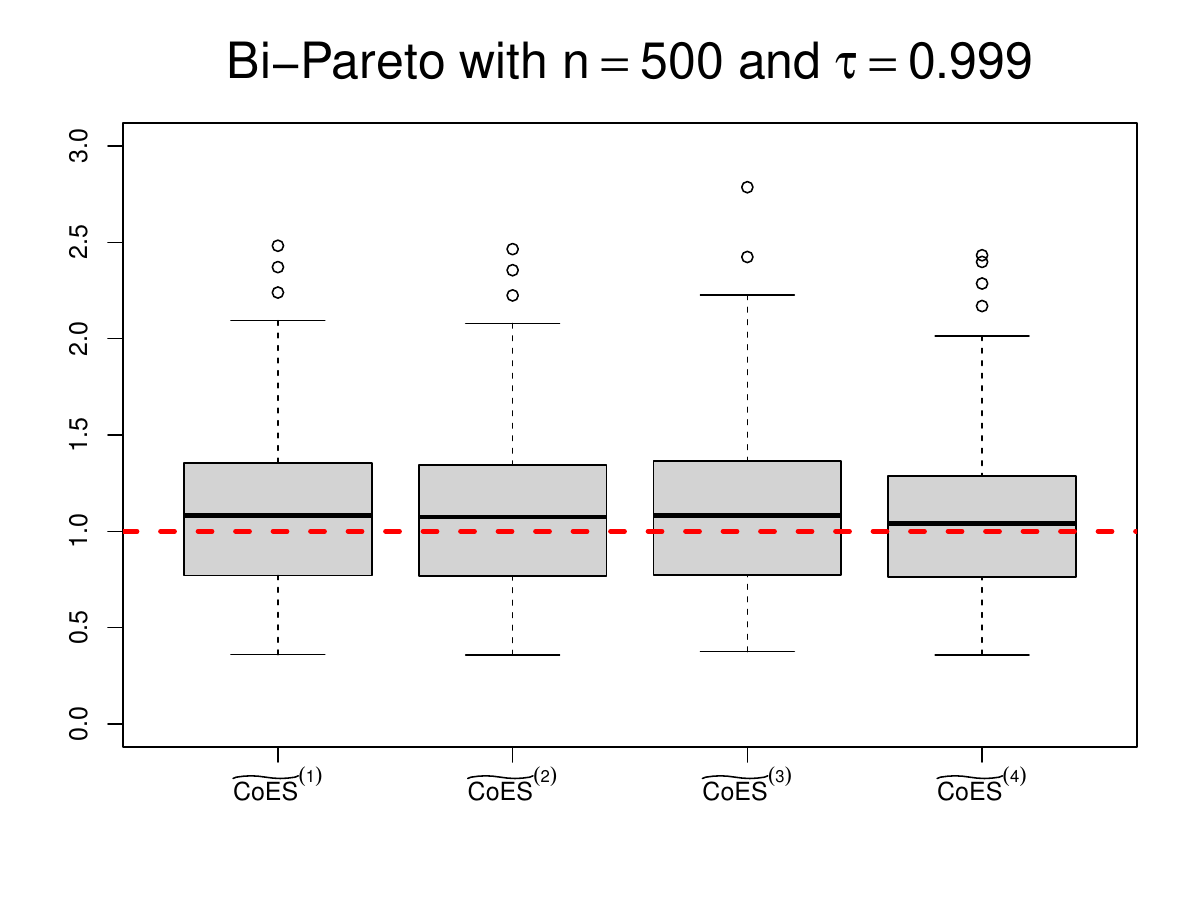}
\end{minipage}
\begin{minipage}[b]{0.24\textwidth}
\includegraphics[width=\textwidth,height = 0.15\textheight]{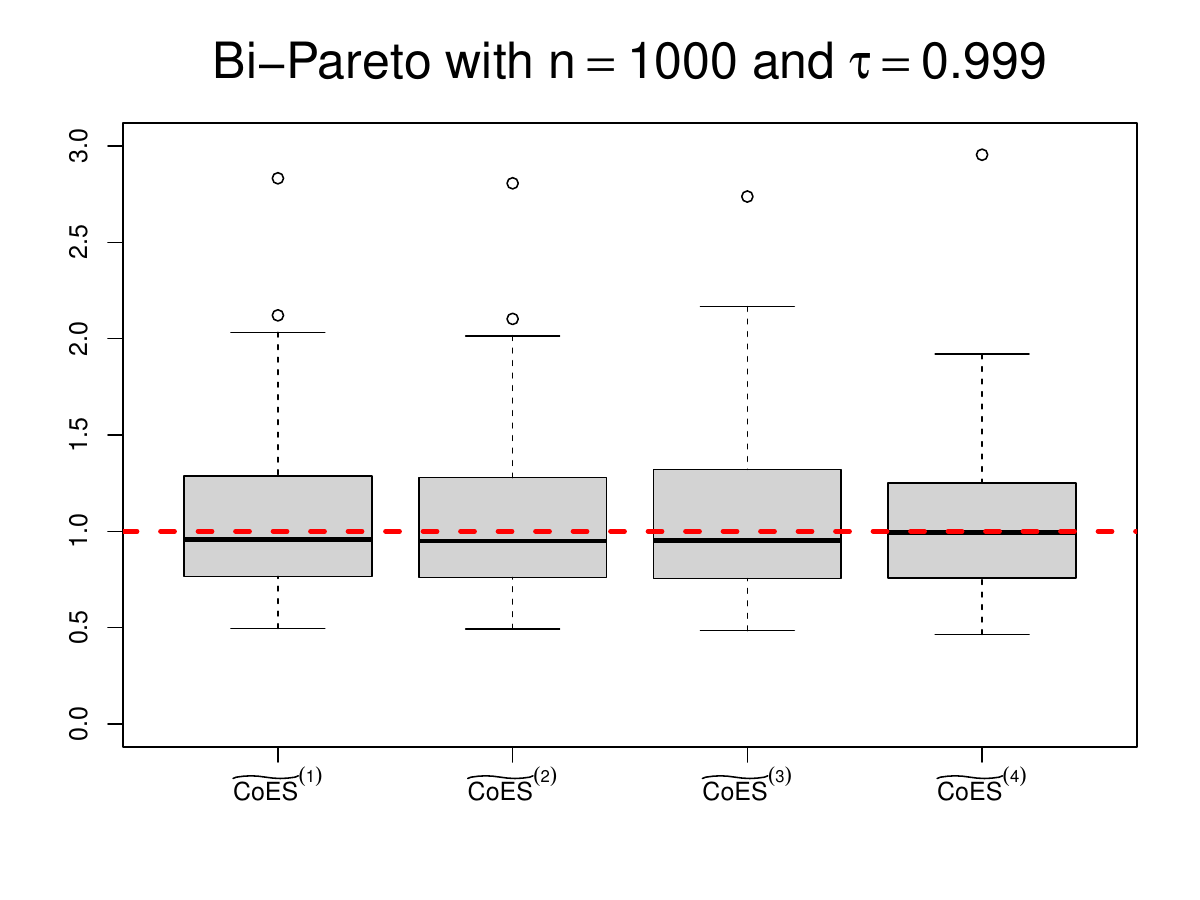}
\end{minipage}
\begin{minipage}[b]{0.24\textwidth}
\includegraphics[width=\textwidth,height = 0.15\textheight]{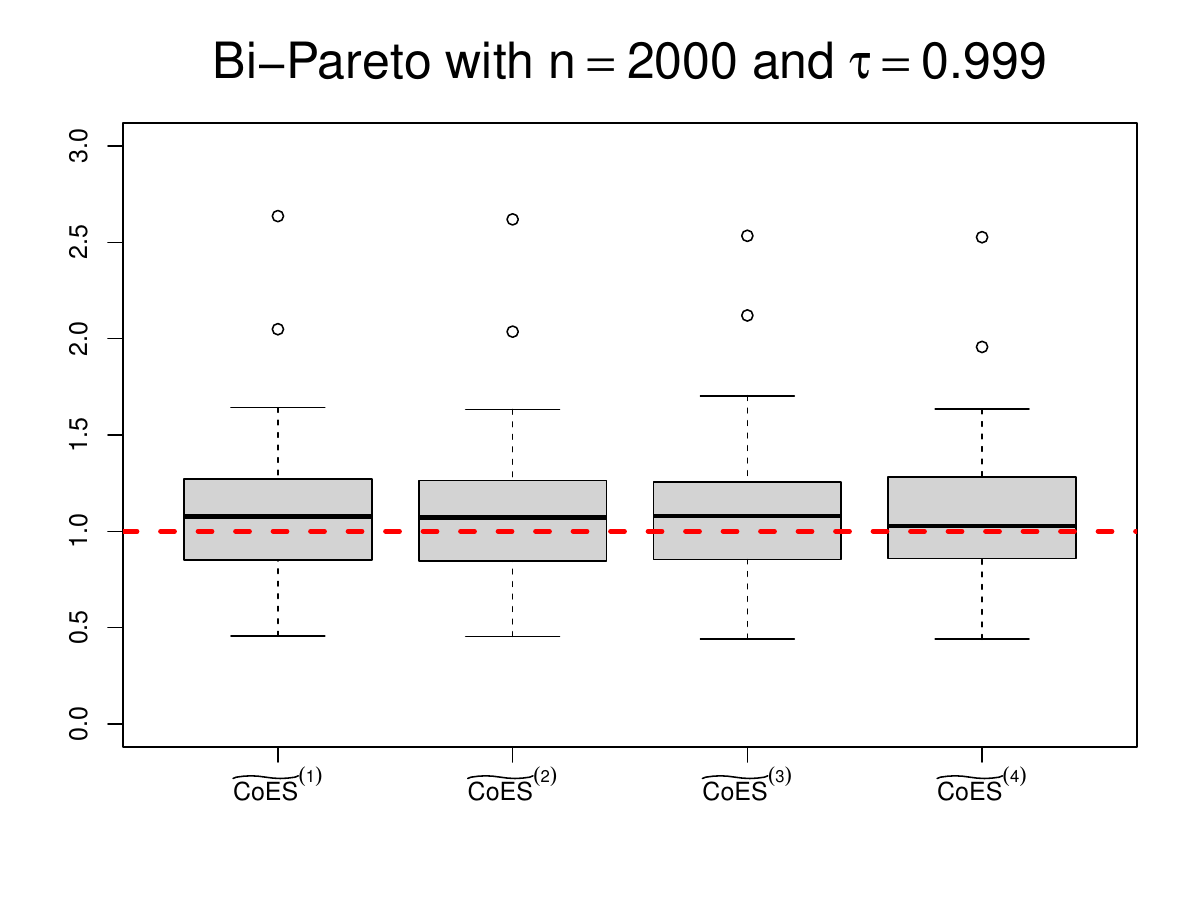}
\end{minipage}
\begin{minipage}[b]{0.24\textwidth}
\includegraphics[width=\textwidth,height = 0.15\textheight]{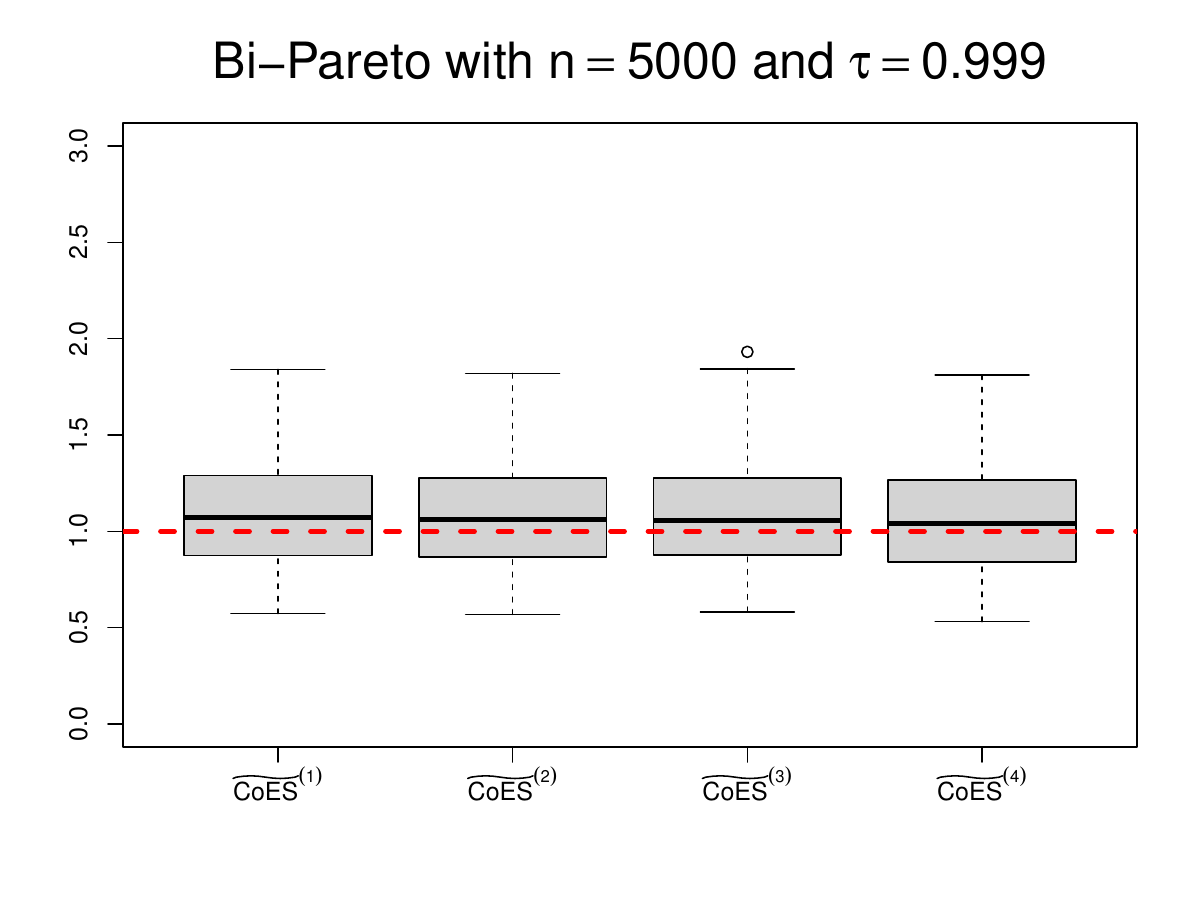}
\end{minipage}
\\
\begin{minipage}[b]{0.24\textwidth}
\includegraphics[width=\textwidth,height = 0.15\textheight]{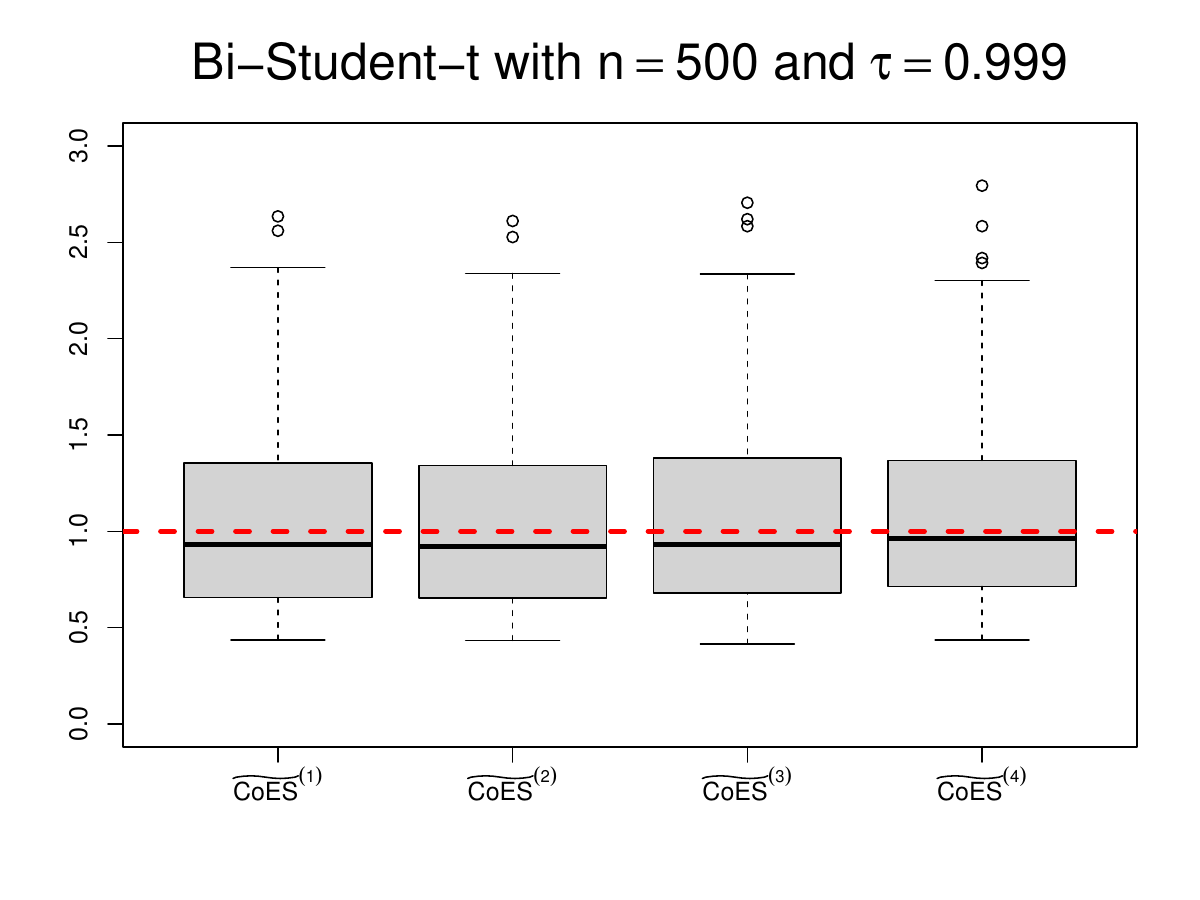}
\end{minipage}
\begin{minipage}[b]{0.24\textwidth}
\includegraphics[width=\textwidth,height = 0.15\textheight]{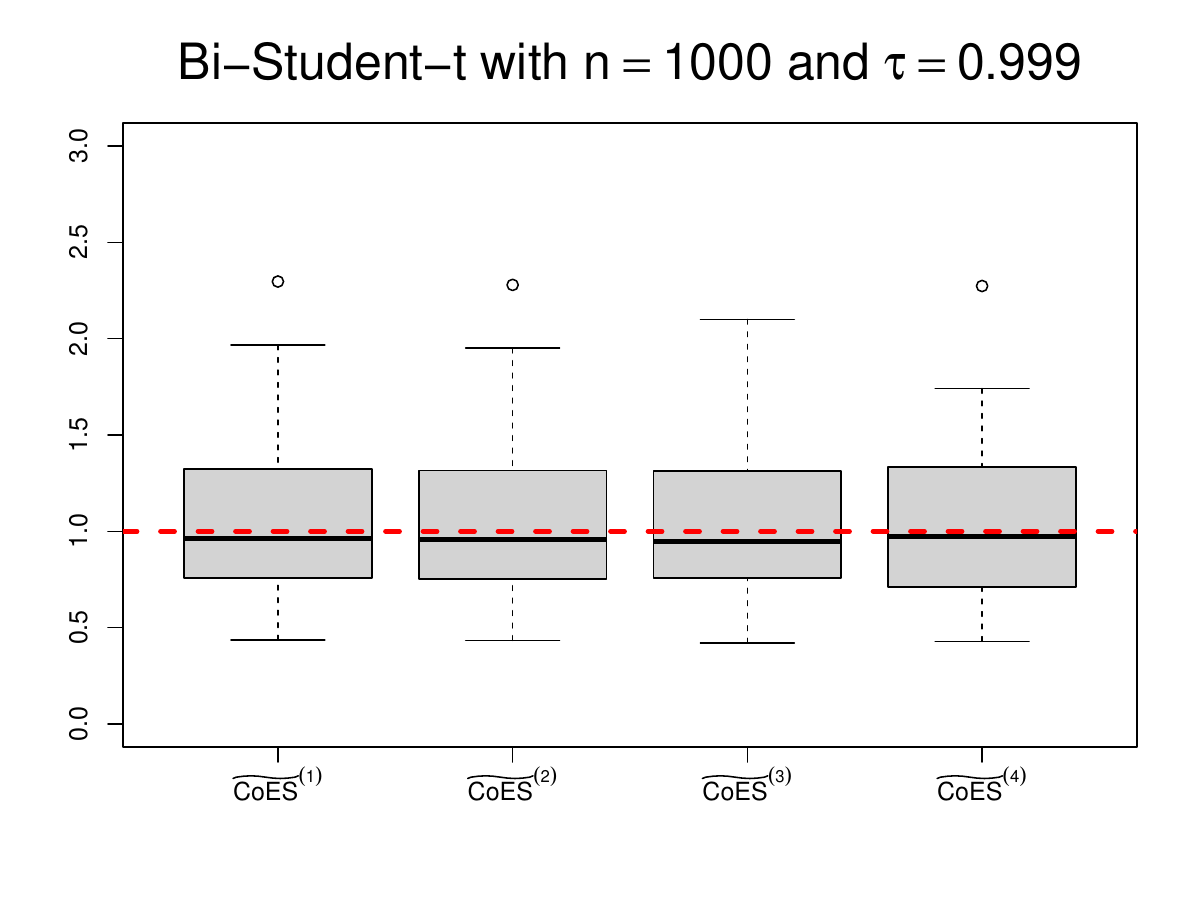}
\end{minipage}
\begin{minipage}[b]{0.24\textwidth}
\includegraphics[width=\textwidth,height = 0.15\textheight]{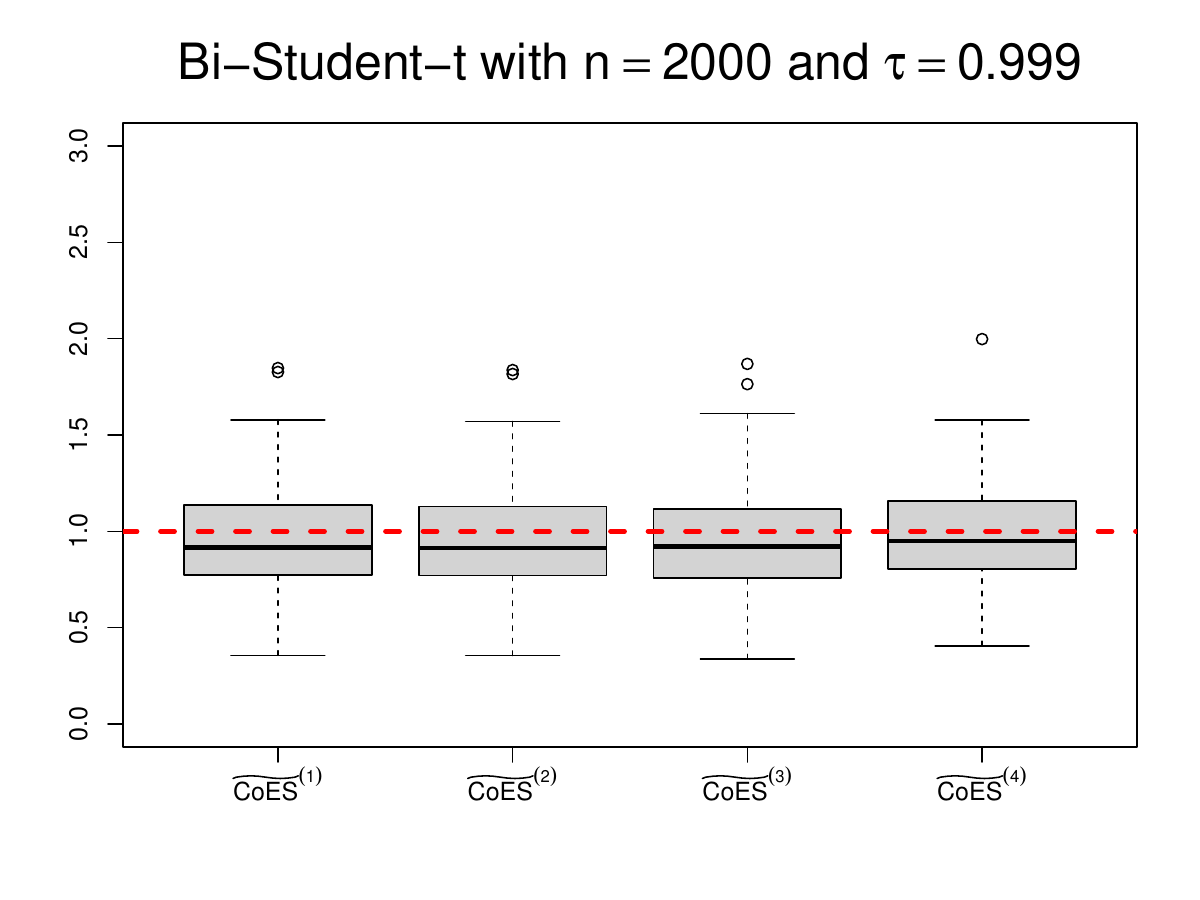}
\end{minipage}
\begin{minipage}[b]{0.24\textwidth}
\includegraphics[width=\textwidth,height = 0.15\textheight]{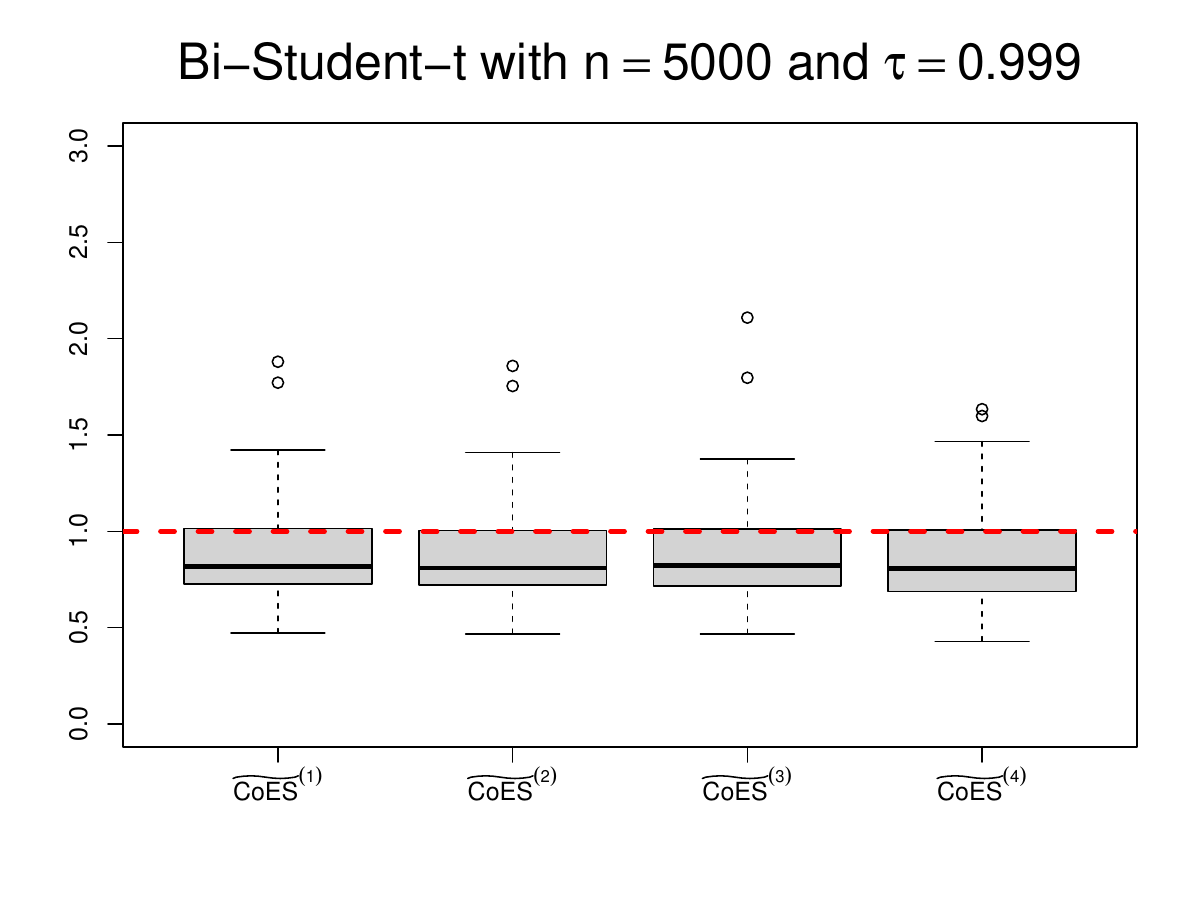}
\end{minipage}
\caption{The boxplots of $\widetilde{\coes}^{(i)}_{X|Y}(\tau'_n)$ for $i=1,2,3,4$ with $\tau'_n = 0.999$, under transformed bivariate Logistic, Cauchy, Pareto and Student-$t$ models. The boxplots from the left panel to the right panel are drawn with $n=500,1000,2000,$ and $5000$, respectively.}
\label{Fig:boxplots_CES_999}
\end{figure}

Tables \ref{tab:msre_99} - \ref{tab:msre_999} collect the values of MSREs and indicate the specific values we choose for intermediate order $k$ for each bivariate distribution. We further present a series of boxplots of the ratios between the estimations $\widetilde{\covar}_{X|Y}^{(i)}(\tau'_n)$, $\widetilde{\coes}_{X|Y}^{(i)}(\tau'_n)$ and the true values $\covar_{X|Y}(\tau'_n)$, $\coes_{X|Y}(\tau'_n)$ in Figures B1 - B3 (for CoVaR) of the supplementary material 
 and Figures \ref{Fig:boxplots_CES_99} - \ref{Fig:boxplots_CES_999} (for CoES), which reflect an intuitive comparison of consistency among these methods. A global observation gives that all these extrapolative methods exhibit a more concentrated pattern, characterized by lower variations and MSREs, as the sample size increases. Moreover, the behaviors of our proposed extrapolations are highly consistent and remain stable across all scenarios. The discrepancies between these methods are negligible, with all methods demonstrating robust performance characteristics, as evidenced by both the MSREs and boxplots. In addition, among the four distributions, the transformed bivariate Cauchy distribution achieves the best performance, as its median ratios are closest to 1 and it exhibits the smallest variations in terms of MSREs. In contrast, the experimental results for the transformed bivariate Logistic model demonstrate the poorest performance across most scenarios, particularly under small sample sizes and very extreme levels. This finding underscores the critical importance of the condition \(\alpha > 1\).

To compare different extrapolative methods, it suffices to note that the distinction between \(\widetilde{\covar}^{(1)}_{X|Y}(\tau'_n)\) and \(\widetilde{\covar}^{(2)}_{X|Y}(\tau'_n)\) lies in \(\hat{\eta}^{(1)}_{1-k/n}\) and \(\hat{\eta}^{(2)}_{1-k/n}\). This implies that the quality of \(\hat{\eta}^{(i)}_{1-k/n}\) significantly governs extrapolation performance. Since \(\hat{\eta}^{(i)}_{1-k/n}\) are intrinsically linked to the estimations of tail dependence \(R\), the accuracy in estimating \(R\) fundamentally determines extrapolation behaviors. Crucially, both \(\widehat{R}^{(i)}_n\) defined in \eqref{eq:nonp_R_emp} and \eqref{eq:nonp_R_rank} provide excellent approximations for \(R\), resulting in both minimal divergence between \(\widetilde{\covar}^{(1)}_{X|Y}(\tau'_n)\) and \(\widetilde{\covar}^{(2)}_{X|Y}(\tau'_n)\), and consistently superior performances. The same holds true for \(\widetilde{\coes}^{(i)}_{X|Y}(\tau'_n)\) with \(i=1,2,3\). This is because their expressions all take the form of \(\widetilde{\covar}^{(i)}_{X|Y}(\tau'_n)\) with \(i=1,2,3\) multiplied by a \(\hat{\gamma}_1\)-dependent term, which implies that the disparities in their empirical performances are entirely determined by \(\widetilde{\covar}^{(i)}_{X|Y}(\tau'_n)\). On the other hand, \(\widetilde{\covar}^{(3)}_{X|Y}(\tau'_n)\) and \(\widetilde{\coes}^{(4)}_{X|Y}(\tau'_n)\) are constructed based on the intermediate estimators \(\widehat{\covar}_{X|Y}(1-k/n)\) and \(\widehat{\coes}_{X|Y}(1-k/n)\). The former enjoys similar empirical features to \(\widetilde{\covar}^{(i)}_{X|Y}(\tau'_n)\) with \(i=1,2\), while the latter shows better properties on minor samples but slightly poorer performances on larger sample sizes.


\begin{table}
\centering
\caption{The MSREs of $\widetilde{\covar}_{X|Y}^{(i)}(\tau'_n)$ ($i=1,2,3$) and $\widetilde{\coes}_{X|Y}^{(i)}(\tau'_n)$ ($i=1,2,3,4$) with $\tau'_n = 0.99$, under transformed bivariate Logistic, Cauchy, Pareto and Student-$t$ models.}
\label{tab:msre_99}
\setlength{\tabcolsep}{5.5pt}
\begin{tabular}{@{}cccccccc@{}}
\hline\hline
\addlinespace[5pt]
 & $\widetilde{\covar}_{X|Y}^{(1)}$ & $\widetilde{\covar}_{X|Y}^{(2)}$ & $\widetilde{\covar}_{X|Y}^{(3)}$ & $\widetilde{\coes}_{X|Y}^{(1)}$ & $\widetilde{\coes}_{X|Y}^{(2)}$ & $\widetilde{\coes}_{X|Y}^{(3)}$ & $\widetilde{\coes}_{X|Y}^{(4)}$ \\[2pt]
\midrule
& \multicolumn{7}{c}{$n=500$, $k=90$ for Student-$t$ and $k=120$ for others} \\[2pt]
\midrule
Bi-Logistic      & 0.08782 & 0.08529 & 0.09128 & 0.14442 & 0.14059 & 0.15022 & 0.13952 \\[3pt]
Bi-Cauchy        & 0.05395 & 0.05209 & 0.05396 & 0.08372 & 0.08093 & 0.08294 & 0.08429 \\[5pt]
Bi-Pareto        & 0.07352 & 0.07155 & 0.08623 & 0.11510 & 0.11218 & 0.13379 & 0.10801 \\[5pt]
Bi-Student-$t$   & 0.09146 & 0.08753 & 0.10143 & 0.15453 & 0.14822 & 0.16873 & 0.15182 \\[5pt]
\midrule
& \multicolumn{7}{c}{$n=1000$, $k=150$} \\[2pt]
\midrule
Bi-Logistic     & 0.06927 & 0.06665 & 0.07175 & 0.11165 & 0.10777 & 0.11461 & 0.12218 \\[5pt]
Bi-Cauchy       & 0.04070 & 0.03995 & 0.04232 & 0.06156 & 0.06026 & 0.06288 & 0.06401 \\[5pt]
Bi-Pareto       & 0.05583 & 0.05434 & 0.06046 & 0.08725 & 0.08498 & 0.09316 & 0.07976 \\[5pt]
Bi-Student-$t$  & 0.05602 & 0.05388 & 0.05651 & 0.09383 & 0.09049 & 0.09333 & 0.08914 \\[5pt]
\midrule
& \multicolumn{7}{c}{$n=2000$, $k=250$} \\[2pt]
\midrule
Bi-Logistic     & 0.03794 & 0.03674 & 0.03722 & 0.05905 & 0.05728 & 0.05748 & 0.07784 \\[5pt]
Bi-Cauchy       & 0.01873 & 0.01845 & 0.01915 & 0.02853 & 0.02802 & 0.02839 & 0.03249 \\[5pt]
Bi-Pareto       & 0.03071 & 0.02989 & 0.03159 & 0.04649 & 0.04534 & 0.04708 & 0.04604 \\[5pt]
Bi-Student-$t$  & 0.02625 & 0.02541 & 0.02797 & 0.04242 & 0.04109 & 0.04395 & 0.05077 \\[5pt]
\midrule
& \multicolumn{7}{c}{$n=5000$, $k=300$} \\[2pt]
\midrule
Bi-Logistic    & 0.02857 & 0.02720 & 0.02822 & 0.04426 & 0.04232 & 0.04265 & 0.05283 \\[5pt]
Bi-Cauchy      & 0.01776 & 0.01733 & 0.01822 & 0.02736 & 0.02663 & 0.02697 & 0.05510 \\[5pt]
Bi-Pareto      & 0.01902 & 0.01769 & 0.02291 & 0.02911 & 0.02737 & 0.03346 & 0.03961 \\[5pt]
Bi-Student-$t$ & 0.02385 & 0.02283 & 0.02389 & 0.03742 & 0.03589 & 0.03684 & 0.04467 \\[5pt]
\hline\hline
\end{tabular}
\end{table}

\begin{table}
\centering
\caption{The MSREs of $\widetilde{\covar}_{X|Y}^{(i)}(\tau'_n)$ ($i=1,2,3$) and $\widetilde{\coes}_{X|Y}^{(i)}(\tau'_n)$ ($i=1,2,3,4$) with $\tau'_n = 0.995$, under transformed bivariate Logistic, Cauchy, Pareto and Student-$t$ models.}
\label{tab:msre_995}
\setlength{\tabcolsep}{5.5pt}
\begin{tabular}{@{}cccccccc@{}}
\hline\hline
\addlinespace[5pt]
 & $\widetilde{\covar}_{X|Y}^{(1)}$ & $\widetilde{\covar}_{X|Y}^{(2)}$ & $\widetilde{\covar}_{X|Y}^{(3)}$ & $\widetilde{\coes}_{X|Y}^{(1)}$ & $\widetilde{\coes}_{X|Y}^{(2)}$ & $\widetilde{\coes}_{X|Y}^{(3)}$ & $\widetilde{\coes}_{X|Y}^{(4)}$ \\[2pt]
\midrule
& \multicolumn{7}{c}{$n=500$, $k=90$ for Student-$t$ and $k=120$ for others} \\[2pt]
\midrule
Bi-Logistic  & 0.13672 & 0.13304 & 0.14212 & 0.21579 & 0.21046 & 0.22447 & 0.20599 \\[3pt]
Bi-Cauchy    & 0.07833 & 0.07576 & 0.07771 & 0.11793 & 0.11419 & 0.11628 & 0.11634 \\[5pt]
Bi-Pareto    & 0.10950 & 0.10671 & 0.12723 & 0.16450 & 0.16053 & 0.19014 & 0.15317 \\[5pt]
Bi-Student-$t$   & 0.14452 & 0.13859 & 0.15803 & 0.27961 & 0.26857 & 0.30118 & 0.27757\\[5pt]
\midrule
& \multicolumn{7}{c}{$n=1000$, $k=150$} \\[2pt]
\midrule
Bi-Logistic  & 0.10633 & 0.10260 & 0.10920 & 0.16404 & 0.15873 & 0.16758 & 0.17424 \\[5pt]
Bi-Cauchy    & 0.05855 & 0.05742 & 0.05995 & 0.08525 & 0.08340 & 0.08625 & 0.08589 \\[5pt]
Bi-Pareto    & 0.08327 & 0.08111 & 0.08897 & 0.12419 & 0.12105 & 0.13160 & 0.11201 \\[5pt]
Bi-Student-$t$   & 0.08825 & 0.08509 & 0.08792 & 0.17529 & 0.16936 & 0.17303 & 0.16436 \\[5pt]
\midrule
& \multicolumn{7}{c}{$n=2000$, $k=250$} \\[2pt]
\midrule
Bi-Logistic    & 0.05668 & 0.05497 & 0.05518 & 0.08465 & 0.08223 & 0.08214 & 0.10785 \\[5pt]
Bi-Cauchy      & 0.02711 & 0.02668 & 0.02715 & 0.03960 & 0.03886 & 0.03891 & 0.04269 \\[5pt]
Bi-Pareto      & 0.04482 & 0.04370 & 0.04545 & 0.06447 & 0.06295 & 0.06471 & 0.06215 \\[5pt]
Bi-Student-$t$ & 0.04020 & 0.03894 & 0.04176 & 0.08180 & 0.07918 & 0.08260 & 0.09286 \\[5pt]
\midrule
& \multicolumn{7}{c}{$n=5000$, $k=300$} \\[2pt]
\midrule
Bi-Logistic    & 0.04259 & 0.04070 & 0.04112 & 0.06293 & 0.06032 & 0.05988 & 0.07032 \\[5pt]
Bi-Cauchy      & 0.02610 & 0.02551 & 0.02601 & 0.03812 & 0.03711 & 0.03696 & 0.06884 \\[5pt]
Bi-Pareto      & 0.02811 & 0.02642 & 0.03241 & 0.04054 & 0.03837 & 0.04538 & 0.04998 \\[5pt]
Bi-Student-$t$ & 0.03573 & 0.03429 & 0.03522 & 0.06645 & 0.06316 & 0.06362 & 0.06893 \\[5pt]
\hline\hline
\end{tabular}
\end{table}

\begin{table}
\centering
\caption{The MSREs of $\widetilde{\covar}_{X|Y}^{(i)}(\tau'_n)$ ($i=1,2,3$) and $\widetilde{\coes}_{X|Y}^{(i)}(\tau'_n)$ ($i=1,2,3,4$) with $\tau'_n = 0.999$, under transformed bivariate Logistic, Cauchy, Pareto and Student-$t$ models.}
\label{tab:msre_999}
\setlength{\tabcolsep}{5.5pt}
\begin{tabular}{@{}cccccccc@{}}
\hline\hline
\addlinespace[5pt]
 & $\widetilde{\covar}_{X|Y}^{(1)}$ & $\widetilde{\covar}_{X|Y}^{(2)}$ & $\widetilde{\covar}_{X|Y}^{(3)}$ & $\widetilde{\coes}_{X|Y}^{(1)}$ & $\widetilde{\coes}_{X|Y}^{(2)}$ & $\widetilde{\coes}_{X|Y}^{(3)}$ & $\widetilde{\coes}_{X|Y}^{(4)}$ \\[2pt]
\midrule
& \multicolumn{7}{c}{$n=500$, $k=90$ for Student-$t$ and $k=120$ for others} \\[2pt]
\midrule
Bi-Logistic  & 0.33333 & 0.32574 & 0.34671 & 0.49183 & 0.48146 & 0.51226 & 0.46031 \\[3pt]
Bi-Cauchy    & 0.17202 & 0.16687 & 0.16904 & 0.25426 & 0.24695 & 0.24933 & 0.24463 \\[5pt]
Bi-Pareto    & 0.24254 & 0.23699 & 0.27902 & 0.34426 & 0.33672 & 0.39562 & 0.31782 \\[5pt]
Bi-Student-$t$   & 0.30857 & 0.29777 & 0.33215 & 0.29428 & 0.28583 & 0.31654 & 0.28327 \\[5pt]
\midrule
& \multicolumn{7}{c}{$n=1000$, $k=150$} \\[2pt]
\midrule
Bi-Logistic  & 0.24884 & 0.24140 & 0.25332 & 0.35783 & 0.34783 & 0.36376 & 0.36478 \\[5pt]
Bi-Cauchy    & 0.12177 & 0.11908 & 0.12229 & 0.17295 & 0.16887 & 0.17271 & 0.16710 \\[5pt]
Bi-Pareto    & 0.18191 & 0.17745 & 0.19167 & 0.25467 & 0.24855 & 0.26783 & 0.22757 \\[5pt]
Bi-Student-$t$   & 0.17675 & 0.17156 & 0.17400 & 0.15428 & 0.15102 & 0.15177 & 0.14115 \\[5pt]
\midrule
& \multicolumn{7}{c}{$n=2000$, $k=250$} \\[2pt]
\midrule
Bi-Logistic  & 0.12536 & 0.12198 & 0.12147 & 0.17484 & 0.17037 & 0.16951 & 0.21248 \\[5pt]
Bi-Cauchy    & 0.05659 & 0.05552 & 0.05510 & 0.08065 & 0.07897 & 0.07777 & 0.08115 \\[5pt]
Bi-Pareto    & 0.09166 & 0.08962 & 0.09135 & 0.12261 & 0.12000 & 0.12160 & 0.11489 \\[5pt]
Bi-Student-$t$   & 0.07400 & 0.07220 & 0.07544 & 0.06780 & 0.06739 & 0.07007 & 0.06866 \\[5pt]
\midrule
& \multicolumn{7}{c}{$n=5000$, $k=300$} \\[2pt]
\midrule
Bi-Logistic  & 0.09212 & 0.08849 & 0.08690 & 0.12656 & 0.12181 & 0.11883 & 0.12963 \\[5pt]
Bi-Cauchy    & 0.05443 & 0.05297 & 0.05216 & 0.07610 & 0.07370 & 0.07179 & 0.11781 \\[5pt]
Bi-Pareto    & 0.05769 & 0.05490 & 0.06323 & 0.07638 & 0.07295 & 0.08271 & 0.08290 \\[5pt]
Bi-Student-$t$   & 0.06697 & 0.06519 & 0.06681 & 0.07978 & 0.08014 & 0.08323 & 0.08681 \\[5pt]
\hline\hline
\end{tabular}
\end{table}

\begin{figure}[htbp]
\centering
\begin{minipage}[b]{0.24\textwidth}
\includegraphics[width=\textwidth,height = 0.15\textheight]{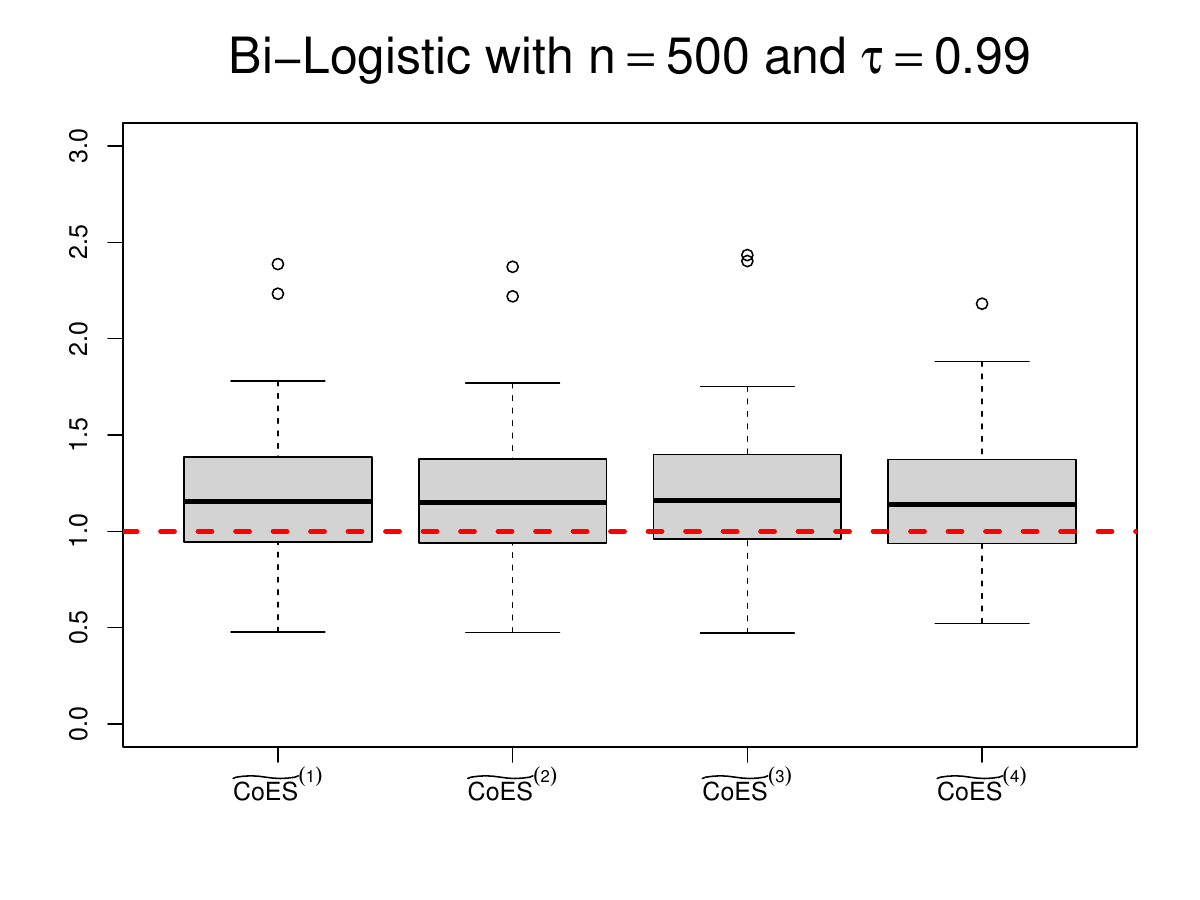}
\end{minipage}
\begin{minipage}[b]{0.24\textwidth}
\includegraphics[width=\textwidth,height = 0.15\textheight]{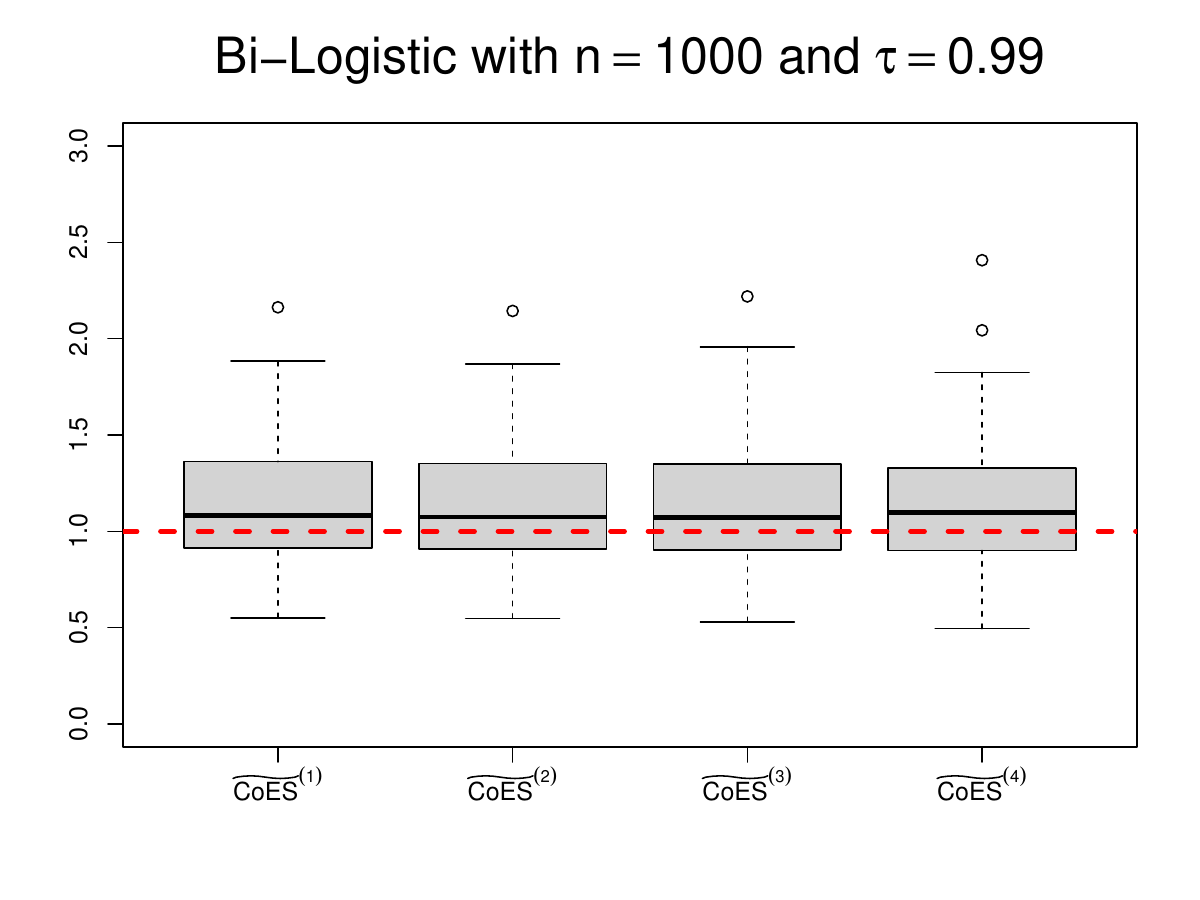}
\end{minipage}
\begin{minipage}[b]{0.24\textwidth}
\includegraphics[width=\textwidth,height = 0.15\textheight]{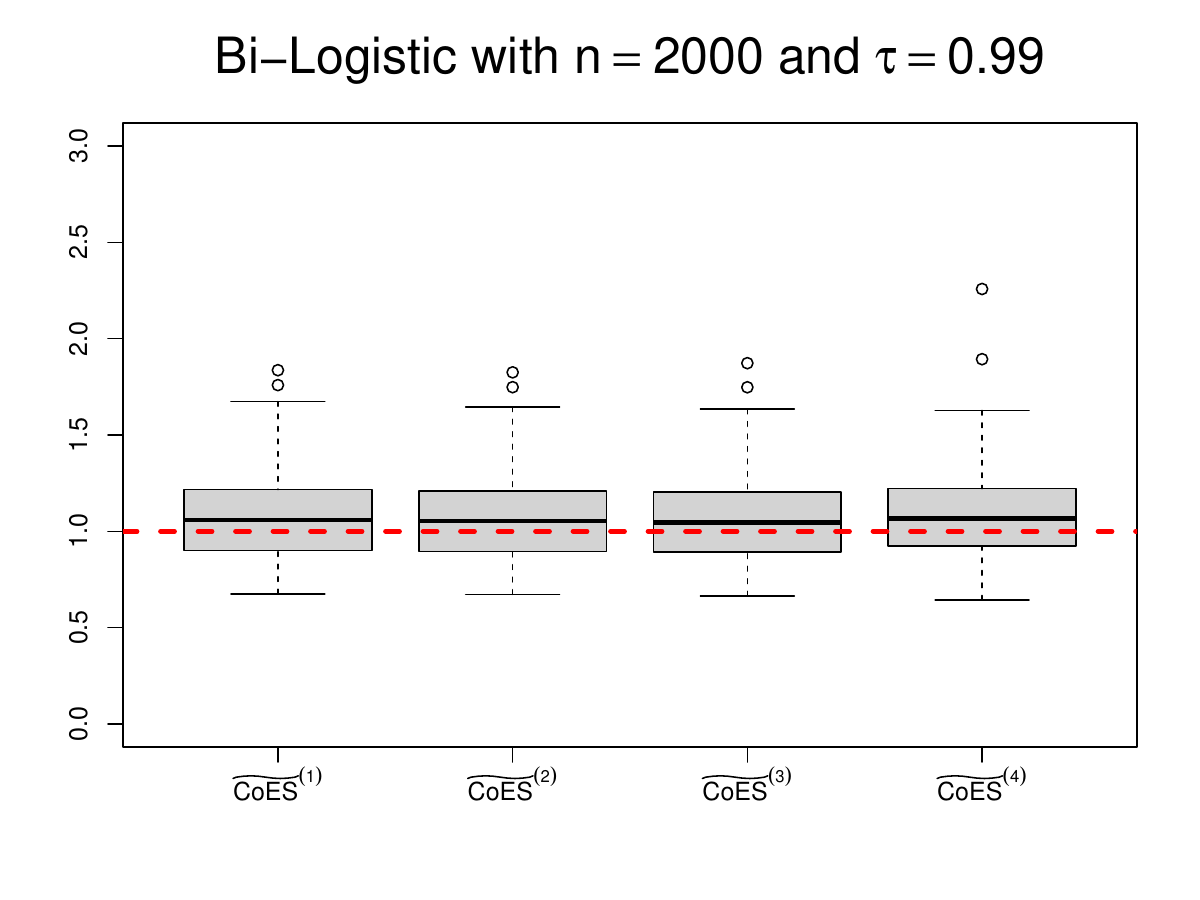}
\end{minipage}
\begin{minipage}[b]{0.24\textwidth}
\includegraphics[width=\textwidth,height = 0.15\textheight]{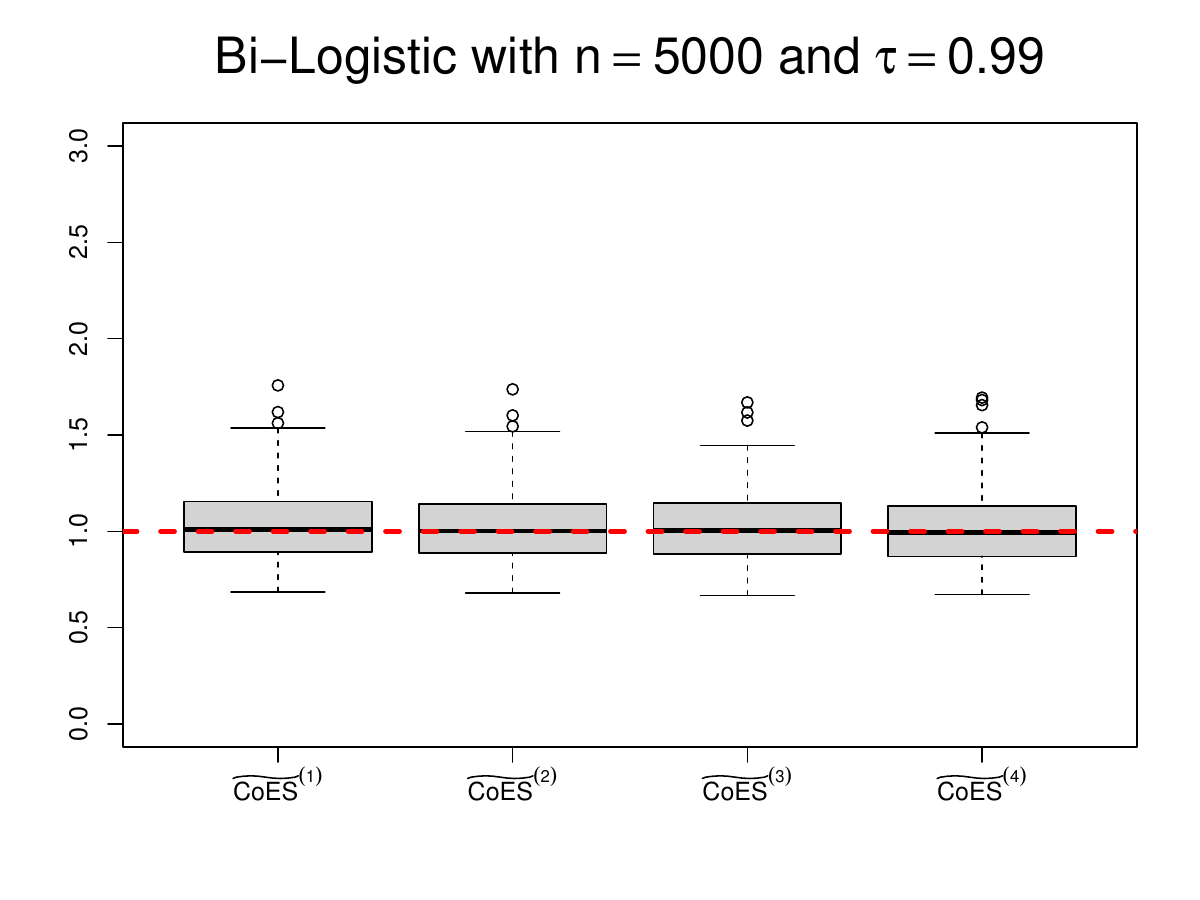}
\end{minipage}
\\
\begin{minipage}[b]{0.24\textwidth}
\includegraphics[width=\textwidth,height = 0.15\textheight]{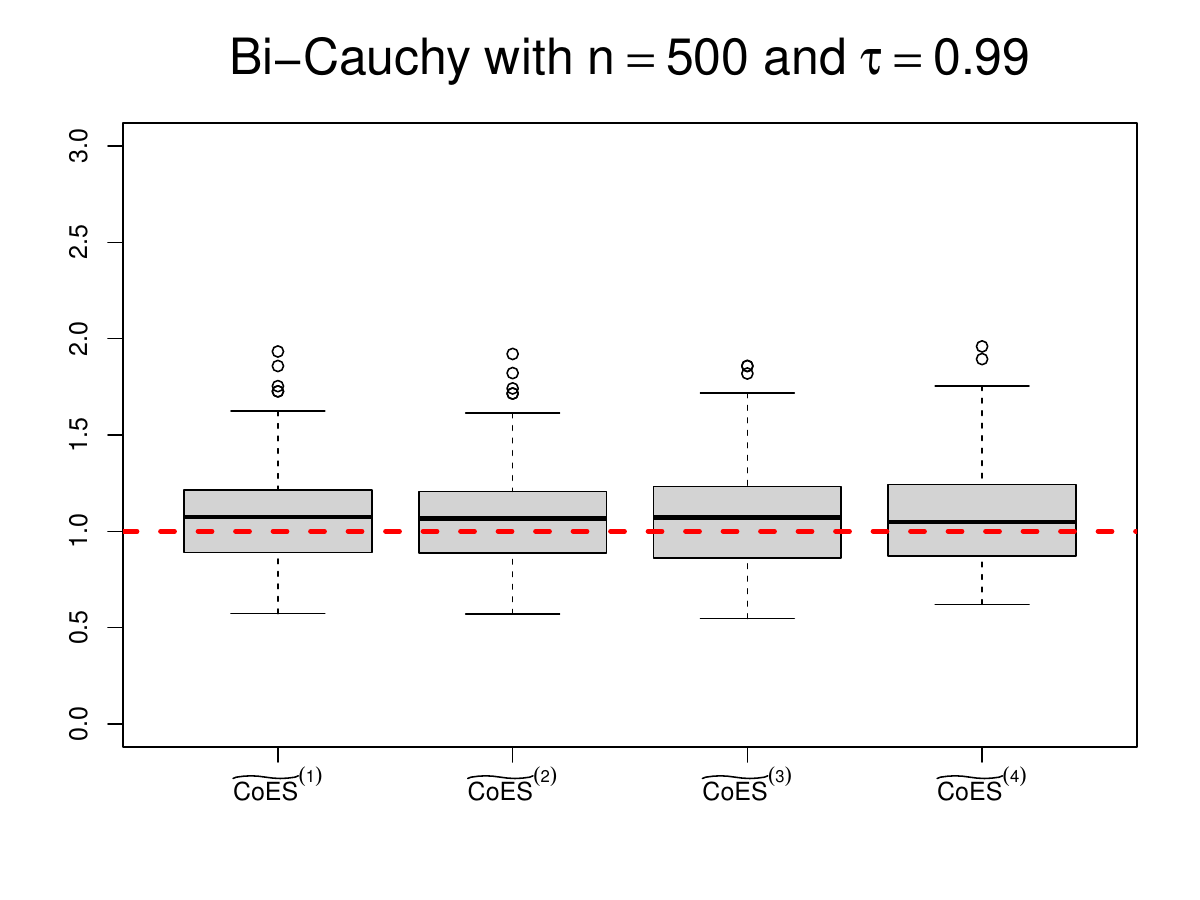}
\end{minipage}
\begin{minipage}[b]{0.24\textwidth}
\includegraphics[width=\textwidth,height = 0.15\textheight]{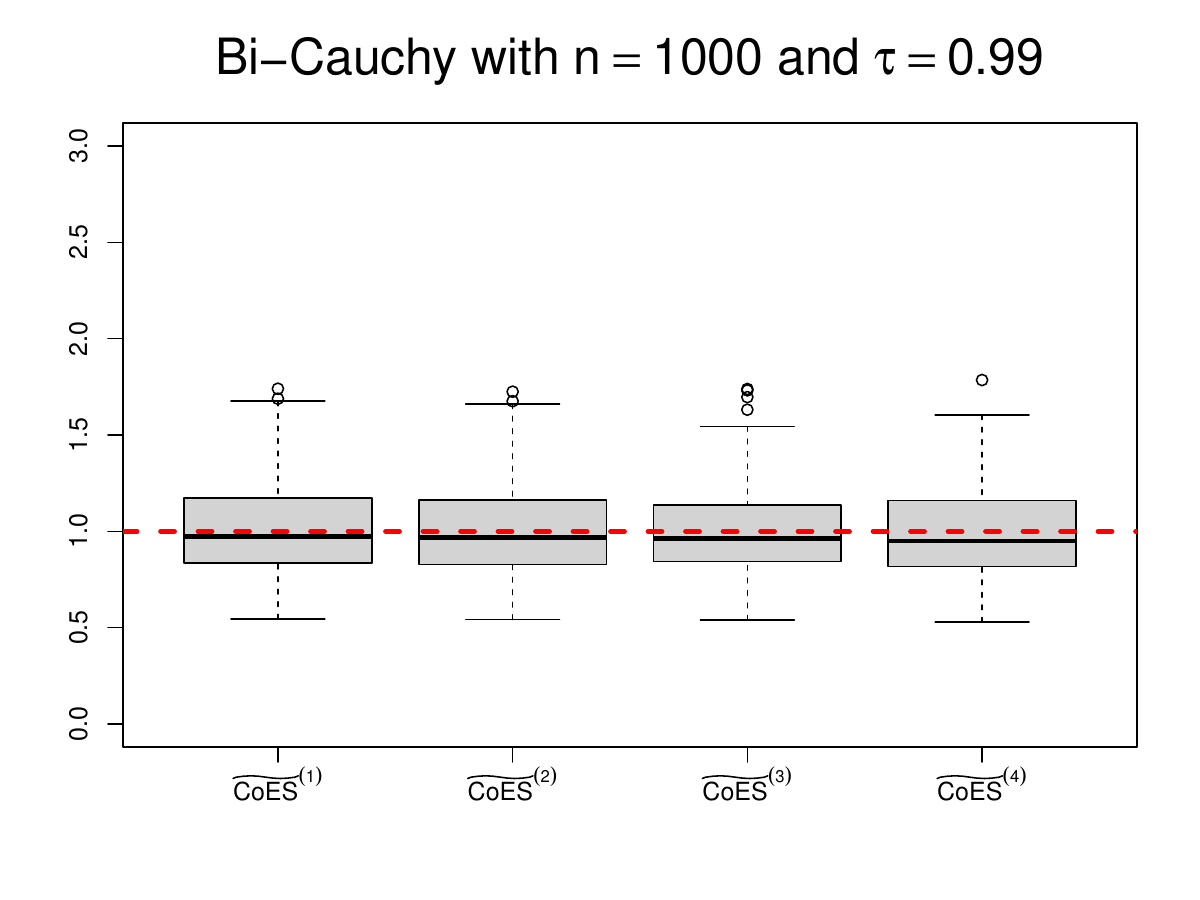}
\end{minipage}
\begin{minipage}[b]{0.24\textwidth}
\includegraphics[width=\textwidth,height = 0.15\textheight]{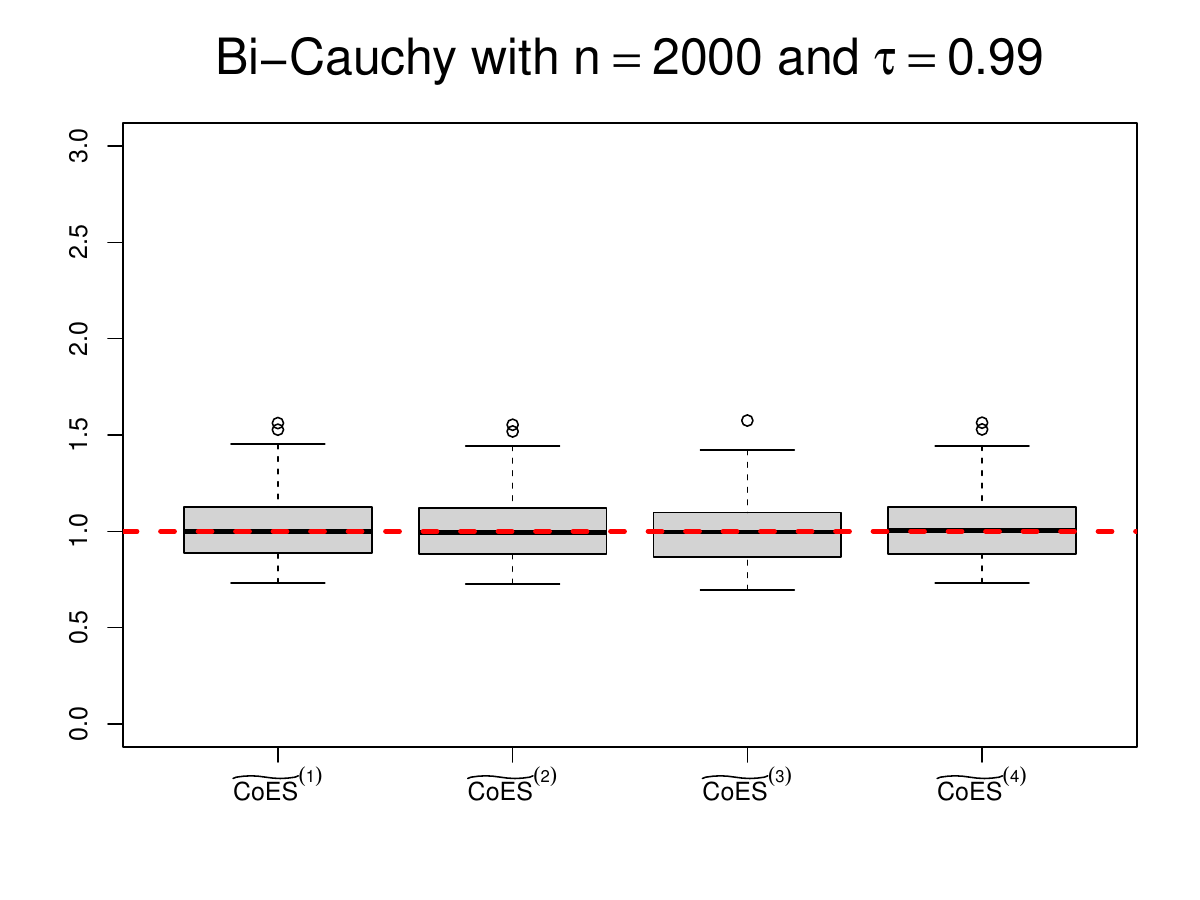}
\end{minipage}
\begin{minipage}[b]{0.24\textwidth}
\includegraphics[width=\textwidth,height = 0.15\textheight]{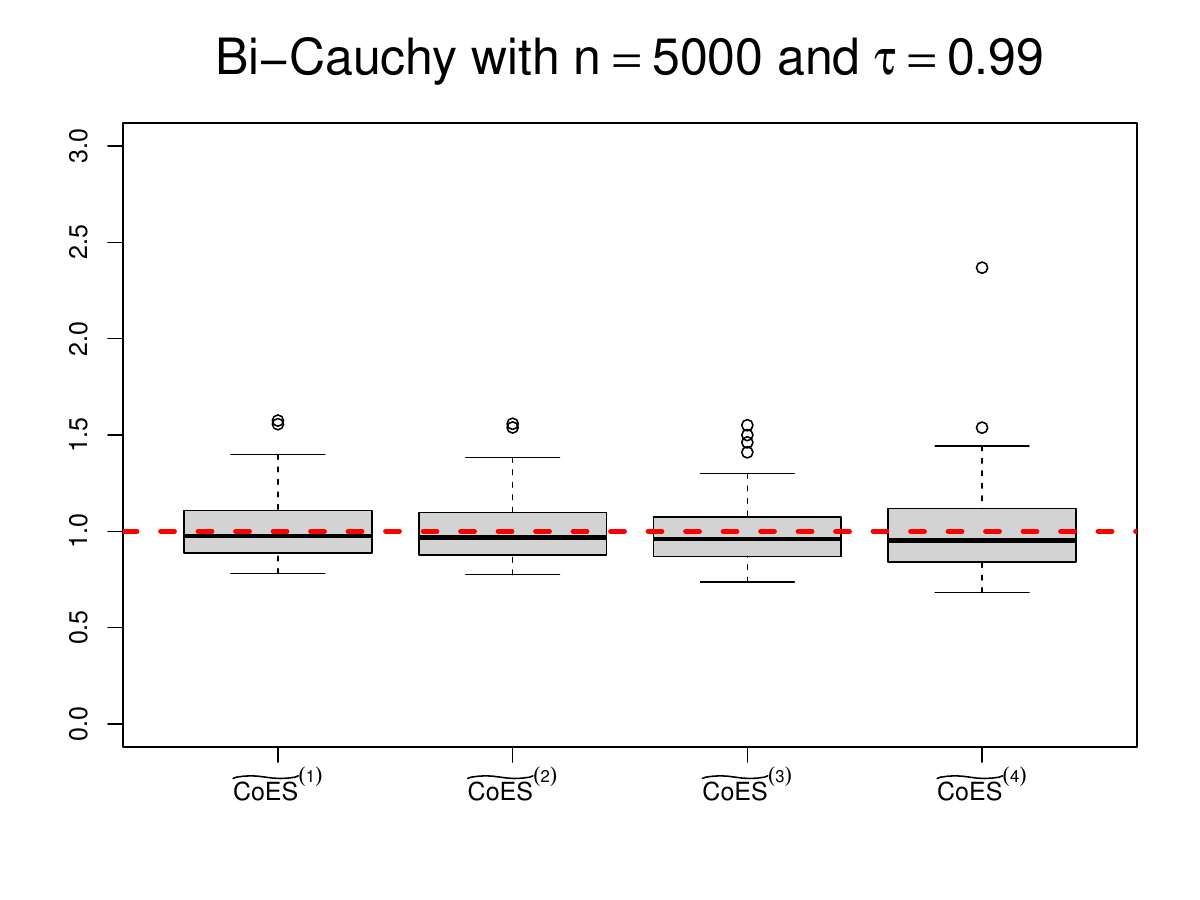}
\end{minipage}
\\
\begin{minipage}[b]{0.24\textwidth}
\includegraphics[width=\textwidth,height = 0.15\textheight]{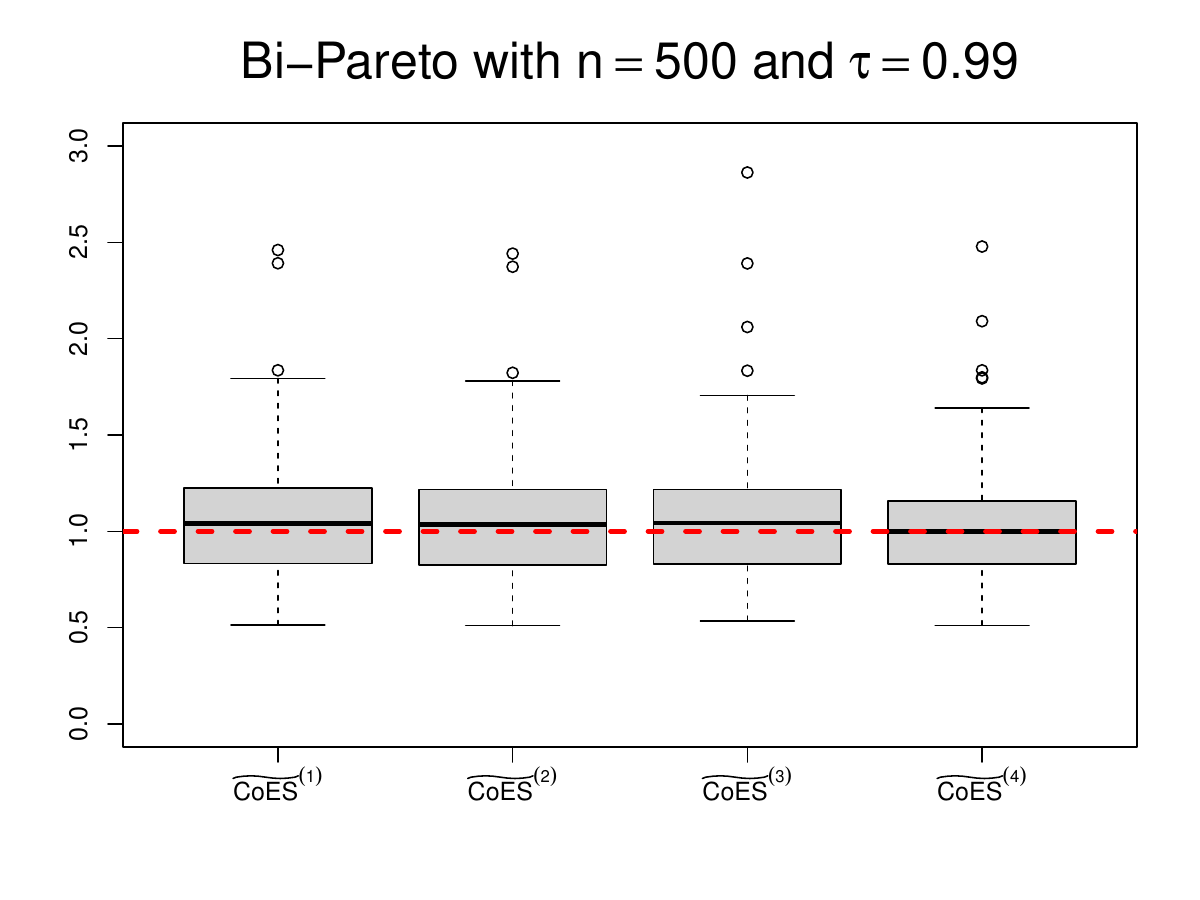}
\end{minipage}
\begin{minipage}[b]{0.24\textwidth}
\includegraphics[width=\textwidth,height = 0.15\textheight]{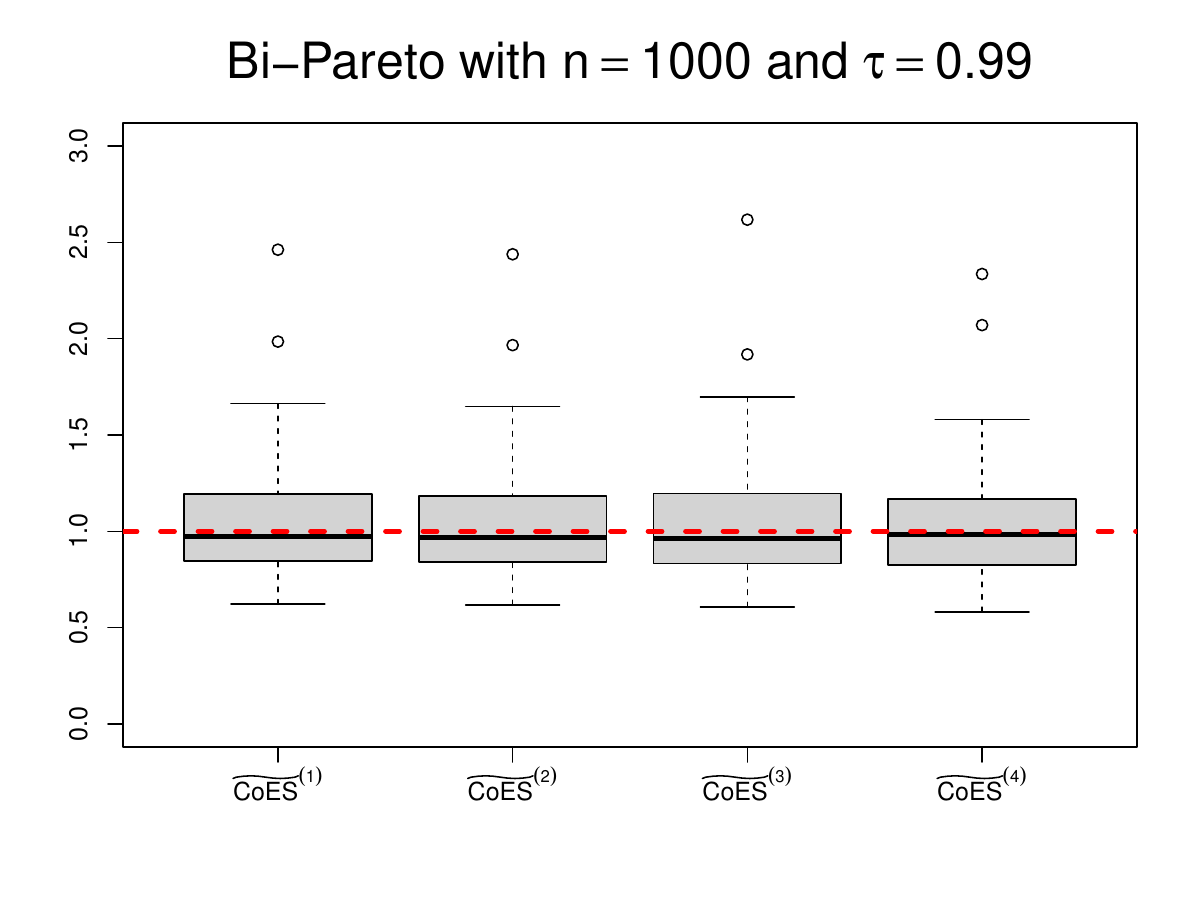}
\end{minipage}
\begin{minipage}[b]{0.24\textwidth}
\includegraphics[width=\textwidth,height = 0.15\textheight]{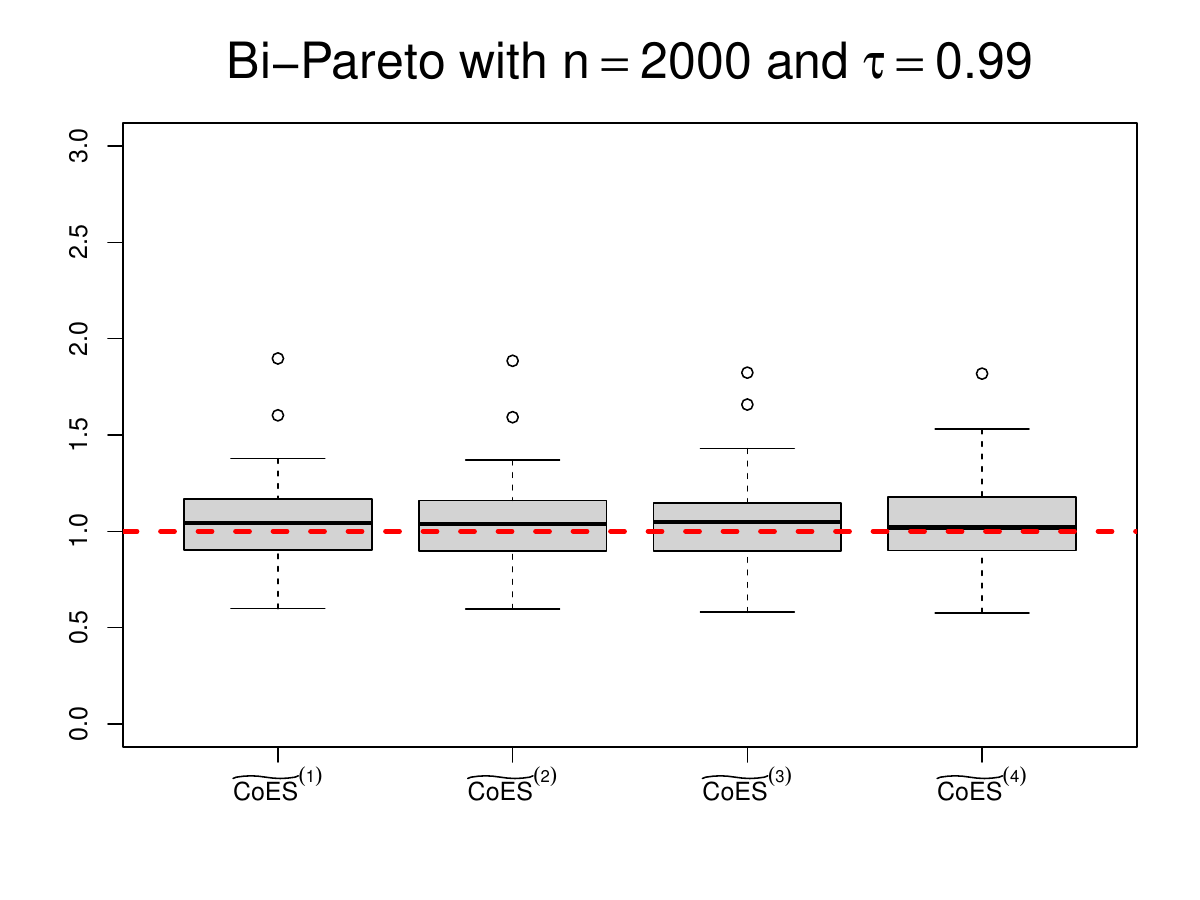}
\end{minipage}
\begin{minipage}[b]{0.24\textwidth}
\includegraphics[width=\textwidth,height = 0.15\textheight]{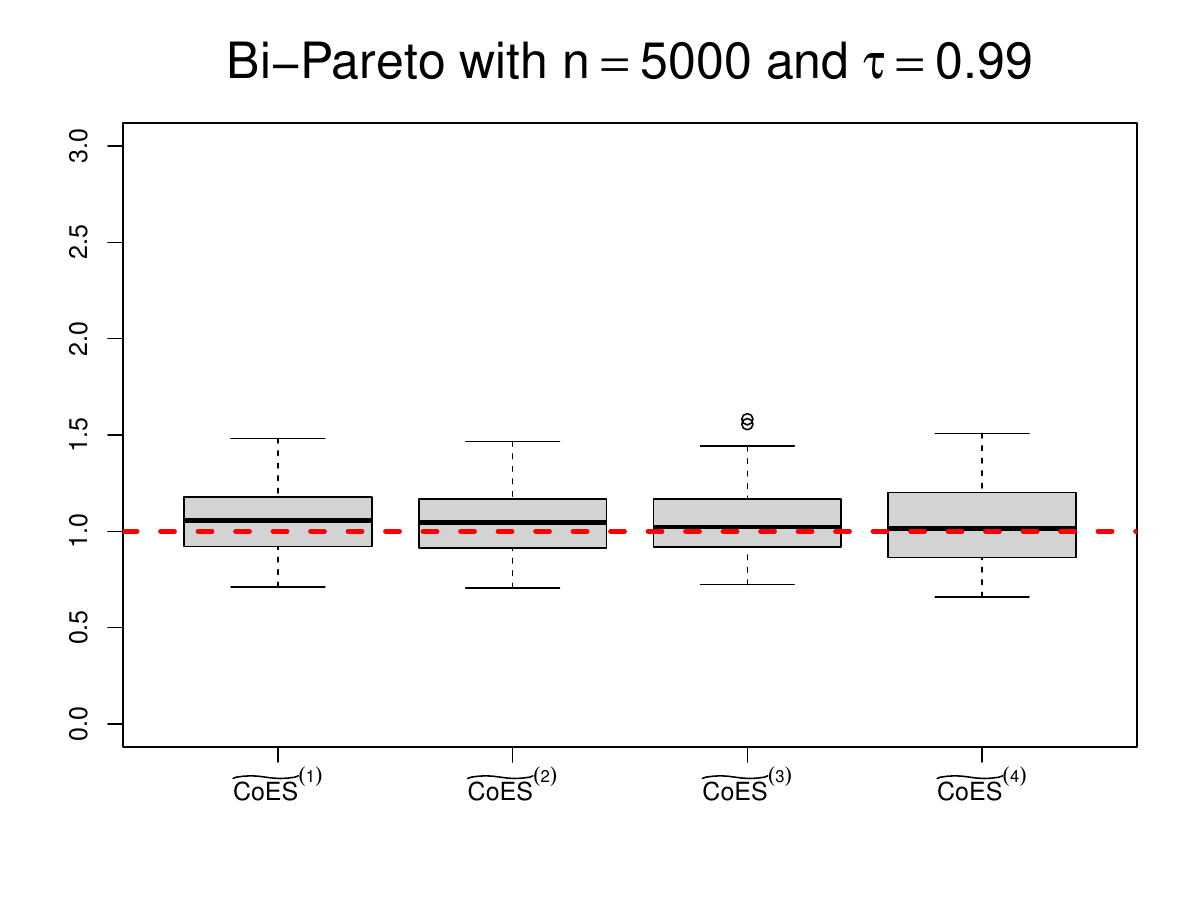}
\end{minipage}
\\
\begin{minipage}[b]{0.24\textwidth}
\includegraphics[width=\textwidth,height = 0.15\textheight]{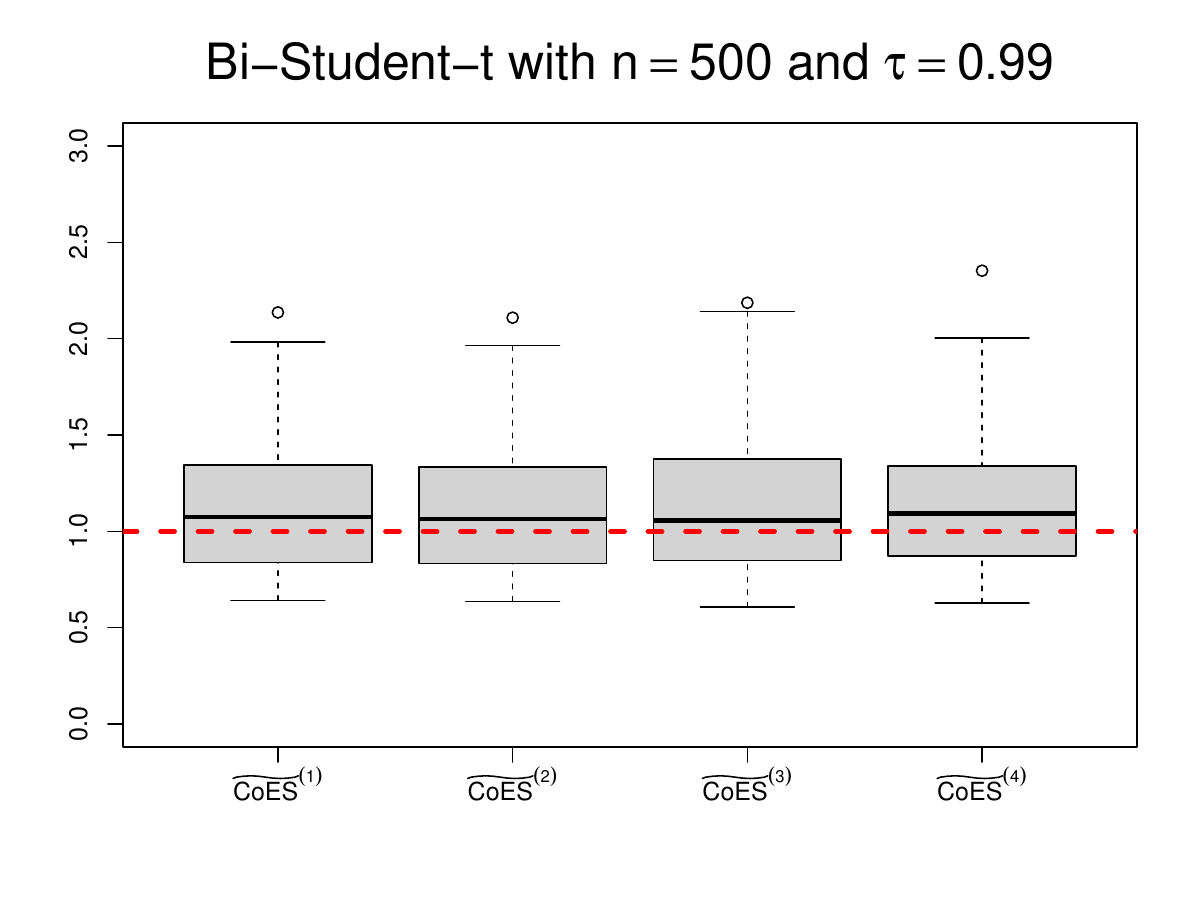}
\end{minipage}
\begin{minipage}[b]{0.24\textwidth}
\includegraphics[width=\textwidth,height = 0.15\textheight]{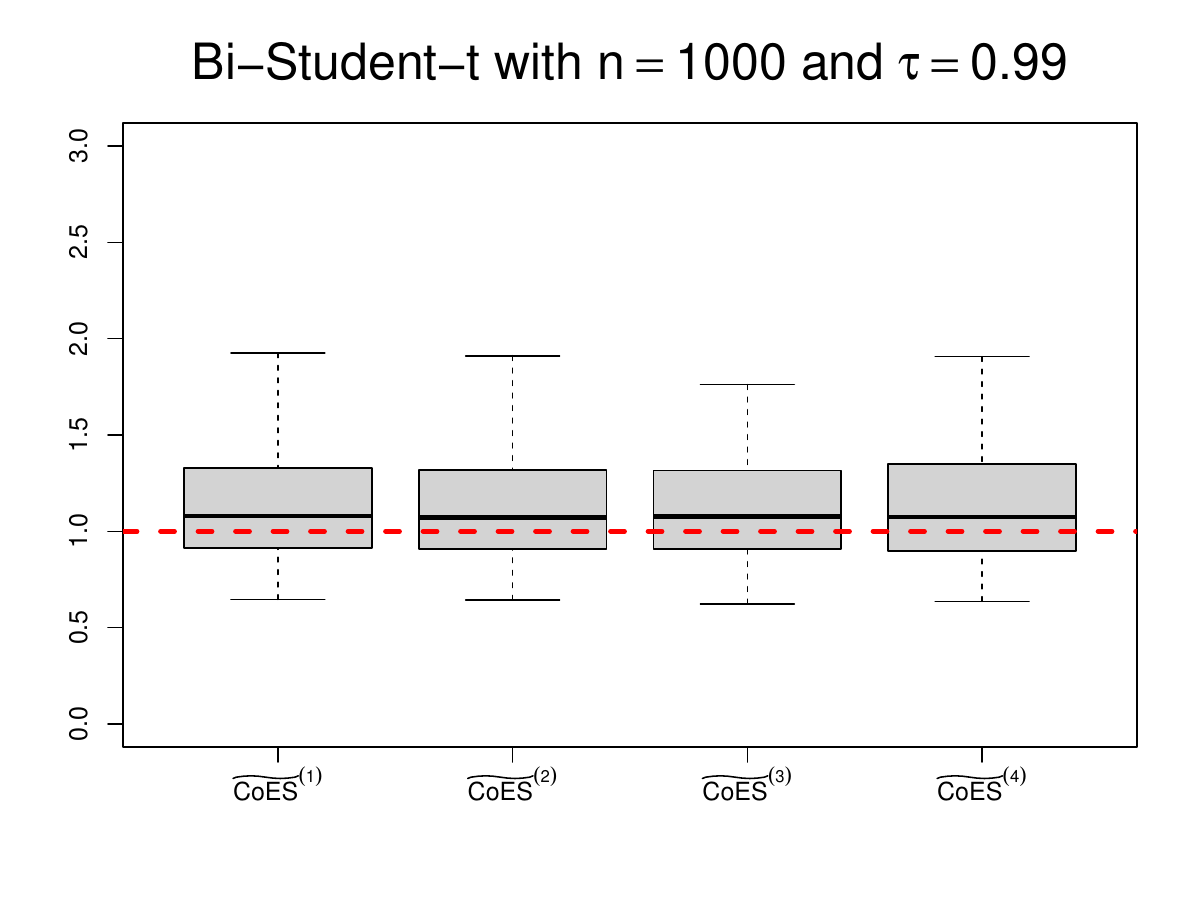}
\end{minipage}
\begin{minipage}[b]{0.24\textwidth}
\includegraphics[width=\textwidth,height = 0.15\textheight]{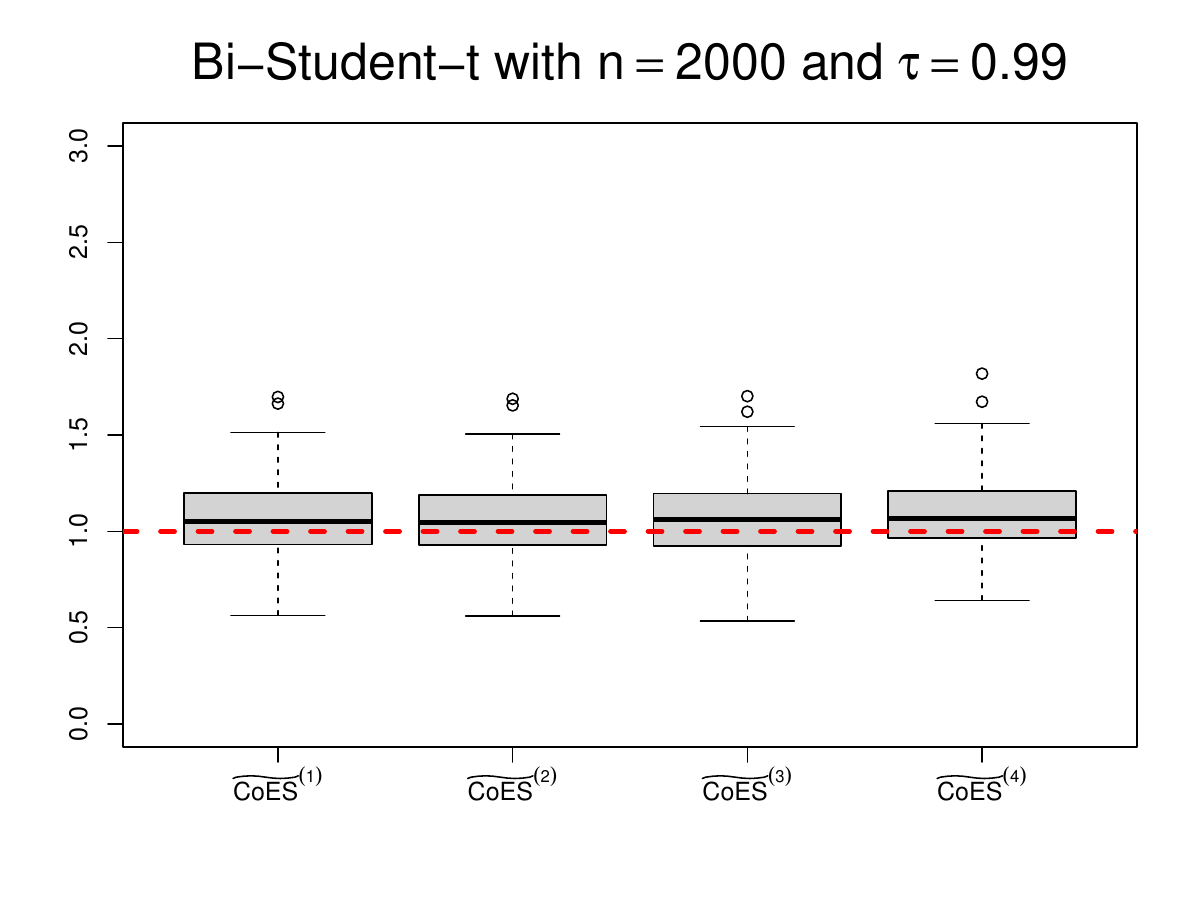}
\end{minipage}
\begin{minipage}[b]{0.24\textwidth}
\includegraphics[width=\textwidth,height = 0.15\textheight]{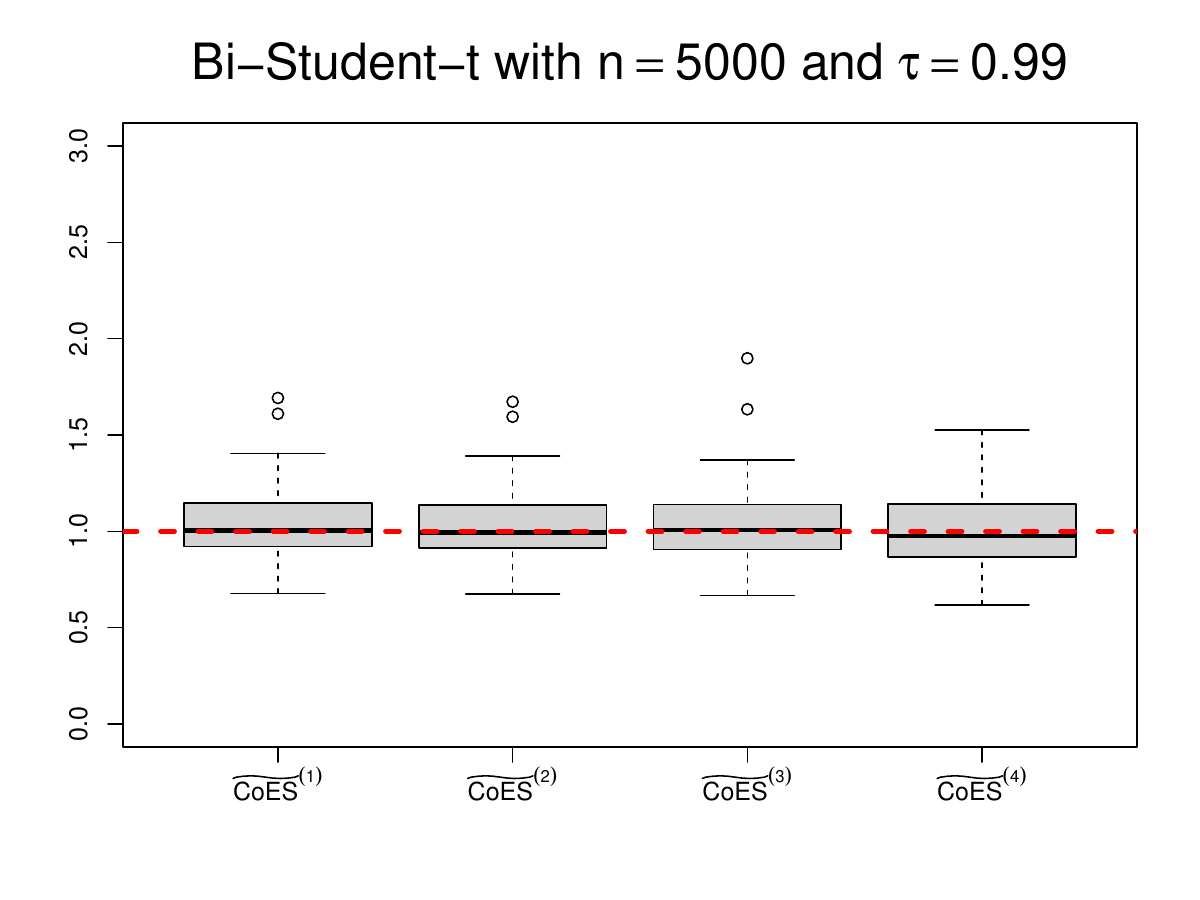}
\end{minipage}
\caption{The boxplots of $\widetilde{\coes}^{(i)}_{X|Y}(\tau'_n)$ for $i=1,2,3,4$ with $\tau'_n = 0.99$, under transformed bivariate Logistic, Cauchy, Pareto and Student-$t$ models. The boxplots from the left panel to the right panel are drawn with $n=500,1000,2000,$ and $5000$, respectively.}
\label{Fig:boxplots_CES_99}
\end{figure}

\section{Real Data Analysis}\label{sec:realanalysis}

\begin{figure}[htbp]
\centering
\begin{minipage}[b]{0.38\textwidth}
\includegraphics[width=\textwidth,height = 0.2\textheight]{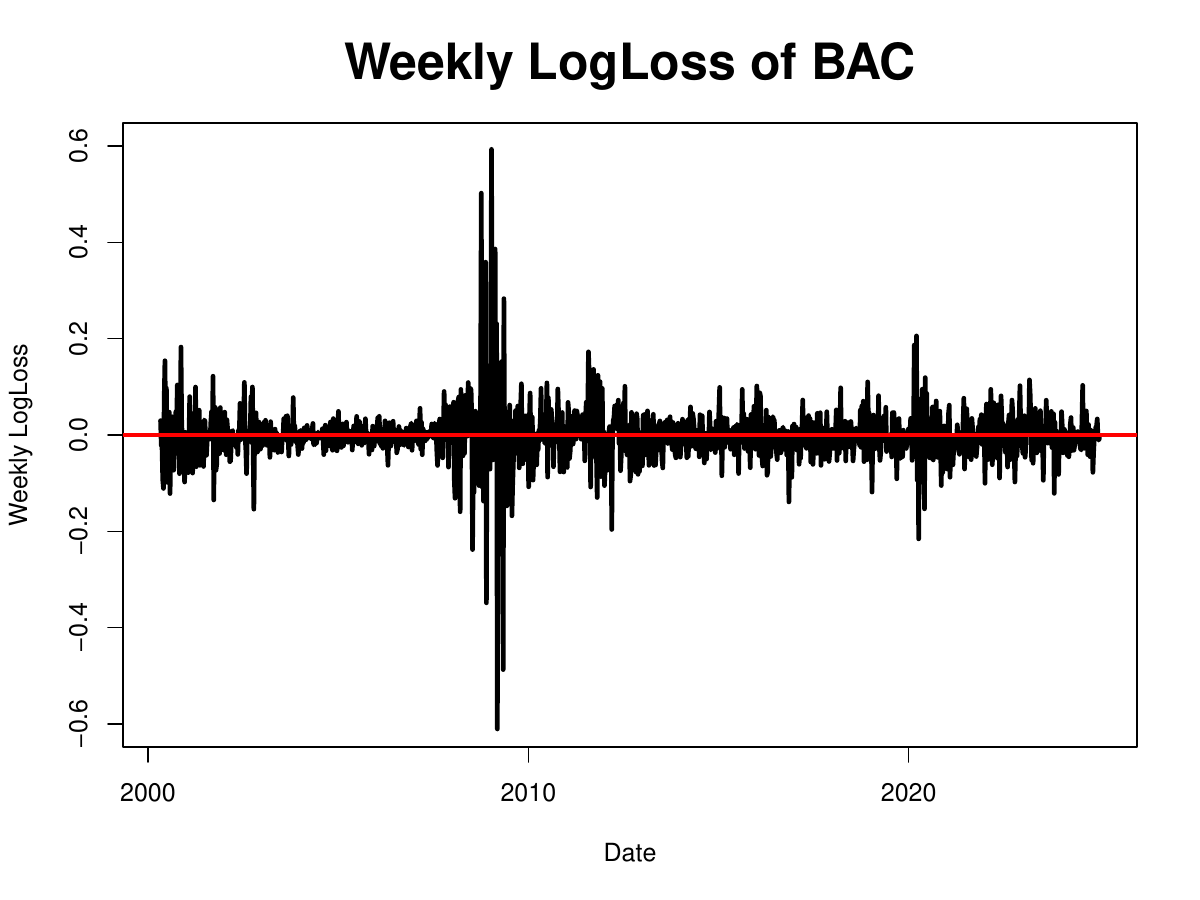}
\end{minipage}
\begin{minipage}[b]{0.38\textwidth}
\includegraphics[width=\textwidth,height = 0.2\textheight]{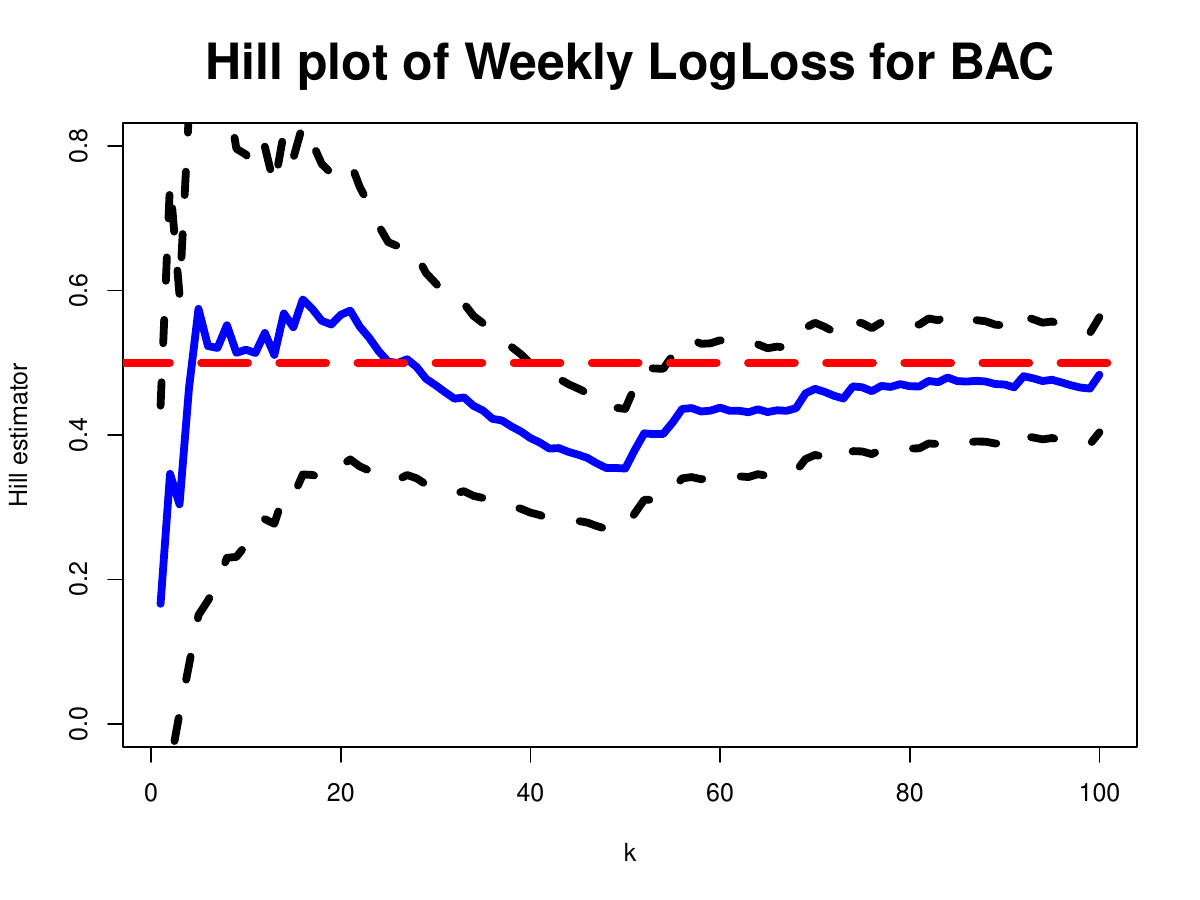}
\end{minipage}
\\
\begin{minipage}[b]{0.38\textwidth}
\includegraphics[width=\textwidth,height = 0.2\textheight]{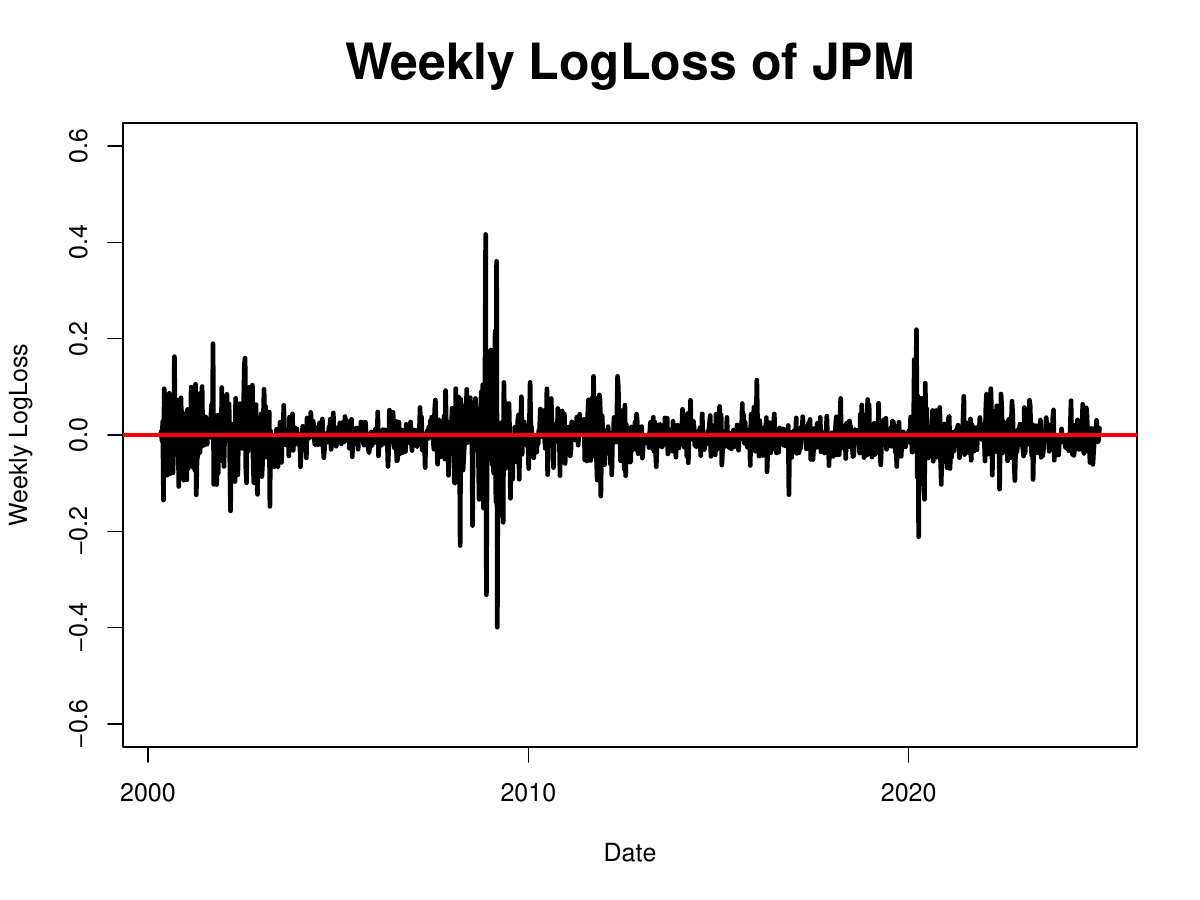}
\end{minipage}
\begin{minipage}[b]{0.38\textwidth}
\includegraphics[width=\textwidth,height = 0.2\textheight]{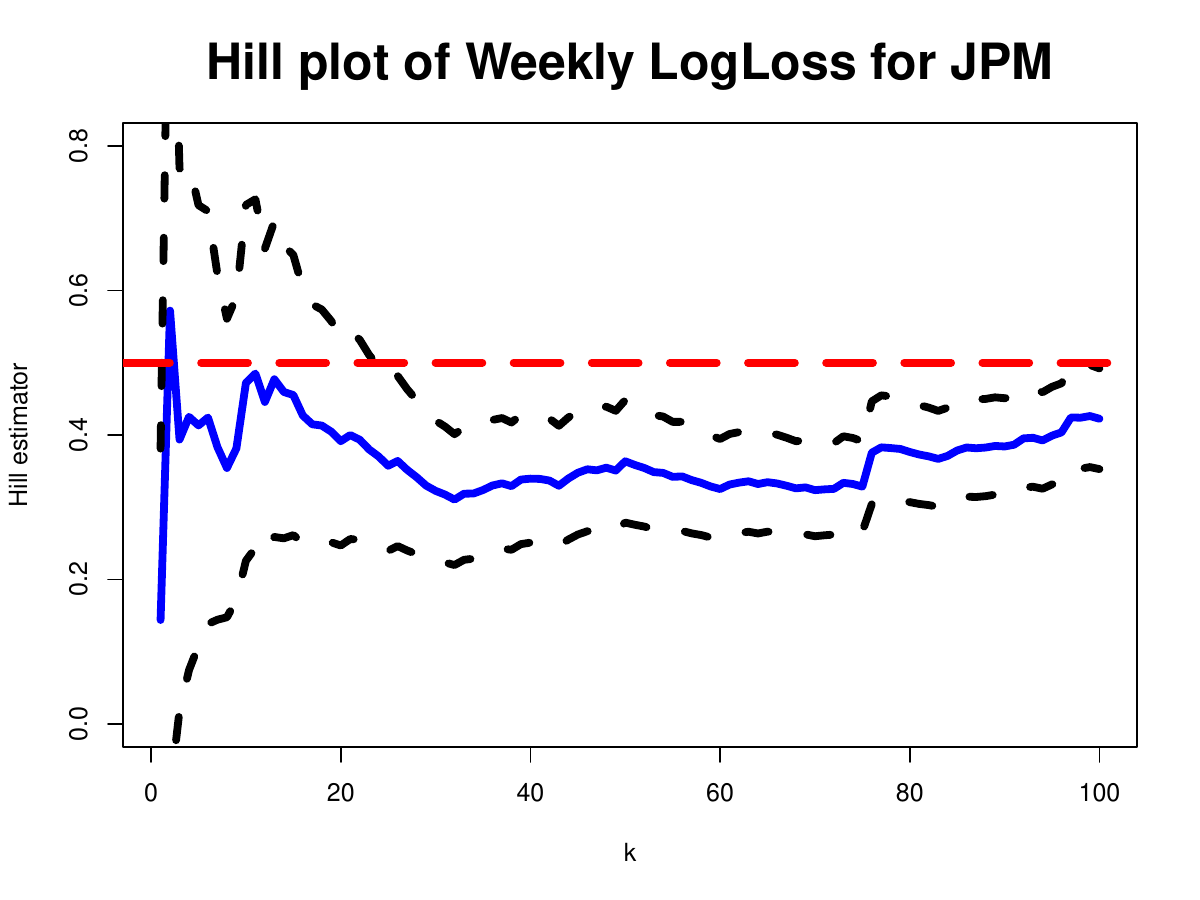}
\end{minipage}
\\
\begin{minipage}[b]{0.38\textwidth}
\includegraphics[width=\textwidth,height = 0.2\textheight]{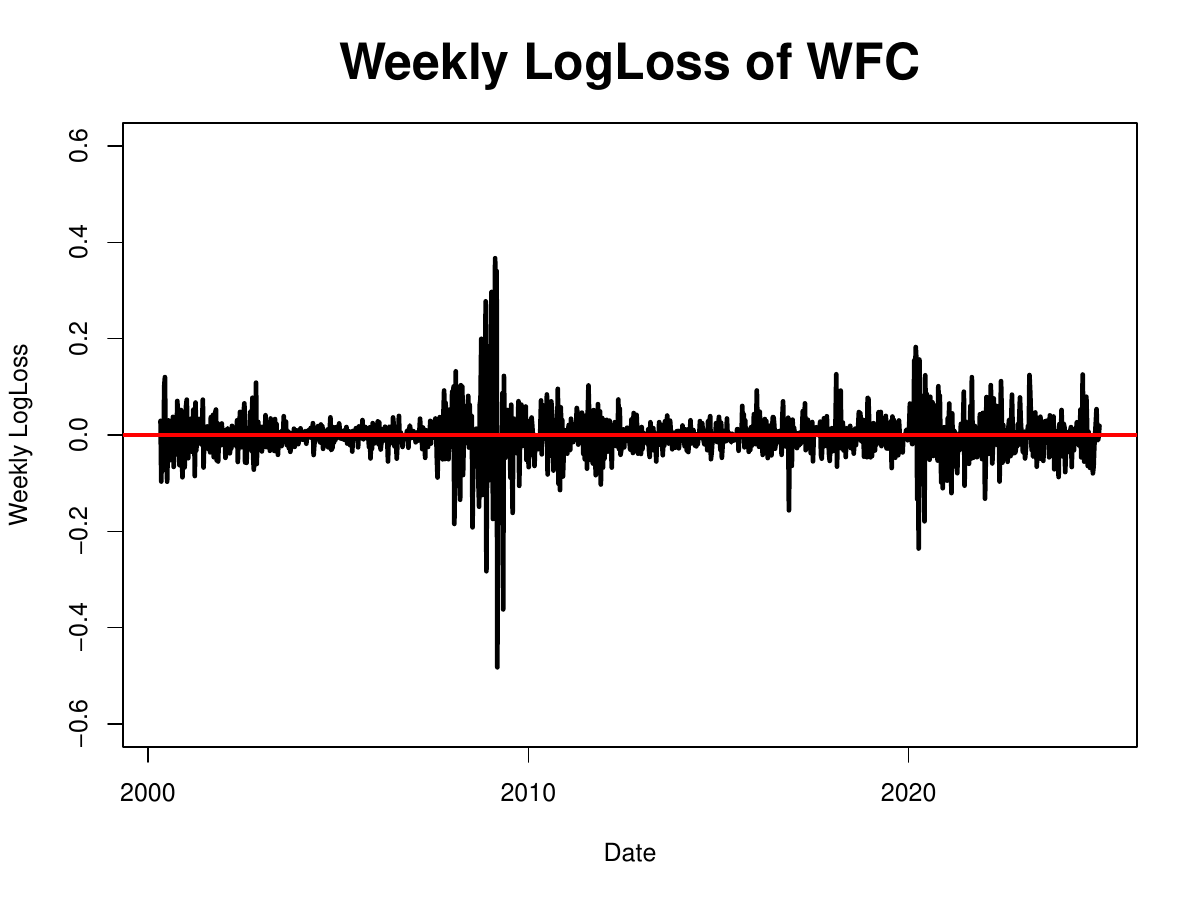}
\end{minipage}
\begin{minipage}[b]{0.38\textwidth}
\includegraphics[width=\textwidth,height = 0.2\textheight]{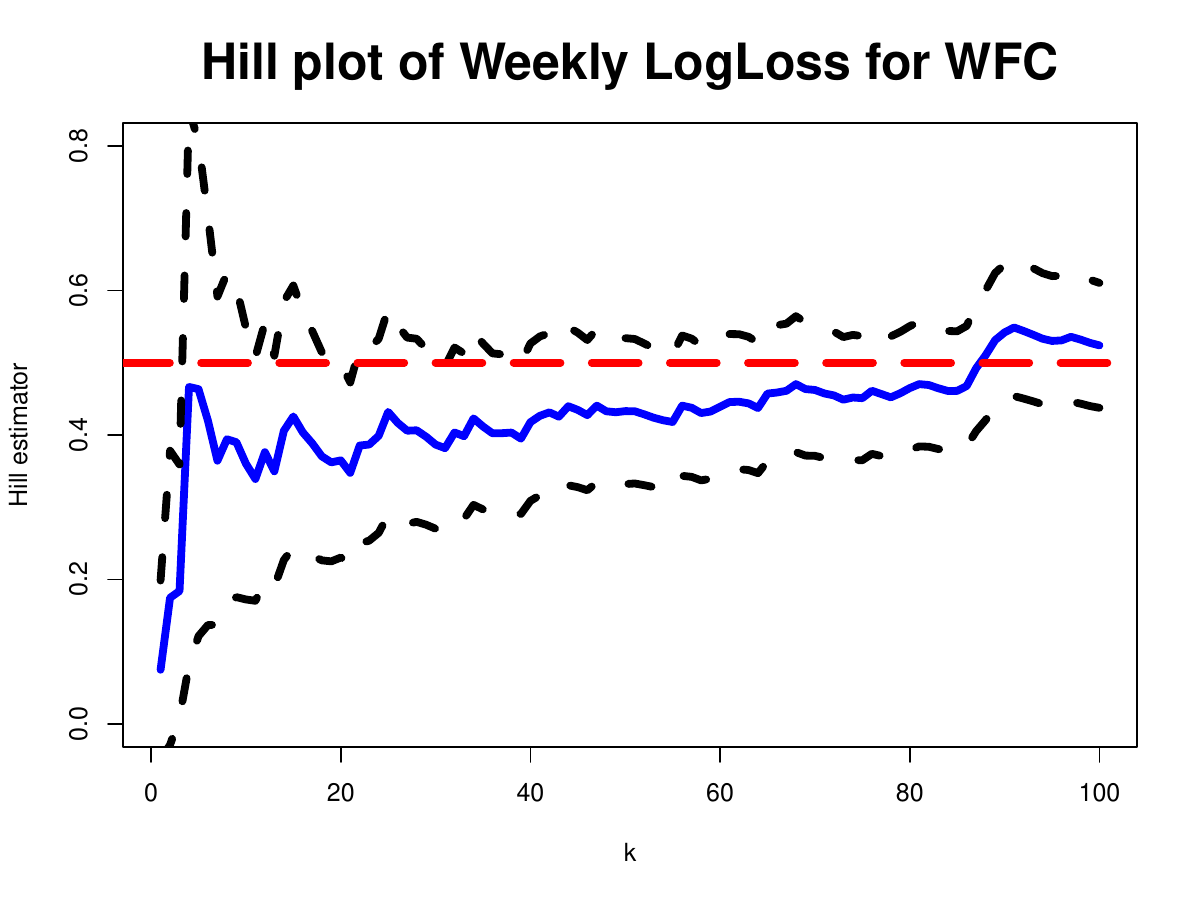}
\end{minipage}
\\
\begin{minipage}[b]{0.38\textwidth}
\includegraphics[width=\textwidth,height = 0.2\textheight]{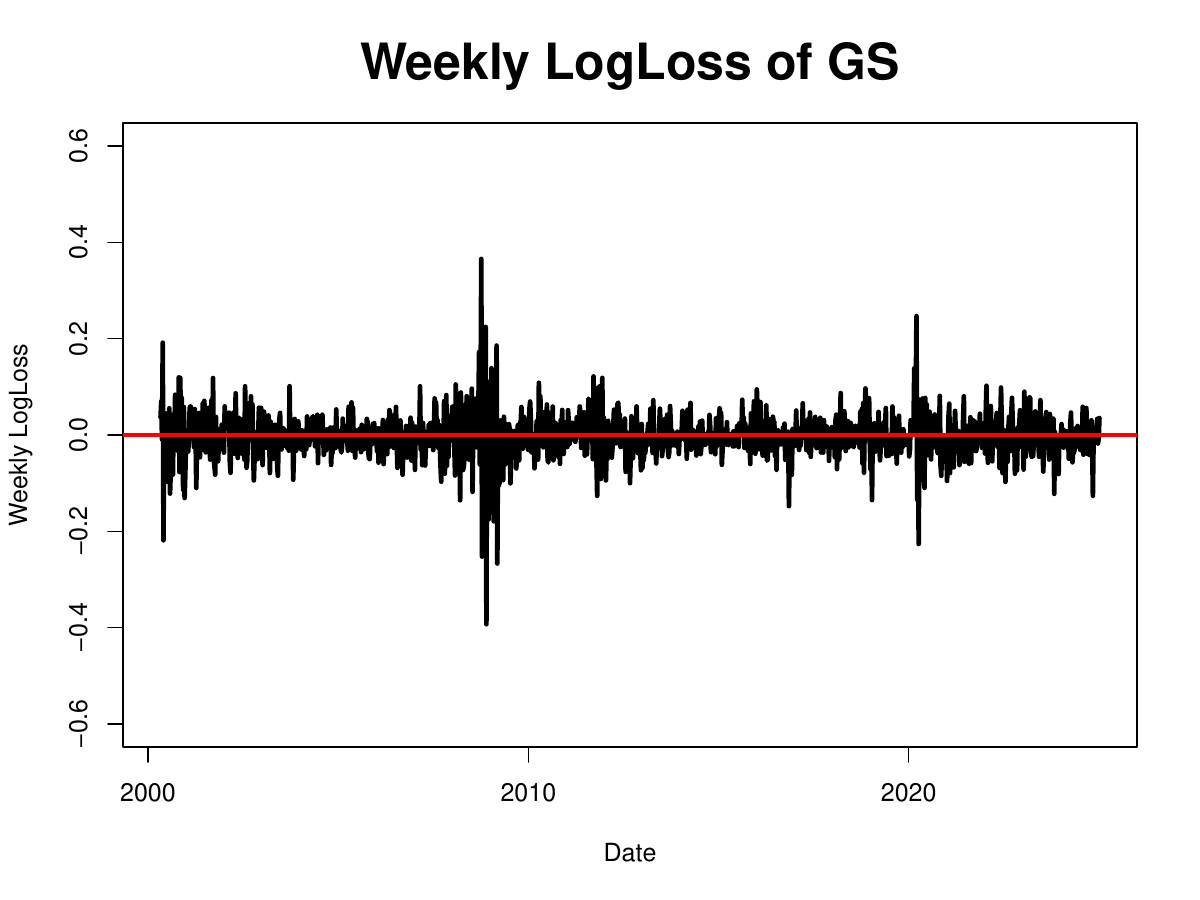}
\end{minipage}
\begin{minipage}[b]{0.38\textwidth}
\includegraphics[width=\textwidth,height = 0.2\textheight]{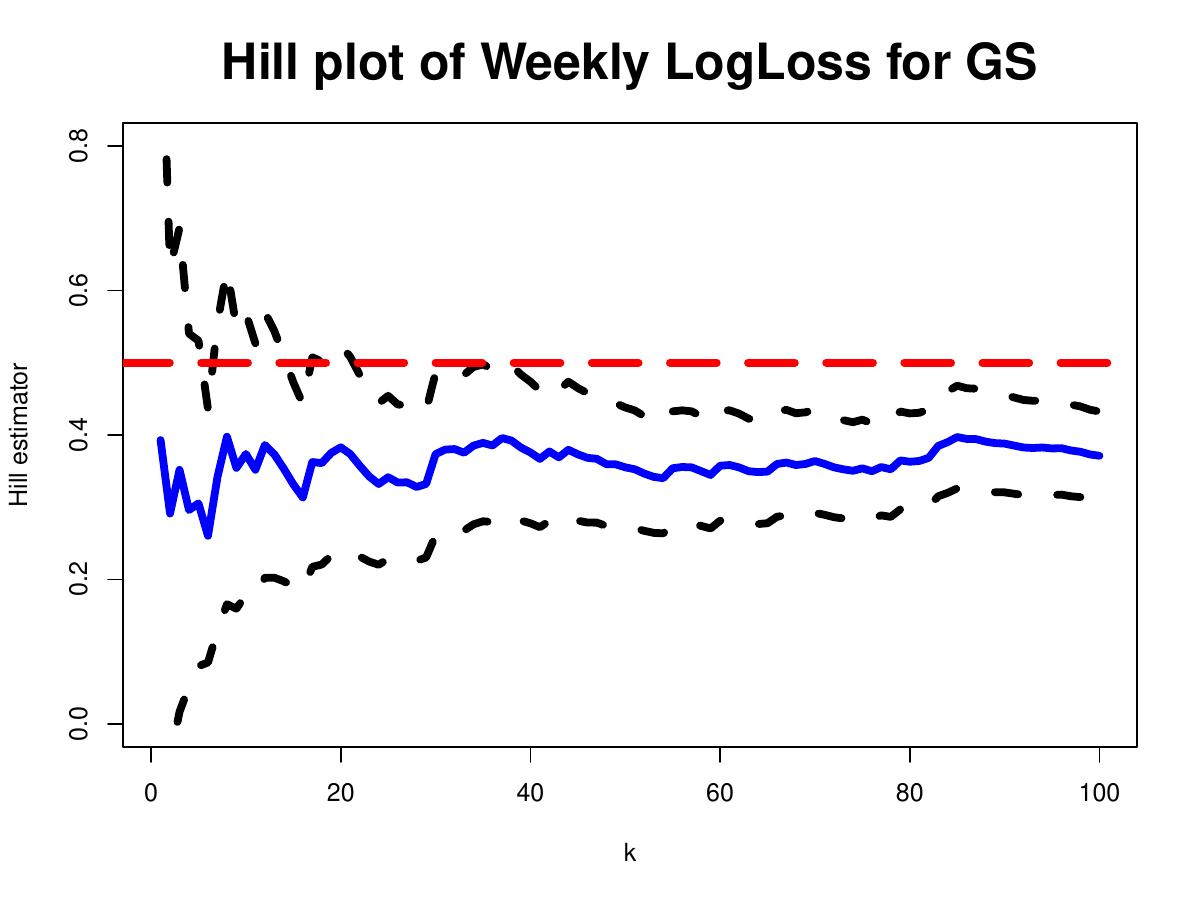}
\end{minipage}
\caption{Left panel: times series plots for weekly losses of the four financial institutions: (from top to bottom) BAC, JPM, WFC, and GS. Right panel: Hill plots for weekly losses of the four financial institutions: (from top to bottom) BAC, JPM, WFC, and GS, with the upper and lower dashed lines are the 90\% confidence bounds.}
\label{Fig:logloss_week}
\end{figure}

In this section, we further illustrate the empirical performances of the proposed extrapolative estimators through a real-world data analysis. In practice, financial and actuarial data typically exhibit asymptotic dependence. We thus consider the following four financial institutions in the USA: Bank of America Corporation (BAC), JPMorgan Chase \& Company (JPM), Wells Fargo \& Company (WFC), and Goldman Sachs Group Inc (GS). All of these institutions achieved a market capitalization in excess of US\$5 billion at the end of June 2007 and made a significant systemic contribution to the financial market. Moreover, the S\&P 500 Index will be used as the system proxy in this analysis. Since the theories for statistical methodologies are derived for independent and identically distributed samples, we utilize the weekly historical adjusted closing prices of their stocks to reduce the potential serial dependence. The data spans from April 24th, 2000, to January 6th, 2025, consisting of 1290 trading records. 
The weekly losses were calculated as negative log returns. Left panel of Figure \ref{Fig:logloss_week} presents the time series plots of weekly losses for the four institutions. All of the plots exhibit typical characteristics of financial time series, notably periods of volatility clustering, such as those observed during the global financial crisis in 2008 and the COVID-19 pandemic in 2020.

\begin{figure}[htbp]
\centering
\begin{minipage}[b]{0.38\textwidth}
\includegraphics[width=\textwidth,height = 0.2\textheight]{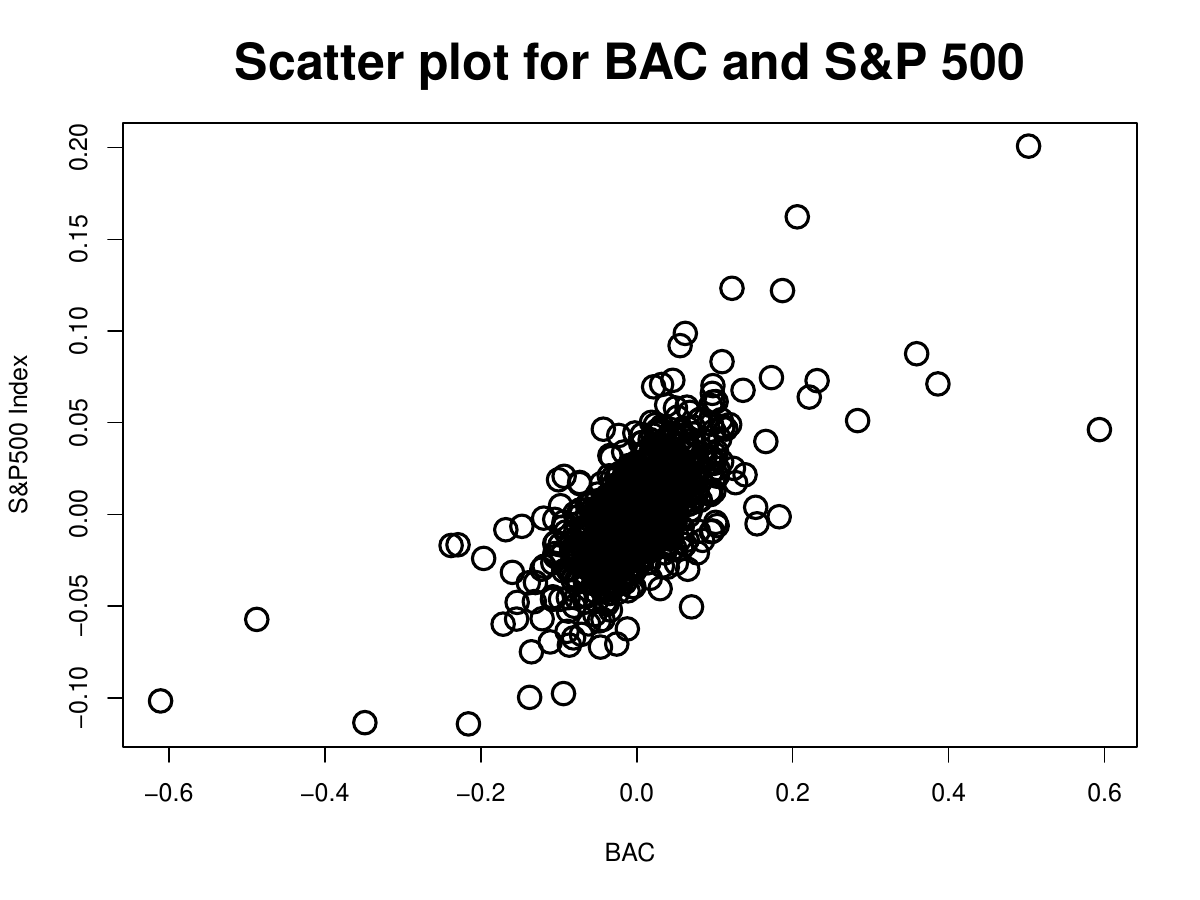}
\end{minipage}
\begin{minipage}[b]{0.38\textwidth}
\includegraphics[width=\textwidth,height = 0.2\textheight]{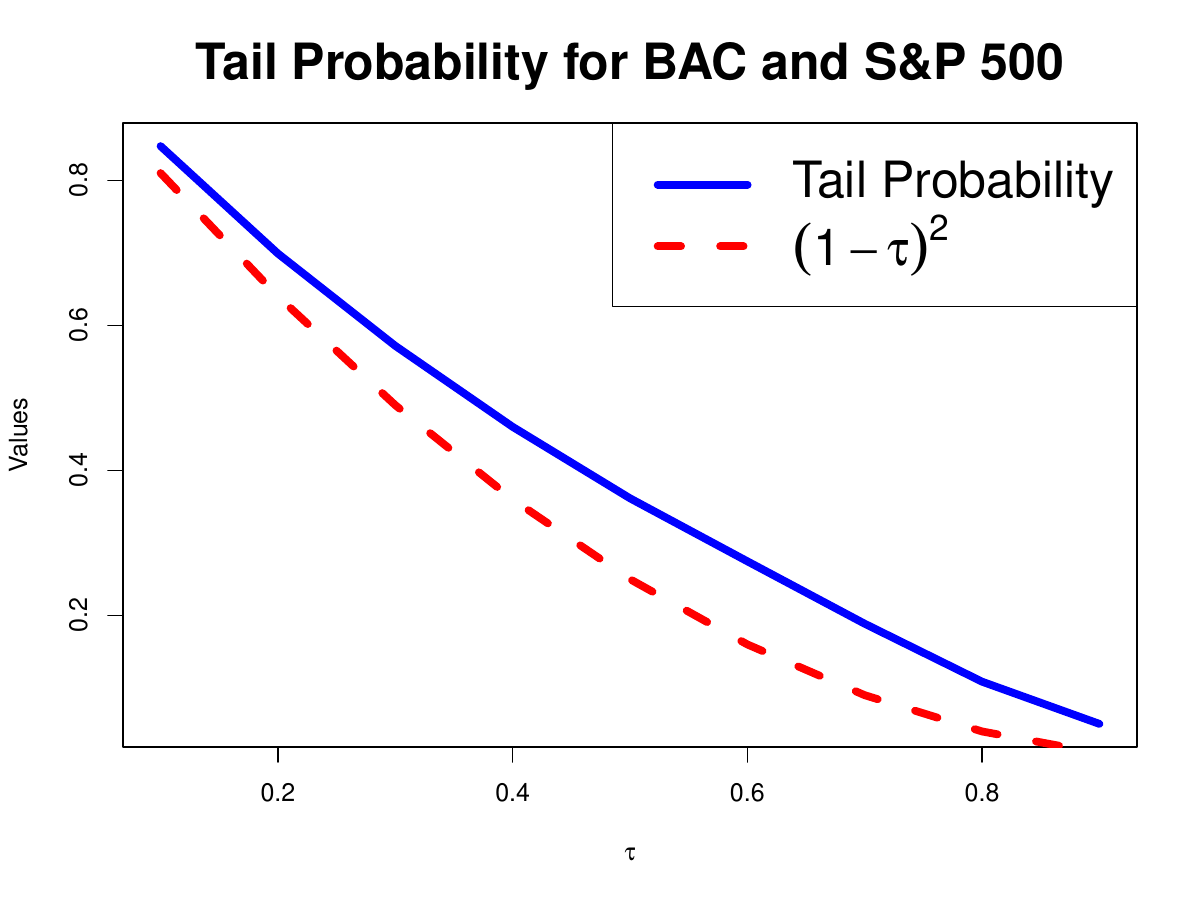}
\end{minipage}
\\
\begin{minipage}[b]{0.38\textwidth}
\includegraphics[width=\textwidth,height = 0.2\textheight]{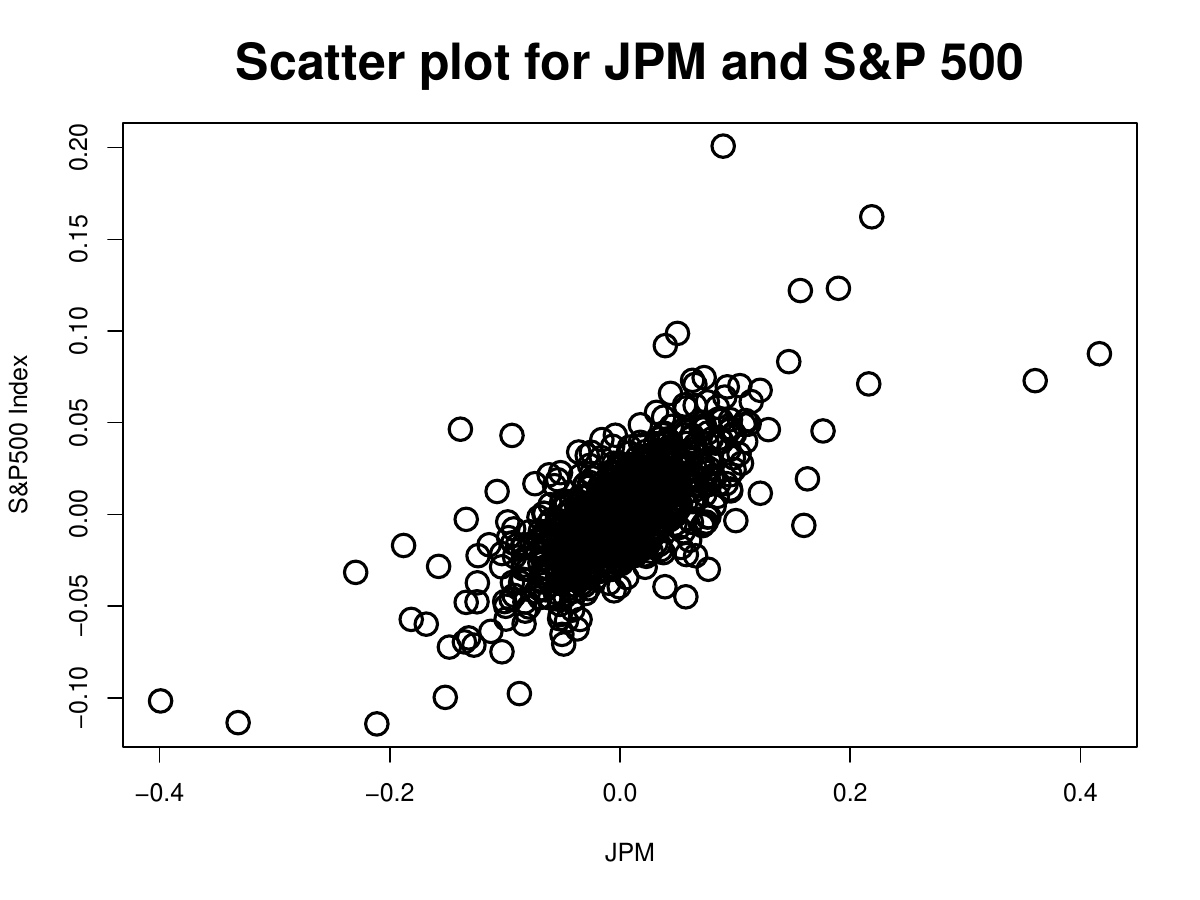}
\end{minipage}
\begin{minipage}[b]{0.38\textwidth}
\includegraphics[width=\textwidth,height = 0.2\textheight]{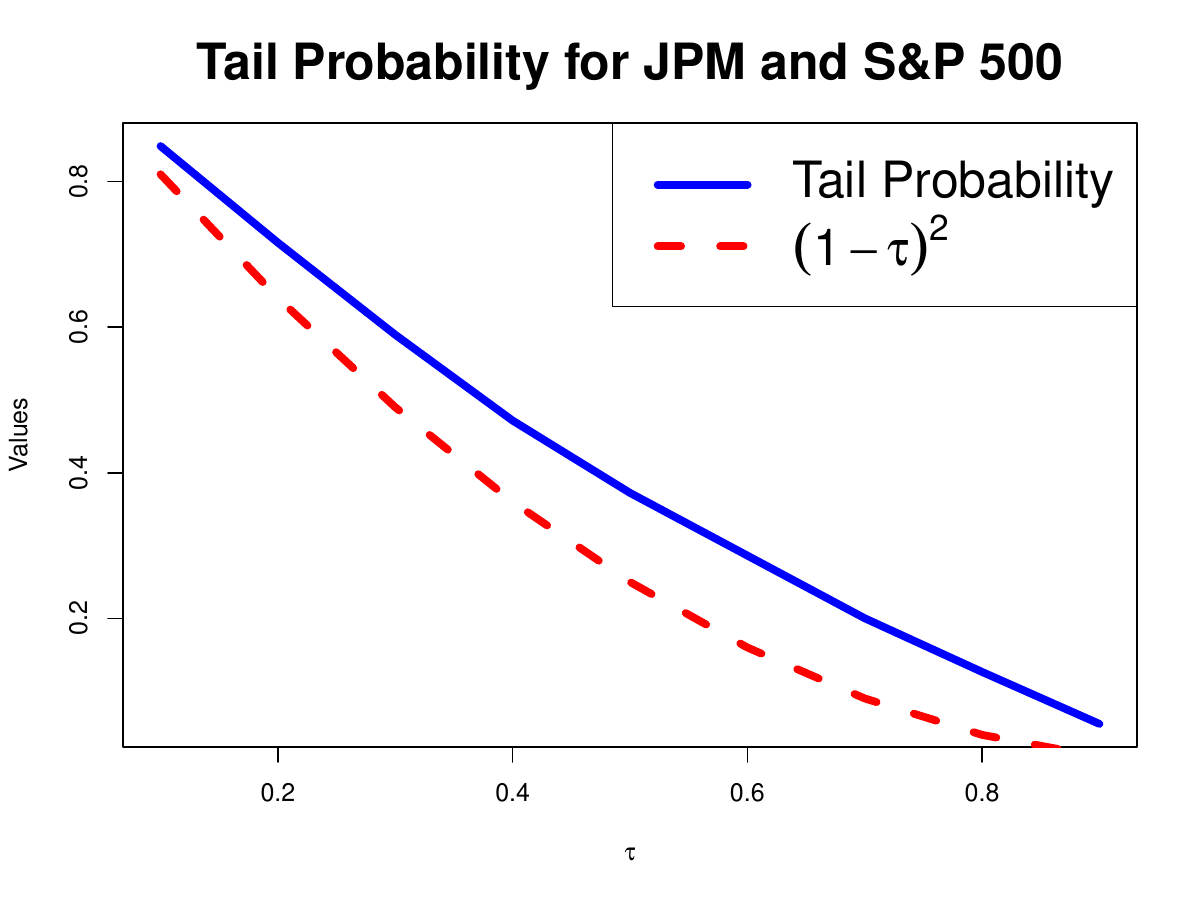}
\end{minipage}
\\
\begin{minipage}[b]{0.38\textwidth}
\includegraphics[width=\textwidth,height = 0.2\textheight]{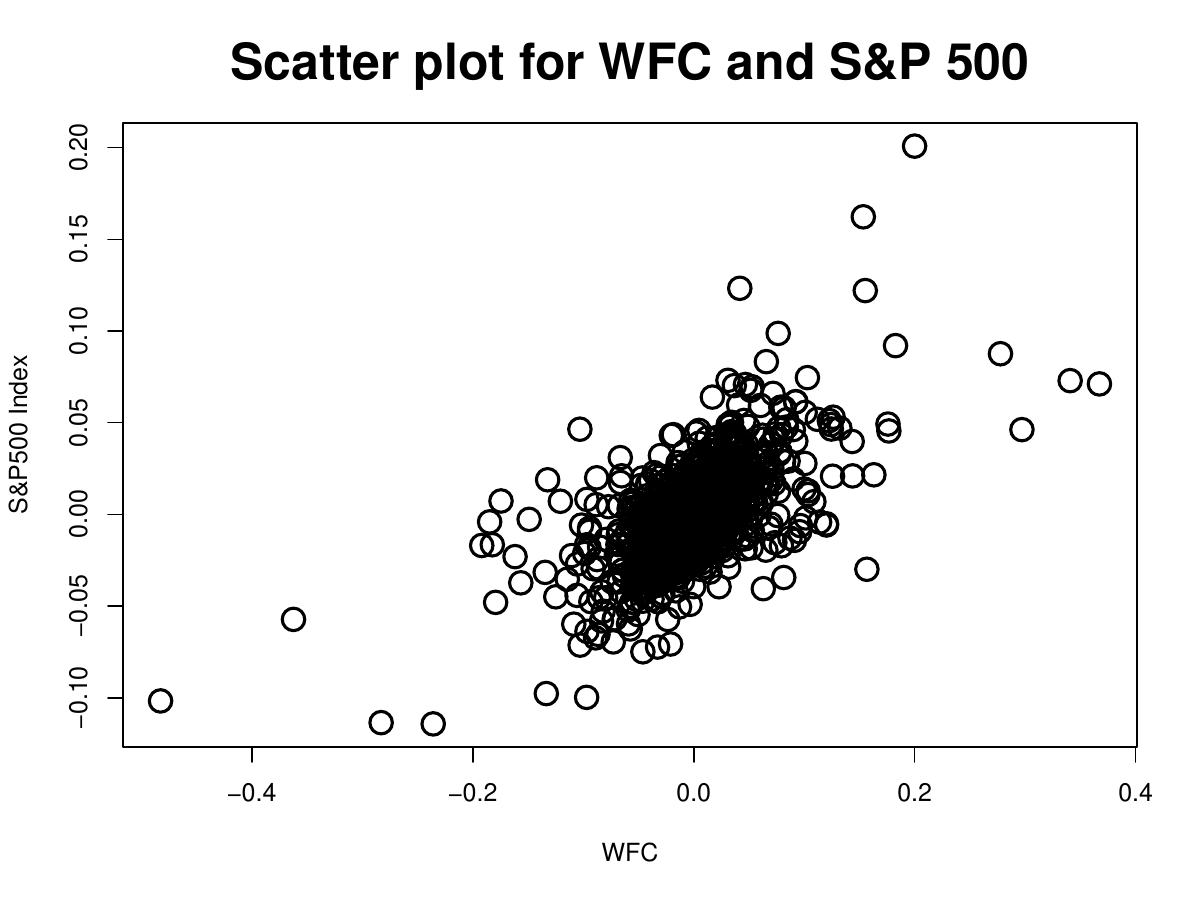}
\end{minipage}
\begin{minipage}[b]{0.38\textwidth}
\includegraphics[width=\textwidth,height = 0.2\textheight]{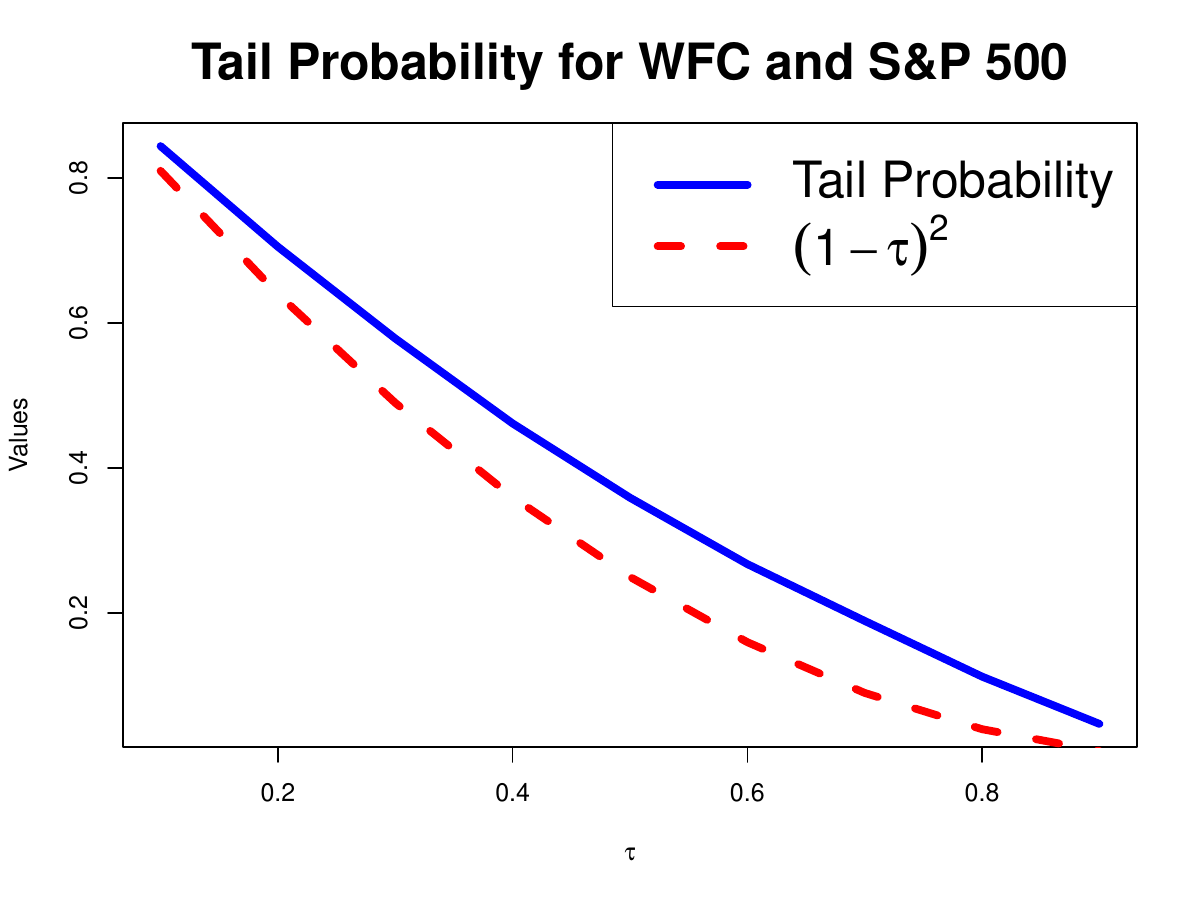}
\end{minipage}
\\
\begin{minipage}[b]{0.38\textwidth}
\includegraphics[width=\textwidth,height = 0.2\textheight]{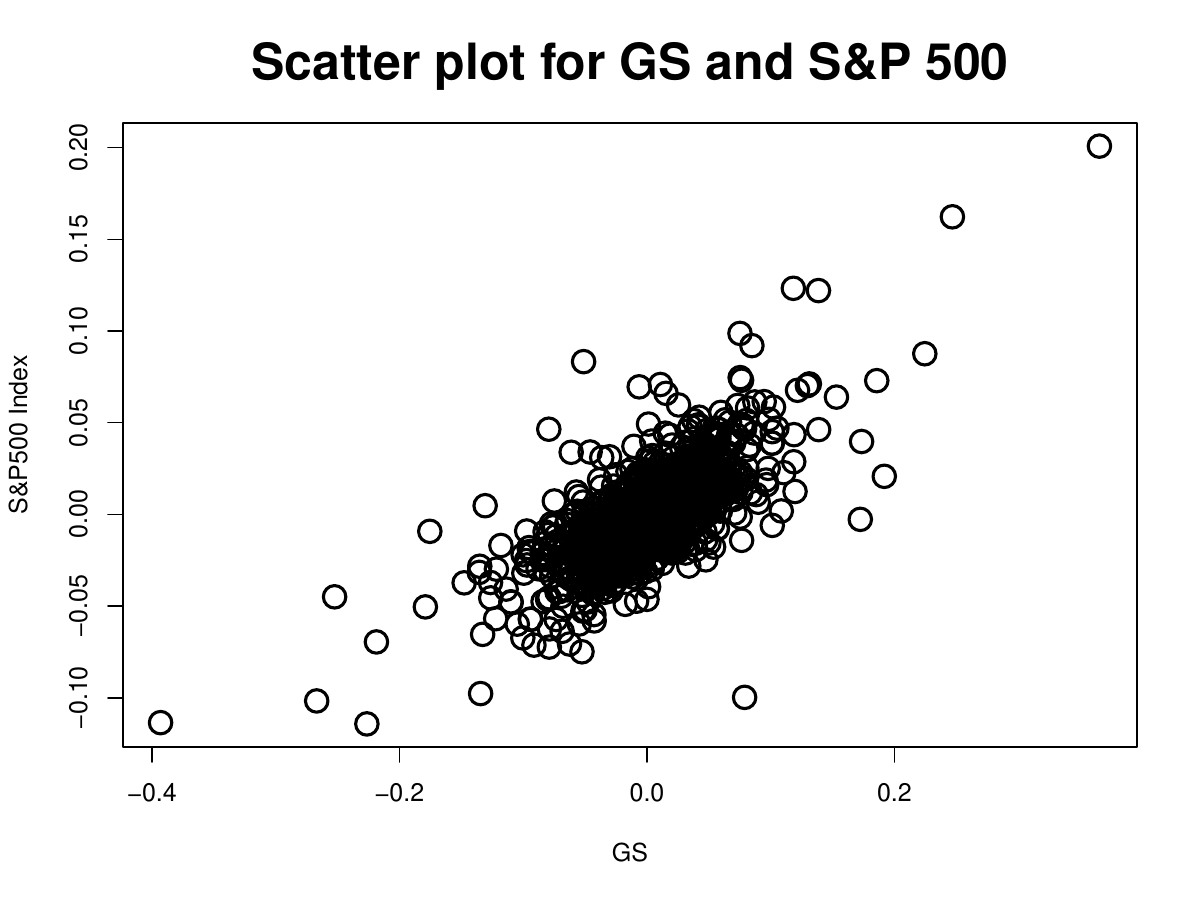}
\end{minipage}
\begin{minipage}[b]{0.38\textwidth}
\includegraphics[width=\textwidth,height = 0.2\textheight]{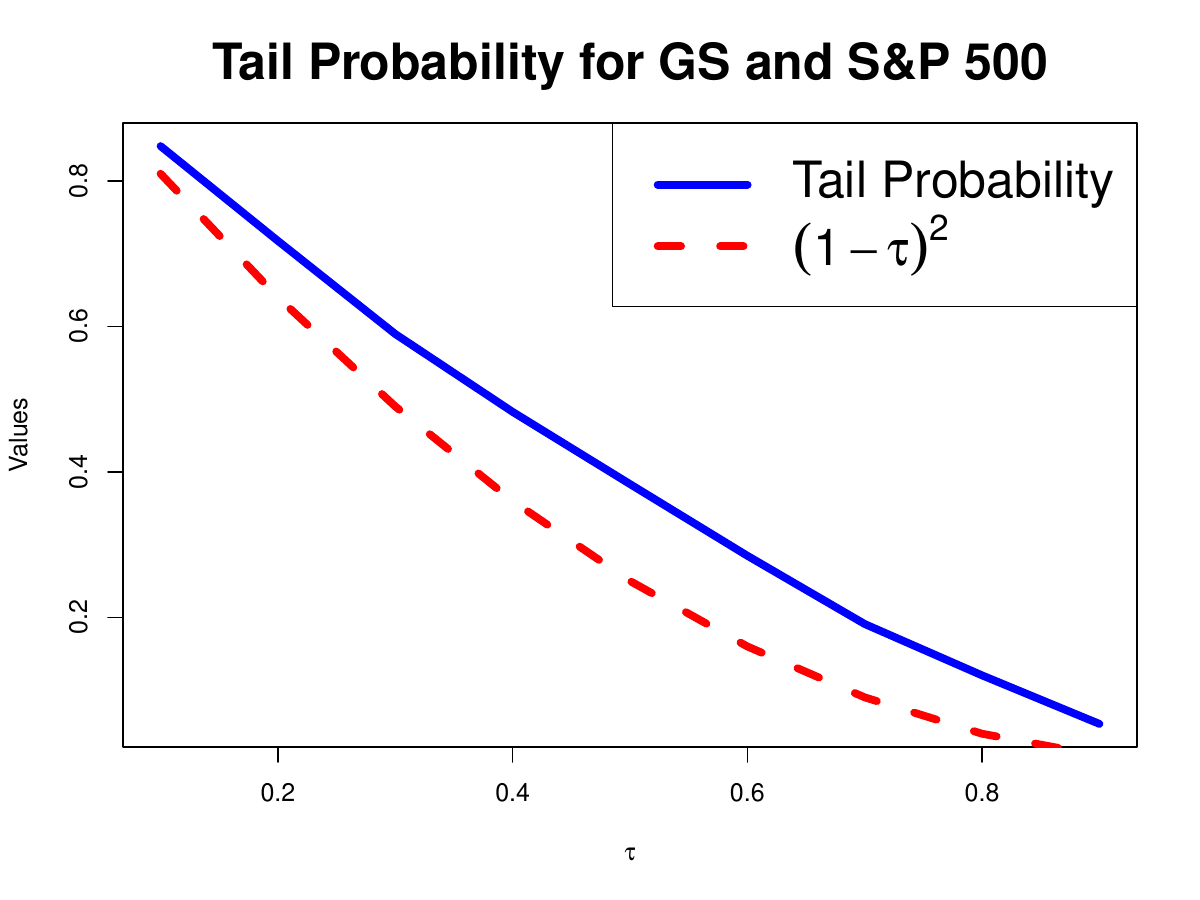}
\end{minipage}
\caption{Left panel: scatter plots for the weekly losses of the four financial institutions: (from top to bottom) BAC, JPM, WFC, and GS with respect to S\&P 500 Index. Right panel: comparison between the empirical tail probability $\bP(X \ge \var_X(\tau), Y \ge \var_Y(\tau))$ with the value $(1-\tau)^2$ for the four financial institutions: (from top to bottom) BAC, JPM, WFC, GS and S\&P 500 Index.}
\label{Fig:scatter_tp_week}
\end{figure}

\begin{figure}[htbp]
\centering
\begin{minipage}[b]{0.42\textwidth}
\includegraphics[width=\textwidth,height = 0.2\textheight]{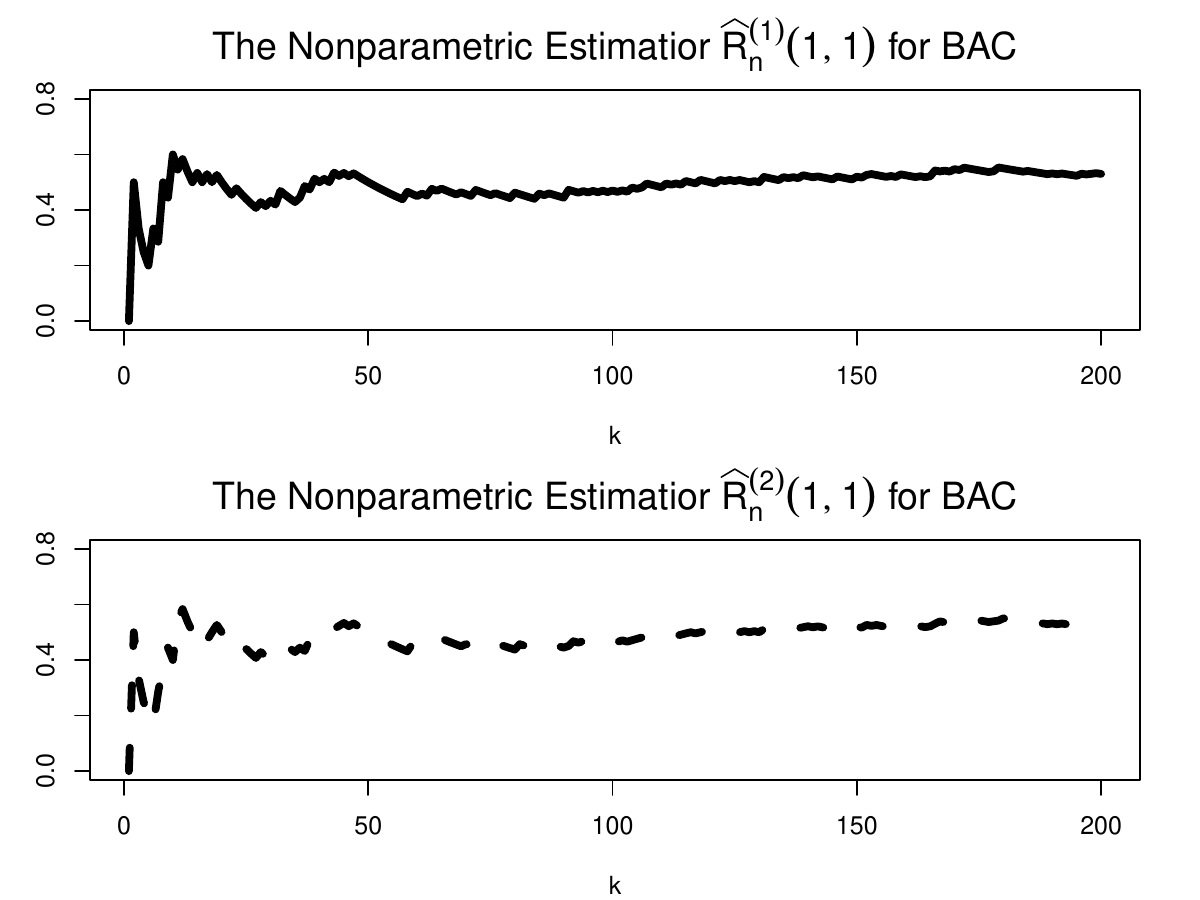}
\end{minipage}
\begin{minipage}[b]{0.42\textwidth}
\includegraphics[width=\textwidth,height = 0.2\textheight]{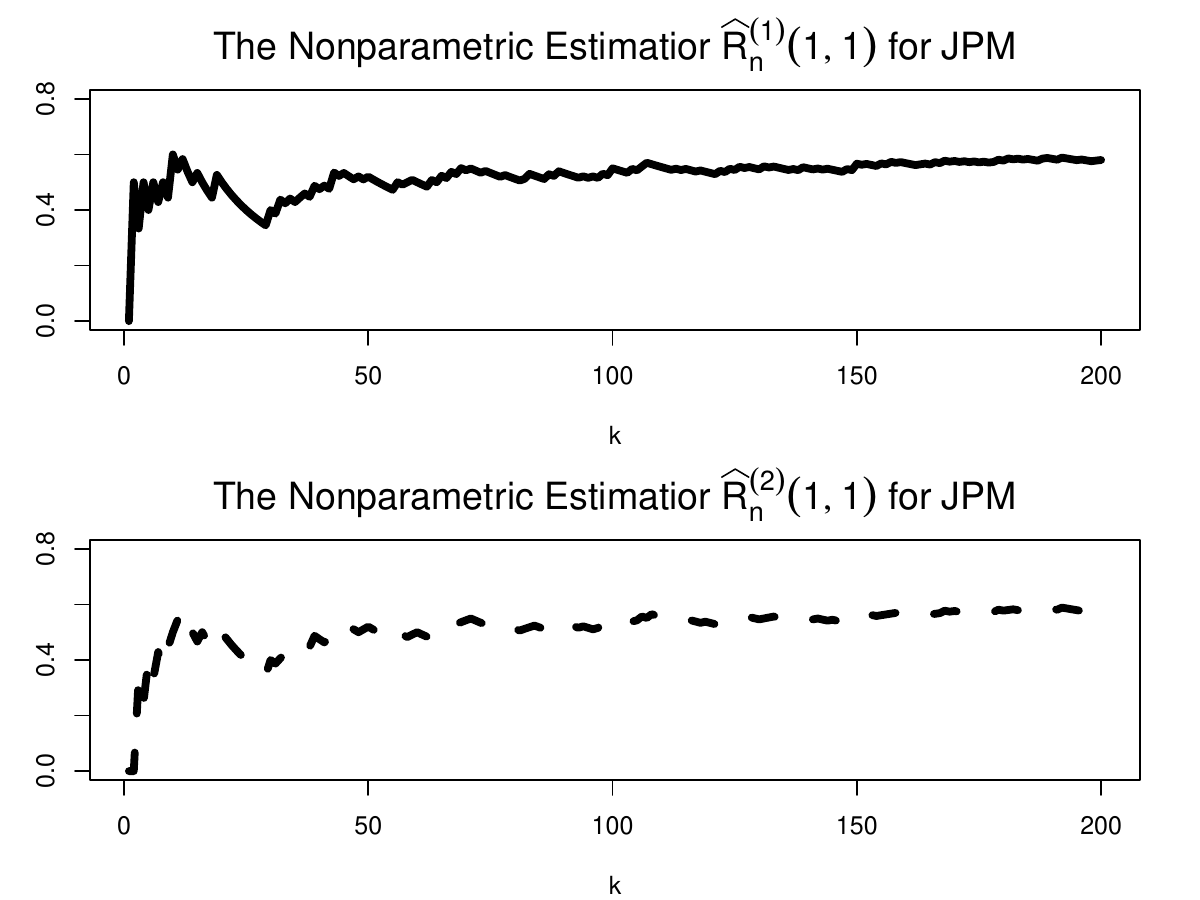}
\end{minipage}
\\
\begin{minipage}[b]{0.42\textwidth}
\includegraphics[width=\textwidth,height = 0.2\textheight]{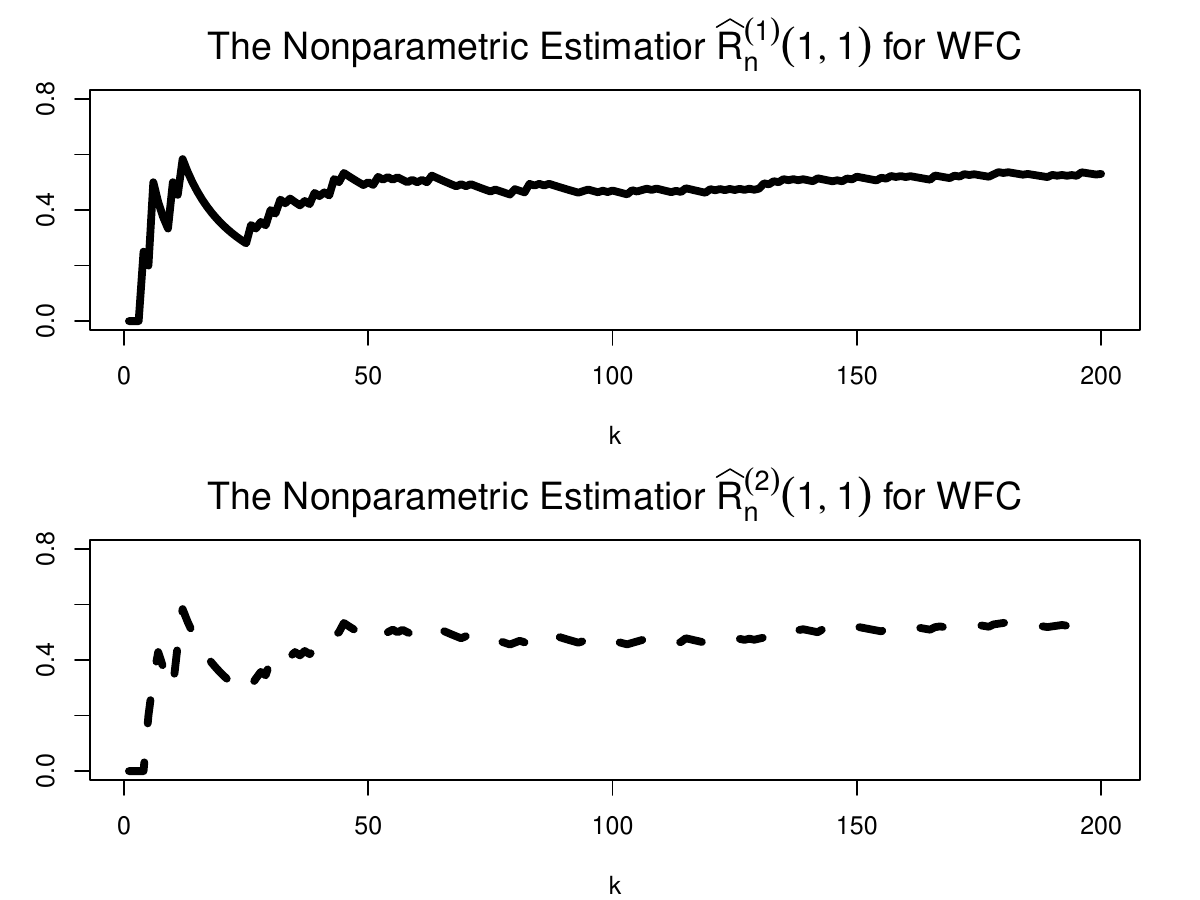}
\end{minipage}
\begin{minipage}[b]{0.42\textwidth}
\includegraphics[width=\textwidth,height = 0.2\textheight]{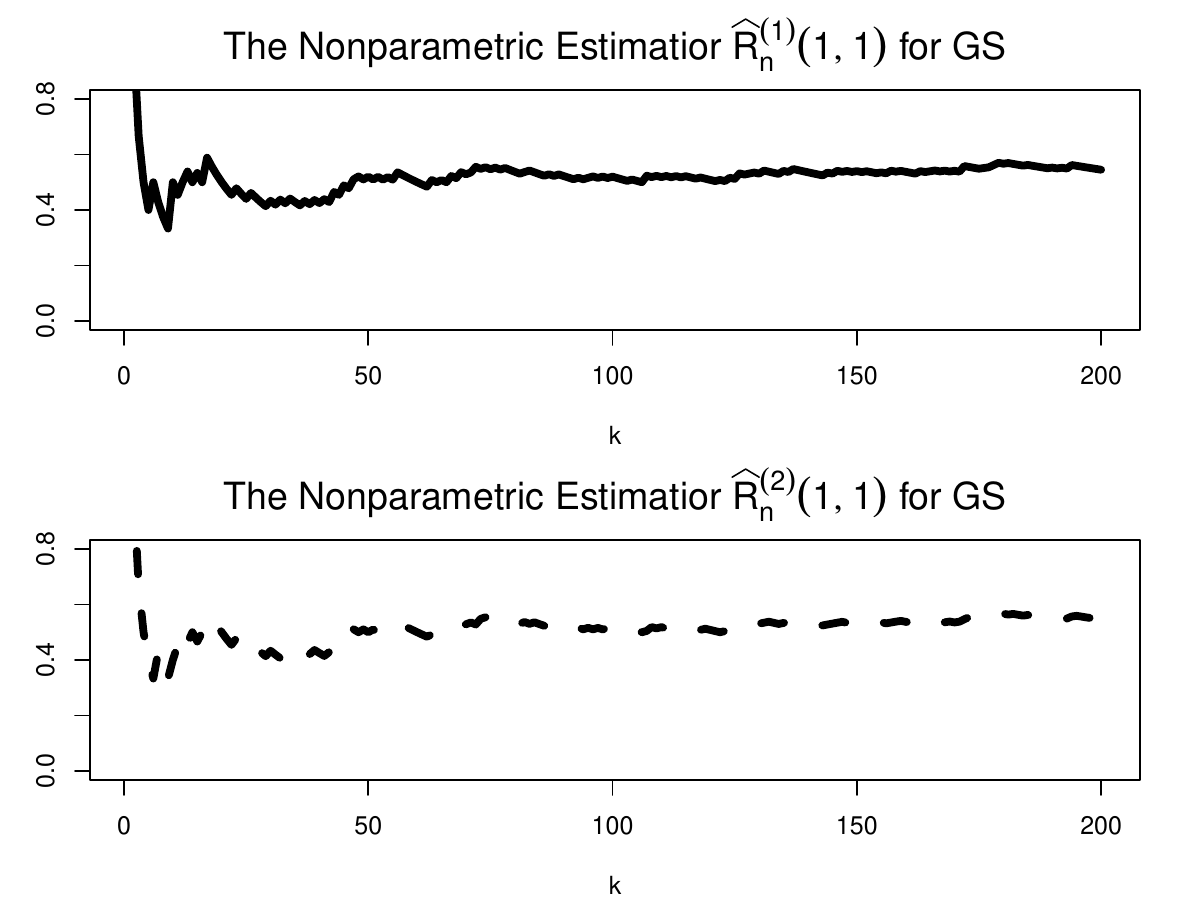}
\end{minipage}
\caption{$\widehat{R}_n^{(1)}(1,1)$ (solid line) and $\widehat{R}_n^{(2)}(1,1)$ (dashed line) for the weekly losses of the four individual institutions: BAC, JPM, WFC, GS and S\&P 500 Index.}
\label{Fig:Rest_week}
\end{figure}

Before implementing the proposed methods, there are some conditions that need to be verified. As a requirement in Proposition \ref{pro:inter_covar_coes}, we first check the assumption $\gamma_1 \in (0,1/2)$, which can be confirmed by the Hill plots in the right panel of Figure \ref{Fig:logloss_week}. Secondly, it is important to check that high values of $Y$ also imply high values of $X$, as well as the equivalent condition for the existence and uniqueness of $\eta_\tau$, that is, $\bP\left( X \ge \var_X(\tau), Y \ge \var_Y(\tau) \right) > (1-\tau)^2$, as judged in Proposition \ref{pro:ext_uni}.
It is because our statistical methodologies are established based on the framework of approximation in \eqref{eq:rutd}. An intuitive empirical evidence is presented in Figure \ref{Fig:scatter_tp_week}, where the left panel plots the losses of the S\&P 500 Index against those of the equity prices of the four institutions, respectively. As we can see, from the upper parts of the plots, the large values of the index are mostly associated with large values of the institutions. The right panel of Figure \ref{Fig:scatter_tp_week} plots the cruves of empirical tail probability $\frac{1}{n}\sum_{i=1}^{n}I\left( X_i \ge \widehat\var_X(\tau), Y_i \ge \widehat\var_Y(\tau) \right)$ and $(1-\tau)^2$ against $\tau$, displaying that empirical tail probability is always larger than the values of $(1-\tau)^2$. This provides powerful supporting evidence for the equivalent condition of existence and uniqueness, that is, $\bP\left( X \ge \var_X(\tau), Y \ge \var_Y(\tau) \right) > (1-\tau)^2$. Furthermore, we check the tail dependence condition by depicting in Figure \ref{Fig:Rest_week} the estimated tail dependence coefficient $\widehat{R}^{(i)}_n(1,1)$ given in \eqref{eq:nonp_R_emp} and \eqref{eq:nonp_R_rank} against $k$. Certainly, all the plots show that the two estimations are positive, which provides strong evidence of right-hand upper asymptotic dependence between the four institutions and the systemic index. Beside of this, we also observe from Figure \ref{Fig:Rest_week} that, both of the two nonparametric estimators \eqref{eq:nonp_R_emp} and \eqref{eq:nonp_R_rank} tend to stabilize around 0.5 when the value of $k$ exceeds 50.

We now implement our proposed extrapolative estimations for \(\covar_{X|Y}(\tau'_n)\) and \(\coes_{X|Y}(\tau'_n)\) for the four different individual institutions, including BAC, JPM, WFC, and GS related to the S\&P 500 Index. 
In the estimation procedures, a key step is to choose a suitable intermediate \( k \), which needs to ensure that both the Hill estimator \(\hat\gamma_1\) and the tail dependence coefficient \(\widehat{R}_n^{(i)}\) perform well. Note also that another constraint on \( k \) lies in \( k = O(n^{\iota}) \) with \(\iota > 2/3\) in Assumption \ref{ass:conditions} (d). As plotted in Figures \ref{Fig:logloss_week} and \ref{Fig:Rest_week}, by balancing the potential estimation bias and variance, a usual practice is to choose \( k \) from the first stable region of the plots. In this case, we choose \( k \in [80,100] \) for BAC, JPM, GS and $k \in [60,80]$ for WFC. To gain stability in all proposed extrapolations, we take the average of the estimations corresponding to those \( k \)-values and regard them as the estimations of \(\covar_{X|Y}(\tau'_n)\) and \(\coes_{X|Y}(\tau'_n)\). Following the setting in the simulation with extreme levels \(\tau'_n = 0.99, 0.995, 0.999\), the results are reported in Table \ref{tab:CRCES_week}. These numbers represent the average weekly losses for a market crisis within a period of nearly 25 years.

Table~\ref{tab:CRCES_week} shows that, as the risk level becomes increasingly extreme, all estimates $\widetilde{\covar}^{(i)}_{X|Y}(\tau'_n)$ and $\widetilde{\coes}^{(i)}_{X|Y}(\tau'_n)$ increase, which is consistent with intuition.  
For $\widetilde{\covar}^{(i)}_{X|Y}(\tau'_n)$, the estimates with $i=1,2$ are almost identical.  
This occurs because both rely on the same adjustment factor $\eta^{(i)}_{1-k/n}$ and, hence, on the tail-dependence function estimator $\widehat{R}_n^{(i)}$; the two nonparametric estimators of $\widehat{R}_n^{(i)}$ behave almost identically, as shown in Figure~\ref{Fig:Rest_week}.  
By contrast, $\widetilde{\covar}^{(3)}_{X|Y}(\tau'_n)$ slightly underestimates the risk.  
Because of the relationships in \eqref{extcoes_covar} and \eqref{eq:extra3_coes}, the CoES estimates $\widetilde{\coes}^{(i)}_{X|Y}(\tau'_n)$, $i=1,2,3$, follow the same pattern as the corresponding CoVaR estimates.  
Unexpectedly, $\widetilde{\coes}^{(4)}_{X|Y}(\tau'_n)$ is markedly lower, and the gap relative to the other three extrapolations widens as $\tau'_n \uparrow 1$.  
Across institutions, the estimates for the four banks are very similar, with BAC producing the largest values, followed by WFC.  
This finding suggests that BAC is the institution most exposed to overall market tail risk, with WFC being the second most affected.

Finally, we adopt a rolling-window scheme to produce dynamic estimates of $\mathrm{CoVaR}_{X|Y}(\tau'_n)$ and $\mathrm{CoES}_{X|Y}(\tau'_n)$.  
The sample is enlarged to daily observations from 3 January 2005 to 9 March 2020, yielding 3,821 observations.  
We implement the procedure with a 1000-day moving window that is re-estimated each trading day.  
For instance, the estimates for 2 December 2016 are based on the 1000 observations from 13 December 2012 to 1 December 2016.  
This long window captures the evolution of tail-risk interdependence over time.
Figures~B4 and~B5 in the supplementary material 
display the rolling paths of $\widetilde{\mathrm{CoVaR}}^{(i)}_{X|Y}(\tau'_n)$ and $\widetilde{\mathrm{CoES}}^{(i)}_{X|Y}(\tau'_n)$, respectively, for the four financial institutions conditional on S\&P 500 Index.  
All extrapolations move in lockstep, corroborating their similar empirical behaviour.  
Three pronounced spikes—around 2008--2009, 2012--2013 and 2019--2020—point to the global financial crisis, the European sovereign-debt turmoil and the COVID-19 shock, confirming that extreme tail-risk co-movements coincide with major market distress events. Another significant observation gives that the impact of COVID-19 pandemic on these institutes is much smaller than that of the former events.

\begin{table}
\centering
\caption{The estimations $\widetilde{\covar}^{(i)}_{X|Y}(\tau'_n)$ ($i=1,2,3$) and $\widetilde{\coes}^{(i)}_{X|Y}(\tau'_n)$ ($i=1,2,3,4$) of the four individual institutions: BAC, JPM, WFC, and GS conditional on S\&P 500 Index.}
\label{tab:CRCES_week}
\setlength{\tabcolsep}{5.5pt}
\begin{tabular}{@{}cccccccc@{}}
\hline\hline
\addlinespace[5pt]
 & $\widetilde{\covar}_{X|Y}^{(1)}$ & $\widetilde{\covar}_{X|Y}^{(2)}$ & $\widetilde{\covar}_{X|Y}^{(3)}$ & $\widetilde{\coes}_{X|Y}^{(1)}$ & $\widetilde{\coes}_{X|Y}^{(2)}$ & $\widetilde{\coes}_{X|Y}^{(3)}$ & $\widetilde{\coes}_{X|Y}^{(4)}$ \\[2pt]
\midrule
& \multicolumn{7}{c}{$\tau'_n = 0.99$} \\[2pt]
\midrule
BAC  & 1.47362 & 1.41626 & 1.39725 & 2.79732 & 2.68844 & 2.65198 & 2.54093 \\[3pt]
JPM  & 0.77394 & 0.75240 & 0.74094 & 1.27592 & 1.24060 & 1.22093 & 1.25080 \\[5pt]
WFC  & 1.24490 & 1.16362 & 1.13763 & 2.28384 & 2.13490 & 2.08690 & 1.62535 \\[5pt]
GS   & 0.67250 & 0.65645 & 0.63852 & 1.08896 & 1.06298 & 1.03364 & 1.03046 \\[5pt]
\midrule
& \multicolumn{7}{c}{$\tau'_n = 0.995$} \\[2pt]
\midrule
BAC  & 2.83964 & 2.72911 & 2.69220 & 5.39078 & 5.18095 & 5.11015 & 4.89578 \\[5pt]
JPM  & 1.33462 & 1.29765 & 1.27722 & 2.20214 & 2.14145 & 2.10638 & 2.15684 \\[5pt]
WFC  & 2.33866 & 2.18611 & 2.13703 & 4.29119 & 4.01159 & 3.92097 & 3.05293 \\[5pt]
GS   & 1.14260 & 1.11534 & 1.08460 & 1.85056 & 1.80640 & 1.75611 & 1.75059  \\[5pt]
\midrule
& \multicolumn{7}{c}{$\tau'_n = 0.999$} \\[2pt]
\midrule
BAC  & 13.02616 & 12.51910 & 12.34678 & 24.73299 & 23.77020 & 23.43968 & 22.45217 \\[5pt]
JPM  & 4.74285 & 4.61281 & 4.53451 & 7.84162 & 7.62778 & 7.49319 & 7.66354 \\[5pt]
WFC  & 10.11625 & 9.45768 & 9.24302 & 18.56991 & 17.36233 & 16.96623 & 13.20117 \\[5pt]
GS   & 3.91469 & 3.82126 & 3.71382 & 6.34312 & 6.19169 & 6.01595 & 5.99598  \\[5pt]
\hline\hline
\end{tabular}
\end{table}

\section{Conclusion}

In this paper we study the estimations of CoVaR and CoES at extreme quantile levels under a regime of upper-tail dependence. 
In the literature, \cite{Nolde2022} proposed a semi-parametric estimator of extreme CoVaR based on an adjusted factor~$\eta_{\tau}$, yet the accompanying asymptotic theory remains incomplete. We fill this gap by introducing two nonparametric estimators of the adjustment factor that exploit tail-dependence modelling. The resulting extrapolation formulae exploit the limiting relationship between CoVaR and CoES and come with full asymptotic justification. Simulation study demonstrate that the proposed procedures are robust and perform well in finite samples. 


\section*{Supplementary Material}

\begin{description}
\item[Title:] {\bf Supplementary Material for ``Nonparametric Inference for Extreme CoVaR and CoES"} \newline
This document includes all theoretical proofs and some extra experimental figures. \end{description}


\newpage

\begin{thebibliography}{99}

    \small

    \bibitem[\protect\citeauthoryear{Acharya et al.}{2017}]{Acharya2017}
    Acharya, V.V., Pedersen, L.H., Philippon, T., and Richardson, M. 2017. Measuring systemic risk. \emph{The review of financial studies}, 30(1), pp.2-47.


    \bibitem[\protect\citeauthoryear{Adrian and Brunnermeier}{2016}]{Adrian2016}
    Adrian, T., and Brunnermeier, M.K. 2016. CoVaR. \emph{The American Economic Review}, 106(7), pp.1705-1741.

    \bibitem[\protect\citeauthoryear{Beirlant et al.}{2006}]{Beirlant2006}
    Beirlant, J., Goegebeur, Y., Segers, J., and Teugels, J.L. 2006. Statistics of Extremes: Theory and Applications. \emph{John Wiley $\&$ Sons}.



    \bibitem[\protect\citeauthoryear{Cai et al.}{2015}]{Cai2015}
    Cai, J.J., Einmahl, J.H., de Haan, L., and Zhou, C. 2015. Estimation of the marginal expected shortfall: the mean when a related variable is extreme. \emph{Journal of the Royal Statistical Society Series B: Statistical Methodology}, 77(2), pp.417-442.


    \bibitem[\protect\citeauthoryear{Capponi and Rubtsov}{2022}]{Capponi2022}
    Capponi, A., and Rubtsov, A. 2022. Systemic risk-driven portfolio selection. \emph{Operations Research}, 70(3), pp.1598-1612.


    \bibitem[\protect\citeauthoryear{de Haan and Ferreira}{2006}]{dehaan2006}
    de Haan, L., and Ferreira, A. 2006. Extreme Value Theory: An Introduction. \emph{Springer}.

    \bibitem[\protect\citeauthoryear{Drees and Huang}{1998}]{Drees1998}
    Drees, H., and Huang, X. 1998. Best attainable rates of convergence for estimators of the stable tail dependence function. \emph{Journal of Multivariate Analysis}, 64(1), pp.25-46.

    \bibitem[\protect\citeauthoryear{Einmahl et al.}{2006}]{Einmahl2006}
    Einmahl, J.H., de Haan, L., and Li, D. 2006. Weighted approximations of tail copula processes with application to testing the bivariate extreme value condition. \emph{The Annals of Statistics}, 34(4), pp.1987-2014.

    \bibitem[\protect\citeauthoryear{Einmahl et al.}{2001}]{Einmahl2001}
    Einmahl, J.H., Piterbarg, V.I., and De Haan, L. 2001. Nonparametric estimation of the spectral measure of an extreme value distribution. \emph{The Annals of Statistics}, 29(5), pp.1401-1423.

    \bibitem[\protect\citeauthoryear{Einmahl et al.}{2008}]{Einmahl2008}
    Einmahl, J.H., Krajina, A., and Segers, J. 2008. A method of moments estimator of tail dependence. \emph{Bernoulli}, 14(4), pp.1003-1026.

    \bibitem[\protect\citeauthoryear{Einmahl et al.}{2012}]{Einmahl2012}
    Einmahl, J.H., Krajina, A., and Segers, J. 2012. An M-estimator for tail dependence in arbitrary dimensions. \emph{The Annals of Statistics}, 40(3), pp.1764-1793.

    \bibitem[\protect\citeauthoryear{Fissler and Hoga}{2024}]{Fissler2024}
    Fissler, T., and Hoga, Y. 2024. Backtesting systemic risk forecasts using multi-objective elicitability. \emph{Journal of Business $\&$ Economic Statistics}, 42(2), pp.485-498.

    \bibitem[\protect\citeauthoryear{Fougères et al.}{2015}]{Fougères2015}
    Fougères, A.L., De Haan, L., and Mercadier, C. 2015. Bias correction in multivariate extremes. \emph{The Annals of Statistics}, 43(2), pp.903-934.

    \bibitem[\protect\citeauthoryear{Francq and Zakoïan}{2025}]{Francq2025}
    Francq, C., and Zakoïan, J.M. 2025. Inference on dynamic systemic risk measures. \emph{Journal of Econometrics}, 247, pp.105936.

    \bibitem[\protect\citeauthoryear{Girardi and Ergün}{2013}]{Girardi2013}
    Girardi, G., and Ergün, A.T. 2013. Systemic risk measurement: Multivariate GARCH estimation of CoVaR. \emph{Journal of Banking $\&$ Finance}, 37(8), pp.3169-3180.


    \bibitem[\protect\citeauthoryear{Hill}{1975}]{Hill1975}
    Hill, B.M. 1975. A simple general approach to inference about the tail of a distribution. \emph{The Annals of Statistics}, 3(5), pp.1163-1174.

    \bibitem[\protect\citeauthoryear{Huang and Uryasev}{2018}]{Huang2018}
    Huang, W.Q., and Uryasev, S.P. 2018. The CoCVaR approach: Systemic risk contribution measurement. \emph{Journal of Risk}, 20(4).


    \bibitem[\protect\citeauthoryear{Kozt et al.}{2000}]{Kozt2000}
    Kozt, S., Balakrishnan, N., and Norman, L.J. 2000. Continuousmultivariate distributions: Models and Applications. \emph{Wiley}, New York.


    \bibitem[\protect\citeauthoryear{Nolde and Zhang}{2020}]{Nolde2020}
    Nolde, N., and Zhang, J. 2020. Conditional extremes in asymmetric financial markets. \emph{Journal of Business $\&$ Economic Statistics}, 38(1), pp.201-213.

    \bibitem[\protect\citeauthoryear{Nolde et al.}{2022}]{Nolde2022}
    Nolde, N., Zhou, C., and Zhou, M. 2022. An extreme value approach to CoVaR estimation. \emph{Available at arXiv:2201.00892} \url{https://doi.org/10.48550/arXiv.2201.00892}.




\end{thebibliography}
\end{document}